\documentclass[lettersize]{emulateapj}

\usepackage{natbib}
\usepackage{amsmath}
\usepackage{multirow}
\usepackage{float}
\usepackage{textcomp}

\def\be{\begin{equation}}
\def\ee{\end{equation}}
\def\bea{\begin{eqnarray}}
\def\eea{\end{eqnarray}}



\def \logTd6 {\hbox{log$( T/6 \kev)$} }

\def\myputfigure#1#2#3#4#5%
{\vskip#5pt\makebox[0pt]{\hskip#2in
\includegraphics[width=#3\textwidth]{#1}}\vskip#4pt\hfill}

\def \etal      {et~al.\ }

\def \kev       {{\rm\ keV}}

\newcommand{\nhi}{\ensuremath{N_{\rm HI}}}

\newcommand{\mpc}{\ensuremath{\, h^{-1}\,\mathrm{Mpc} }}

\newcommand{\kms}{\, \mathrm{km \; s^{-1}}}
\newcommand{\persqcm}{\,\mathrm{cm^{-2}}}
\newcommand{\snr}{\ensuremath{\mathrm{S/N}}}
\newcommand{\lya}{Ly$\alpha$}
\newcommand{\lyb}{Ly$\beta$}

\newcommand{\heii}{\ion{He}{2}}

\newcommand{\scien}[2]{#1  \times 10^{#2}} 
\newcommand{\beq}{\begin{equation}}
\newcommand{\eeq}{\end{equation}}
\newcommand{\bc}{\begin{center}}
\newcommand{\ec}{\end{center}}
\newcommand{\bfig}{\begin{figure}}
\newcommand{\efig}{\end{figure}}

\newcommand{\fmean}{\ensuremath{\langle F \rangle}}
\newcommand{\fcont}{\ensuremath{\langle F \rangle_\mathrm{cont}}}
\newcommand{\fmeanlya}{\ensuremath{\langle F \rangle_\mathrm{Ly\alpha} }}
\newcommand{\feff}{\ensuremath{\langle F \rangle_{\rm eff} }}

\newcommand{\lambrest}{\ensuremath{\lambda_\mathrm{rest}}}

\newcommand{\zq}{\ensuremath{z_\mathrm{qso}}}
\newcommand{\zabs}{\ensuremath{z_\mathrm{abs}}}
\newcommand{\zav}{\ensuremath{\langle z \rangle}}

\newcommand{\ang}{\ensuremath{\mathrm{\AA}}}

\newcommand{\sigcont}{\ensuremath{\sigma_\mathrm{cont}}}

\newcommand{\tref}{\texttt{T\textunderscore REF}}
\newcommand{\tcold}{\texttt{T\textunderscore COLD}}
\newcommand{\thot}{\texttt{T\textunderscore HOT}}

\def \etal {et~al.}

\slugcomment{Accepted for publication by ApJ}
\shorttitle{\lya\ Forest Transmission PDF from BOSS}
\shortauthors{Lee \etal}

\begin{document}

\title{IGM Constraints from the SDSS-III/BOSS DR9 \\ \lya\ Forest Transmission Probability Distribution Function}
\author{Khee-Gan Lee\altaffilmark{1,2}, 
Joseph F. Hennawi\altaffilmark{1},  
David N. Spergel\altaffilmark{2}, 
David H.\ Weinberg\altaffilmark{3},
David W.\ Hogg\altaffilmark{4}, \\
Matteo Viel\altaffilmark{5,6}, 
James S.\ Bolton\altaffilmark{7}, 
Stephen Bailey\altaffilmark{8}, 
Matthew M.\ Pieri\altaffilmark{9},
William~Carithers\altaffilmark{8}, \\
David J.\ Schlegel\altaffilmark{8}, 
Britt Lundgren\altaffilmark{10},
Nathalie~Palanque-Delabrouille\altaffilmark{11},
Nao~Suzuki\altaffilmark{12}, \\
Donald P. Schneider\altaffilmark{13,14},
Christophe~Y\`eche\altaffilmark{11} }
\altaffiltext{1}{Max Planck Institute for Astronomy, K\"{o}nigstuhl 17, D-69117 Heidelberg, Germany}
\altaffiltext{2}{Department of Astrophysical Sciences, Princeton University, Princeton, New Jersey 08544, USA}
\altaffiltext{3}{Department of Astronomy and Center for Cosmology and Astro-Particle Physics, Ohio State University, Columbus, OH 43210, USA}
\altaffiltext{4}{Center for Cosmology and Particle Physics, New York University, 4 Washington Place, Meyer Hall of Physics, New York, NY 10003, USA}
\altaffiltext{5}{INAF, Osservatorio Astronomico di Trieste, Via G. B. Tiepolo 11, 34131 Trieste, Italy}
\altaffiltext{6}{INFN/National Institute for Nuclear Physics, Via Valerio 2, I-34127 Trieste, Italy}
\altaffiltext{7}{School of Physics and Astronomy, University of Nottingham, University Park, Nottingham NG7 2RD, UK}

\altaffiltext{8}{E.O. Lawrence Berkeley National Lab, 1 Cyclotron Rd., Berkeley, CA, 94720, USA}
\altaffiltext{9}{Institute of Cosmology \& Gravitation, University of Portsmouth, Dennis Sciama Building, Portsmouth PO1 3FX, UK}
\altaffiltext{10}{Department of Astronomy, University of Wisconsin, Madison, WI 53706, USA}
\altaffiltext{11}{CEA, Centre de Saclay, Irfu/SPP, F-91191 Gif-sur-Yvette, France} 
\altaffiltext{12}{Kavli Institute for the Physics and Mathematics of the Universe (IPMU), The University of Tokyo, Kashiwano-ha 5-1-5, Kashiwa-shi, Chiba, Japan}
\altaffiltext{13}{Department of Astronomy and Astrophysics, The Pennsylvania State University, University Park, PA 16802, USA}
\altaffiltext{14}{Institute for Gravitation and the Cosmos, The Pennsylvania State University, University Park, PA 16802, USA}
\email{lee@mpia.de}

\begin{abstract}
The \lya\ forest transmission probability distribution function (PDF) is an established probe
of the intergalactic medium (IGM) astrophysics, especially the temperature-density relationship of the IGM.
We measure the transmission PDF from 3393 Baryon Oscillations
Spectroscopic Survey (BOSS) quasars from SDSS Data Release 9, and compare with mock spectra
that include careful modeling of the noise, continuum, and astrophysical uncertainties.
The BOSS transmission PDFs, measured at $\langle z \rangle = [2.3,2.6,3.0]$, 
are compared with PDFs created from mock spectra drawn from a suite of hydrodynamical simulations
that sample the IGM temperature-density relationship, $\gamma$, and temperature at mean-density, $T_0$, 
where $T(\Delta) = T_0 \Delta^{\gamma-1}$. 
We find that a significant population of partial Lyman-limit systems with a column-density distribution slope of 
$\beta_\mathrm{pLLS} \sim -2$ are required to explain the data at the low-transmission end of transmission PDF, 
while uncertainties in the mean \lya\ forest transmission affect the high-transmission end.
After modelling the LLSs and marginalizing over 
mean-transmission uncertainties, we find that $\gamma=1.6$ best describes the data over our
entire redshift range, although constraints on $T_0$ are affected by systematic uncertainties. 
 {Within our model framework, isothermal or inverted temperature-density relationships 
 ($\gamma \leq 1$) are disfavored at a significance of over 4$\sigma$, although this could 
 be somewhat weakened by cosmological and astrophysical uncertainties that we did not model.}
\end{abstract}

\keywords{intergalactic medium --- quasars: emission lines --- 
quasars: absorption lines --- methods: data analysis}


\section{Introduction}
Remarkably soon after the discovery of the first high-redshift ($\zq \gtrsim 2$) quasars \citep{schmidt:1965},
\citet{gunn:1965} realized
that the amount of resonant Lyman-$\alpha$ (\lya) scattering off neutral hydrogen structures
observed in the spectra of these quasars could be used to constrain the state of the
inter-galactic medium (IGM) at high-redshifts:
they deduced that the hydrogen in the inter-galactic medium had to be
highly photo-ionized (neutral fractions of $n_{\rm HI}/n_{\rm H} < 10^{-4}$) and hot 
(temperatures, $T > 10^4\; \mathrm{K}$). 

\citet{lynds:1971} then discovered that this \lya\ absorption could be separated into discrete absorption
lines, i.e.\ the \lya\ ``forest''. Over the next two decades, it was recognized that
the individual \lya\ forest lines have Voigt absorption profiles corresponding to Doppler-broadened
systems with $T \sim \scien{1-3}{4}\; \mathrm{K}$ 
\citep[see, e.g.,][]{rauch:1992,ricotti:2000, schaye:2000,mcdonald:2001,tytler:2004,
lidz:2010,becker:2011} and neutral column densities
of $N \sim 10^{13}- 10^{17} \mathrm{cm^{-2}}$ \citep{petitjean:1993,penton:2000,janknecht:2006,rudie:2013}, and increasingly
precise measurements of mean \lya\ forest transmission have been carried out \citep{theuns:2002,bernardi:2003,faucher-giguere:2008a,becker:2013}. 
However, the exact physical nature of these absorbers was unclear for many years
\citep[see][for a historical review of the field]{rauch:1998}. 

Beginning in the 1990s, detailed hydrodynamical simulations of the intergalactic medium
 led to the current physical picture of the \lya\ forest arising from
baryons in the IGM which trace fluctuations in the dark matter field induced by gravitational collapse, 
in ionization balance with a uniform
 ultraviolet ionizing background
 \citep[see, e.g.,][]{cen:1994,miralda-escude:1996,croft:1998,dave:1999,theuns:1998}. 
 A physically-motivated analytic description of this picture is the fluctuating Gunn-Peterson 
 approximation \citep[FGPA, ][]{croft:1998,hui:1997}, 
 in which the \lya\ optical depth, $\tau$, scales with
  underlying matter density, $\rho$, through a polynomial relationship:
 \beq \label{eq:fgpa}
 \tau \propto \frac{T^{-0.7}}{\Gamma} \Delta^2 
    \propto \frac{T_0^ {-0.7}}{\Gamma}  \Delta^{2- 0.7(\gamma - 1)},
 \eeq
 where $\Gamma$ is the background photoionization rate, and $\Delta\equiv \rho/\langle \rho \rangle $ 
 is the matter density relative to the mean density of the universe at the given epoch. 
 In the second proportionality above, we have made the assumption that the local temperature of the gas has
 a polynomial relationship with the local density, 
 \beq
 T = T_0 \Delta^{\gamma-1},
 \eeq 
 where
 $T_0$ is the gas temperature at mean-density and $\gamma$ parametrizes the
 temperature-density relation, which encodes the thermal history of the IGM (e.g., \citealt{hui:1997a}, \citealt{schaye:1999}, 
 \citealt{ricotti:2000},
 \citealt{mcdonald:2001}, \citealt{hui:2003}; see \citealt{meiksin:2009} for a detailed overview on the relevant physics).

{Over the the past decade-and-a-half,} the 2000-2008 Sloan Digital Sky Survey 
\citep[SDSS-I and -II,][\url{http://www.sdss.org}]{york:2000, stoughton:2002}
spectroscopic data has represented a dramatic improvement in the statistical power available to 
\lya\ forest studies: \citet{mcdonald:2006} measured the 1-dimensional \lya\ forest transmission power spectrum 
from $\approx 3000$ SDSS quasar sightlines. This measurement
was used to place significant constraints on cosmological parameters and large-scale structure
\citep[see, e.g.,][]{mcdonald:2005a,seljak:2005,viel:2006}.
 
The \citet{mcdonald:2006} quasar sample, which in its time represented a $\sim 100$ 
increase in sample size over previous data sets, 
is superseded by the Baryon Oscillations Sky Survey 
\citep[BOSS, part of SDSS-III;][]{eisenstein:2011, dawson:2013} quasar survey. 
This spectroscopic survey, which operated between fall 2009 and spring 2014, 
is aimed at taking spectra of $\sim 150,000$ $\zq \gtrsim 2.2$ quasars \citep{dawson:2013} 
with the goal of constraining dark
energy at $z > 2$ using transverse correlations of \lya\ forest absorption \citep[see, e.g.,][]{slosar:2011} 
to measure the 
baryon acoustic oscillation (BAO) scale\footnote{There is also a simultaneous
effort to observe $\sim 1.5$ million luminous red galaxies, to measure the BAO at $z \sim 0.5$.
See, e.g., \citet{anderson:2014}. }.
At time of writing, the full BOSS survey is complete, with $\sim 170,000$ high-redshift quasars observed,
although 
this paper is based on the earlier sample of $\sim 50,000$ BOSS 
quasars from SDSS Data Release 9 
\citep[DR9][]{ahn:2012, paris:2012, lee:2013}.

The quality of the individual BOSS \lya\ forest spectra might appear at first glance inadequate for
{studying the astrophysics of the IGM, that have to-date been carried out largely with 
high-resolution, high-S/N spectra}: the typical BOSS
spectrum has
$\snr \sim 2$ per pixel\footnote{All spectral signal-to-noise ratios quoted in this paper 
are per $69\;\kms$ SDSS/BOSS pixel unless noted otherwise} , since the BAO analysis
is optimized with large numbers of low signal-to-noise-ratio sightlines, densely-sampled on the sky
\citep{mcdonald:2007,mcquinn:2011}.
It is therefore interesting to ask whether it is possible to model the various instrumental and astrophysical
effects seen in
the BOSS \lya\ forest spectra, to sufficient accuracy level to exploit the unprecedented statistical
power.

In this paper, we will measure the probability distribution function (PDF) of the \lya\ forest transmission, 
$F \equiv \exp(-\tau)$, from BOSS.
This one-point statistic, which was first studied by \citet{jenkins:1991}, is sensitive to astrophysical 
parameters such as the amplitude of matter fluctuations and the thermal history of the IGM. 
However, the transmission\footnote{The \lya\ forest transmitted flux fraction 
is sometimes also referred to as 'flux' in the literature; but we do however use the variable $F$ to refer to this quantity.} 
PDF is also highly sensitive to 
effects such as pixel noise level, resolution of the spectra, and 
systematic uncertainties in the placement of the quasar
continuum level, especially in moderate resolution spectra such as SDSS or BOSS.
\citet{desjacques:2007} studied the transmission PDF from a sample of $\sim 3500$ \lya\ forest spectra from 
SDSS Data Release 3 \citep{abazajian:2005}. Using mock spectra generated from a
 log-normal model of the \lya\ forest with parameters tuned to reproduce high-resolution,
high-\snr\ spectra, they fitted for the estimated pipeline noise level
and continuum-fitting errors in the SDSS spectra. 
They concluded that the noise levels reported by the SDSS pipeline were underestimated by
$\sim 10\%$, consistent with the findings of \citet{mcdonald:2006}. They also found that
the quasar
continuum-level was systematically lower by $\sim 10\%$ in comparison with a power-law
extrapolated from redwards of the quasar \lya\ line, with a RMS variance of $\sim 20\%$,
although certain aspects of their study, e.g., the noise modelling and quasar 
continuum model, were rather crude.

We intend to take an approach distrinct from that of \citet{desjacques:2007}: instead of 
treating the noise and continuum as free parameters, we will attempt to measure
the BOSS \lya\ forest transmission PDF using a rigorous treatment of the noise and continuum-fitting,
and then adopt a ``forward-modeling'' approach of trying to model the various 
instrumental effects
as accurately as possible in mock spectra generated from detailed hydrodynamical simulations. 
Using the raw individual exposures and calibration data from BOSS, we will first
implement a novel probabilistic method for co-adding the exposures, which will
yield more accurate noise estimates as well as enable self-consistent noise modelling in mock spectra. 
Similarly, we will use a new method for continuum estimation called
mean-flux regulated/principal component analysis \citep[MF-PCA;][]{lee:2012a}. 
This technique provides unprecedented continuum accuracy
for noisy \lya\ forest spectra: $<10\%$ RMS errors for $\snr \sim 2$ and
$<5\%$ RMS errors for $\snr \gtrsim 5 $ spectra. 

On the modeling side, we will use the detailed hydrodynamical IGM simulations 
of \citet{viel:2013} as a basis. 
The mock spectra are then smoothed to BOSS resolution, have Lyman-limit systems (LLS) and metal contamination added, 
followed by the
introduction of pixel noise based on our improved noise estimates.
We will then self-consistently introduce continuum errors by applying our continuum-estimation 
procedure on the mock spectra.

With the increase in statistical power from the sheer number of BOSS spectra, 
and our improved modeling of the noise and continuum, we expect to significantly
reduce the errors on the measured transmission PDF in comparison with \citet{desjacques:2007}.
This should enable us to place independent
constraints on the shape of the underlying transmission PDF, and the thermal history
of IGM 
as parametrized by the power-law temperature-density relation, $\gamma$ and $T_0$.

The IGM temperature-density relationship is a topic of recent interest, as \citet{bolton:2008} and \citet{viel:2009} 
have found evidence of an inverted 
temperature-density relation, $\gamma < 1$, implying that voids are hotter than 
overdensities,  the IGM at $z \sim 2-3$
from the transmission PDF from high-resolution, 
high-\snr\ \lya\ forest spectra \citep{kim:2007}. 
This result is in contrast with theoretical expectations of $\gamma \approx 1.6$
 \citep{miralda-escude:1994,hui:1997a, theuns:1998,hui:2003},
which arises from the balance between adiabatic cooling in the lower-density IGM and
photoheating in the higher-density regions.
Even inhomogeneous \heii\ reionization, which is
expected to flatten the IGM temperature-density relation
\citep[see, e.g.,][]{furlanetto:2008,bolton:2009,mcquinn:2009}, 
is insufficient to account for
the extremely low values of $\gamma \sim 0.5$ estimated by the aforementioned authors
\citep[although inversions could occur at higher densites, see, e.g.,][]{meiksin:2012}.

Indeed, earlier papers studying the temperature-density relationship using either the 
transmission PDF \citep{mcdonald:2001} or by measuring the Doppler parameters and 
hydrogen column densities of individual forest absorbers 
\citep[the so-called $b-N_\mathrm{HI}$ relation, e.g.,][]{schaye:1999,ricotti:2000, rudie:2012}
have found no evidence of an inverted $\gamma$.
In recent years, the decay of 
blazar gamma rays via plasma instabilities (\citealt{broderick:2012}, \citealt{chang:2012}; although see \citealt{sironi:2014}) 
has been invoked as a possible mechanism to 
 supply the heat necessary to flatten $\gamma$ to the 
 observed levels \citep{puchwein:2012}. 
 
It would be desirable to perform an independent re-analysis of high-resolution data
taking into account continuum-fitting bias \citep{lee:2012}, to place these claims on a firmer
footing. However, \citet{lee:2011} have argued that the complete SDSS DR7 \citep{abazajian:2009} \lya\ forest data set 
could have sufficient statistical power to place interesting constraints on $\gamma$, even assuming
continuum-fitting errors at the $\sim 10\%$ RMS level. 
Therefore, with the current BOSS data, we hope to model noise and resolution, as well 
as astrophysical systematics, at a sufficient precision to place interesting constraints
 on the IGM thermal history. 

This paper is organized as follows: we first give a broad overview of the BOSS
\lya\ forest data set, followed by our measurement of the BOSS transmission PDF
with detailed descriptions of our method of combining multiple raw exposures and
continuum estimation.
We then discuss how we include various instrumental and astrophysical effects into our modeling of the 
transmission PDF starting with hydrodynamical simulations.
The model transmission PDF is then compared with the observed PDF to obtain constraints on the thermal parameters governing the IGM.

\section{Data} \label{sec:data}

\subsection{Summary of BOSS}

BOSS \citep{dawson:2013} is part of SDSS-III (\citealt{eisenstein:2011}; the other surveys are
SEGUE-2, MARVELS, and APOGEE). The primary goal
of the survey is to carry out precision baryon acoustic oscillations at $z \sim 0.5$ and $z \sim 2.5$, from the 
luminous red galaxy distribution and \lya\ forest absorption field, respectively 
\citep[see, e.g.,][]{anderson:2014,busca:2013, slosar:2013}.
Its eventual goal is to obtain spectra of $\sim 1.5$ million luminous red galaxies and $\sim 170,000$ $z > 2.15$ quasars over
4.5 years of operation. 

BOSS is conducted on upgraded versions of the twin SDSS spectrographs \citep{smee:2013} mounted
on the 2.5m Sloan telescope \citep{gunn:2006} at Apache Point Observatory, New Mexico.
One thousand 
optical fibers mounted on a plug-plate at the focal plane (spanning a $3\,^{\circ}$ field of view) feed the incoming flux to the 
two identical spectrographs, of which 160-200 fibers per plate are allocated to quasar targets \citep[see][for a detailed description of the
quasar target selection]{ross:2012,bovy:2011}. 
Both spectrographs split the light into a blue and red camera that cover $3610 - 10140\;\ang$, with 
the dichroic overlap region occurring at around $6000\;\ang$. The resolving power $R \equiv \lambda / \Delta \lambda$
ranges from 1300 at the blue end to the 2600 at the red end. 

Each plate is observed for sufficiently long to achieve the S/N requirements set by the survey goals;
typically, 5 individual exposures of 15 minutes are taken.The data are processed, calibrated, and combined into co-added spectra by
the ``idlspec2d'' pipeline, followed by a pipeline which operates on the 1D spectra to classify objects and assign redshifts \citep{bolton:2012}. 
However, as described later in this paper, 
we will generate our own co-added spectra from the individual exposures and other intermediate data products.

\subsection{Data Cuts} \label{sec:datacuts}

In this paper we use data from the publicly-available SDSS Data Release 9 \citep[DR9][]{ahn:2012}.
This includes 87,822 quasars at all redshifts, that have been confirmed by visual inspection as described
in \citet{paris:2012}. In \citet{lee:2013}, we have defined a further subset of 54,468 quasars with $\zq \geq 2.15$ 
that are suitable for \lya\ forest analysis, and have provided {in individual FITS files for each quasar} 
various products such as sky masks, masks for
damped \lya\ absorbers (DLAs), 
noise corrections, 
and continua; these are designed to ameliorate systematics in the BOSS spectra and aid in \lya\ forest analysis
{(see Table 1 in \citealt{lee:2013} for a full listing)}.
While we use this \citet{lee:2013} catalog as a starting point, in this paper we will generate our own custom co-added spectra and noise
estimates.

The typical signal-to-noise ratio of the BOSS \lya\ forest quasars is low: $\langle \snr \rangle \approx 2$ per pixel within the \lya\ forest;
this criterion is driven by a strategy to ensure a large number of sightlines over a large area in order to optimize the 3D \lya\ forest BAO analysis.
\citep{mcdonald:2007, mcquinn:2011}, rather than increasing the S/N in individual spectra.
However, for our analysis we wish to select a subset of BOSS \lya\ forest sightlines with reasonably
high S/N in order to 
reduce the sensitivity of our PDF measurement to inaccuracies in our modeling
of the noise and continuum of the BOSS spectra. 
We therefore make a cut on S/N, including only sightlines that have a median $\langle \snr \rangle \geq 6$ per 
pixel within the \lya\ 
forest\footnote{Defined as the $1041-1185\;\ang$ region in the quasar restframe}, defined with respect
to the pipeline noise estimate \citep[see][]{lee:2013} --- this selects only $\sim 10\%$ of the spectra with the highest S/N.
 The $1041-1185\;\ang$ \lya\ forest region of
each quasar must also include at least 30 pixels ($\Delta v = 2071\;\kms$) within one of our 
absorption redshift bins of $\langle z \rangle = 2.3$, $\langle z \rangle = 2.6$, and $\langle z \rangle =3.0$, with bin widths
of $\Delta z = 0.3$ (see \S~\ref{sec:bosspdf}).

We discard spectra with identified DLAs in the sightline, as listed in the `DLA Concordance Catalog' 
used in the \citet{lee:2013} sample. 
This DLA catalog (W.\ Carithers 2014, in prep.) includes objects with column densities $N_\mathrm{HI} > 10^{20}\persqcm$; 
however, the completeness of this catalog is uncertain below $N_\mathrm{HI} = 10^{20.3}\persqcm$.
We therefore discard only sightlines containing DLAs with $N_\mathrm{HI} \geq 10^{20.3}\persqcm$, and take into account 
lower column-density absorbers in our subsequent modelling of mock spectra.
At the relatively high S/N that we will work with (see below), 
the detection efficiency of DLAs is essentially 100\% \citep[see, e.g.,][]{prochaska:2005, noterdaeme:2012} 
and thus we expect our rejection of 
$\nhi \geq 10^{20.3}\persqcm$ DLAs to be quite thorough.

Measurements of the \lya\ forest transmission PDF are known to be sensitive to the continuum estimate \citep{lee:2012}, 
but in this paper we use an automated continuum-fitter, MF-PCA \citep{lee:2012}, that is less susceptible to biases introduced by manual continuum
estimation. Moreover, unlike the laborious process of manually-fitting continua on high-resolution spectra, the automated
continuum estimation can be used to explore various biases in continuum estimation.
For this purpose, we will use the same MF-PCA continuum estimation used in \citet{lee:2013}, 
albeit with minor modifications as described in \S~\ref{sec:cont}. 
We select only quasars that appear to be well-described
by the continuum basis templates, based on the goodness-of-fit to the quasar spectrum redwards of \lya. 
{This is flagged by the variable} \verb|CONT_FLAG|$=1$ {as listed
in the \citet{lee:2013} catalog (see Table 3 in that paper)}. 
Broad Absorption Line (BAL) quasars, which are difficult to estimate continua due to broad intrinsic
absorption troughs, have already been discarded from the \citet{lee:2013} sample.

 \begin{figure}
 \epsscale{1.2}
 \plotone{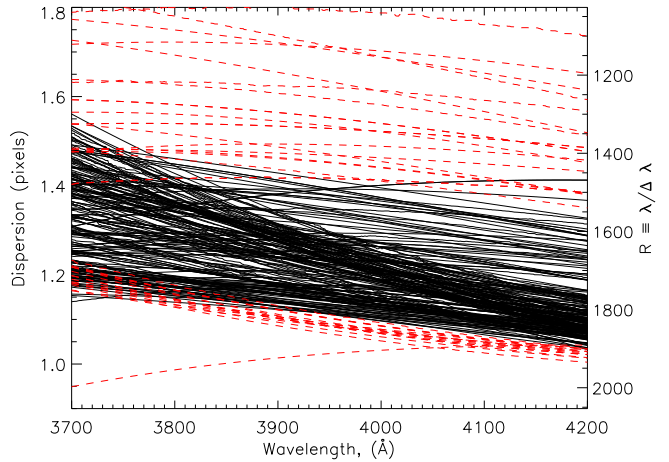}
 \caption{\label{fig:disp}
Wavelength dispersions, $\sigma_{\rm disp}$, for 236 
BOSS quasar spectra randomly-selected from the $\zav = 2.3$, $6 < \snr < 8$ PDF bin.
The ordinate axis on the right shows the equivalent spectral resolution, 
$R \equiv \lambda/\Delta \lambda$.
The dashed-red lines are objects that have been discarded from the analysis on account
of being outliers in spectral dispersion.
 }
 \end{figure}

Another consideration is that the shape of the transmission PDF is affected by the resolution of the spectrum, especially since the BOSS spectrographs
do not resolve the \lya\ forest. 
The exact spectral resolution of a BOSS spectrum at a given wavelength varies 
as a function of both observing conditions and row position on the BOSS CCDs. 
The BOSS pipeline reports the wavelength dispersion at each pixel, $\sigma_{\rm disp}$, 
in units of the co-added wavelength
pixel size ({binned such that} $\ln(10)\; \Delta(\lambda)/\lambda = 10^{-4}$). 
This is related to the resolving power by $R \approx ( 2.35 \times \scien{1}{-4} \ln 10\, \sigma_{\rm disp})^{-1}$.
\citet{palanque-delabrouille:2013a} have recently found, using their own analysis of the width of the arc-lamp lines and bright sky emission lines, 
that the spectral dispersion reported by the pipeline had a bias that depended
on the CCD row and increased with wavelength, up to 10\% 
at $\lambda \approx 6000\;\ang$.
We will correct for this bias when creating mock spectra to compare with the data, as described in \S~\ref{sec:model}.
Figure~\ref{fig:disp} shows the (uncorrected) pixel
 dispersions from 236 BOSS quasars from the $\langle z \rangle =2.3$, $\snr = 6-8$ bin, as a function of wavelength at the 
 blue end ($\lambda = 3700-4200\ang$) of the spectrograph.
At fixed wavelength, there are outliers that contribute to the large spread in $\sigma_{\rm disp}$,
 e.g.,\ ranging from $\sigma_{\rm disp} \approx 0.9 - 1.8$ at $3700\; \ang$.
{We therefore discard spectra with outlying values of $\sigma_{\rm disp}$ based on the following
criterion: we first rank-order the spectra based on their $\sigma_{\rm disp}$ value evaluated at the central
wavelength of each PDF bin (i.e. $\lambda = [4012,4377,4863]\,\ang$ at $\zav = [2.3,2.6,3.0]$), and then discarded
spectra below the 5th percentile and above the 90th percentile. This is illustrated by the red-dashed lines in 
Figure~\ref{fig:disp}.}

Finally, since our noise estimation procedure uses the individual BOSS exposures, we discard objects that have less than 
three individual exposures available.

\begin{figure}
\epsscale{1.2}
\plotone{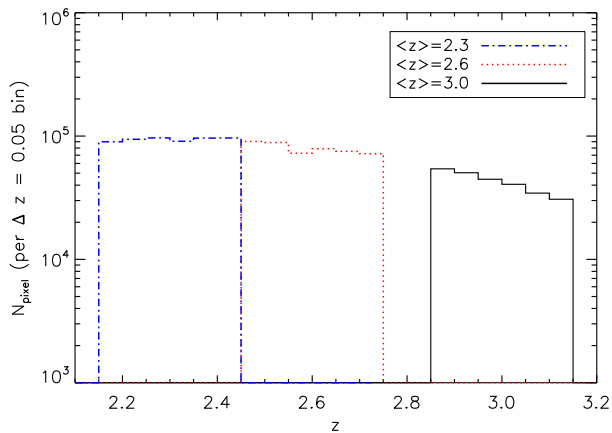}
\caption{\label{fig:hist_zabs}
Pixel distribution of \lya\ absorber redshifts in the BOSS \lya\ forest
sample used in this paper, shown in bin sizes of $\Delta z = 0.05$.
The different colors and line-styles denote the three redshift bins used in this paper.
We have chosen these redshift bins --- with the gap at $2.75 < z < 2.85$ ---
to match the simulation redshifts (\S~\ref{sec:sims}).
}
\end{figure}

Our final data set comprises 3373 unique quasars with redshifts ranging from $\zq = 2.255$ to $\zq = 3.811$, and a median S/N
of $\snr = 8.08$ per pixel. This data set represents only a small subsample of the BOSS DR9 quasar spectra, 
but is over two orders-of-magnitude 
larger than high-resolution quasar samples previously used for transmission PDF analysis.
Table~\ref{tab:pdfbins} summarizes our data sample, and the statistics of the redshifts and S/N bins for which we measure the
transmission PDF. Figure~\ref{fig:hist_zabs} shows histograms of the pixels used in our analysis, as a function of absorption redshift.

\begin{deluxetable*}{c c r r c c c c }
\tablecaption{\label{tab:pdfbins}  Binning of BOSS \lya\ Forest transmission PDFs  }
\tablewidth{0.8\textwidth}
\tablehead{
\colhead{\lya\ Forest}    &  \colhead{$\snr$\tablenotemark{a}} & \colhead{$N_{spec}$\tablenotemark{b}} 
& \colhead{$N_{pix}$\tablenotemark{c}} & \colhead{$\Delta v$\tablenotemark{d}} & \colhead{$\Delta z$\tablenotemark{e}} & \colhead{$\Delta X$\tablenotemark{f}}  \\
 \colhead{Redshift} &  \colhead{(per pixel)} & \colhead{}  & \colhead{}  & \colhead{($\kms$)} & &
}
\startdata
\multirow{3}{*}{$2.15<z<2.45$} 		& 6-8 	&1109 	& 288442  	& $\scien{1.99}{7}$ 	& 219  & 704 \\
					& 8-10 	& 501 	& 129141  	& $\scien{8.90}{6}$ 	& 97.9 &  315  \\
  					& $>10$  &  561 	&  146478		& $\scien{1.01}{7}$   & 111 &  357 \\
					 \\
\multirow{3}{*}{$2.45<z<2.75$} 		& 6-8 	& 1004 	& 229898 	 	& $\scien{1.59}{7}$	& 191 & 646 \\
					& 8-10 	&  490 	& 107001 	 	& $\scien{7.38}{6}$ 	& 88.6 & 300 \\
  					& $>10$	& 604 	& 140843 		& $\scien{9.71}{6}$ 	& 117 &  396 \\	
					\\
\multirow{3}{*}{$2.85<z<3.15$} 		& 6-8 	&   511 	& 108443 	 	& $\scien{7.48}{6}$ & 99.7 & 358 \\
					& 8-10 	&  326 	& 72448 	 	& $\scien{5.00}{6}$ & 66.7  & 239 \\
  					& $>10$ 	& 341 	& 74284 		& $\scien{5.12}{6}$ & 68.3 & 245 \\
					\enddata
\tablenotetext{a}{Median S/N within \lya\ forest.}
\tablenotetext{b}{Number of contributing spectra.}
\tablenotetext{c}{Number of $\Delta v = 69\,\kms$ pixels.}
\tablenotetext{d}{Velocity path length.}
\tablenotetext{e}{Redshift path length.}
\tablenotetext{f}{Absorption distance, where $dX/dz = (1+z)^2 (\Omega_M (1+z)^3 + \Omega_\Lambda)^{-1/2}$.
For this conversion, we assume $\Omega_M=0.3$ and $\Omega_\Lambda = 0.7$.}
\end{deluxetable*}

\section{Measuring the transmission PDF from BOSS} \label{sec:bosspdf}

In this section, we will measure the \lya\ forest transmission PDF from BOSS.
In principle, the transmission PDF is simply the histogram of the transmitted
 flux in the \lya\ forest after dividing by the quasar
continuum. However, with the comparatively noisy BOSS data we need to
ensure an accurate estimate of the pixel noise. We will therefore
first describe a new probabilistic method for co-adding the individual
BOSS exposures that will enable us to have an accurate noise estimate.
We will also describe the continuum-estimation method with which we
normalize the forest transmission.

\subsection{Co-addition of Multiple Exposures and Noise Estimation}\label{sec:mcmc}

Since we intend to model BOSS spectra with modest S/N, we need an accurate 
estimate of the pixel noise that also allows us to separate out the contributions 
from Poisson noise due to the background and sky as well as read noise from the 
the detector. 
In this subsection, we will construct an accurate probabilistic model of the flux
and noise of the BOSS spectrograph, based on the individual exposure
data that BOSS delivers. 

The basic BOSS spectral data consists of a
spectrum of each raw exposure, $f_{\lambda i}$ {(inclusive of noise)}, an estimate of the sky
$s_{\lambda i}$, and a calibration vector $S_{\lambda i}$, where $i$
indicates the exposure of the $n_{\rm exp}$ exposures 
taken\footnote{Typically there are $n_{\rm exp} = 5$ exposures of 15 minutes each, 
although this can vary due to the requirements to achieve a given $(\snr)^2$
 over each individual plug-plate, as determined by the overall BOSS survey strategy \citep[see][]{dawson:2013}.}. 
The quantity $s_{\rm
  \lambda i}$ is the actual sky model that was subtracted from the
fiber spectra in the extraction. The calibration vector is defined as
$S_{\lambda i} \equiv f_{\lambda i}\slash f_{N i}$, with $f_{N i}$
being the flux of exposure $i$ in units of photoelectrons. 
The idlspec2d pipeline then estimates the {co-added spectrum of the true object flux}, 
$\mathcal{F}_\lambda$, from the raw individual exposures, sky estimates, and calibration vectors.

\begin{figure}
\epsscale{1.2}
\plotone{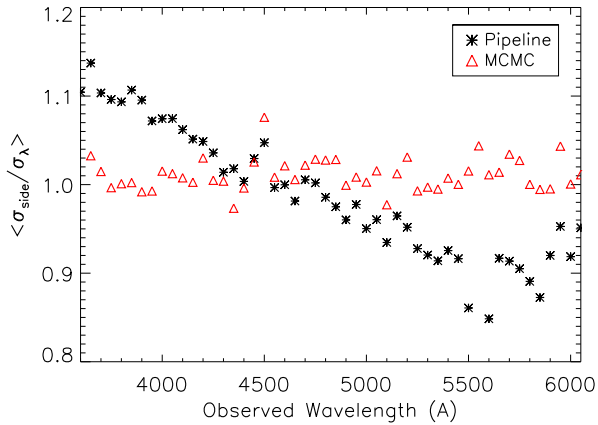}
\caption{\label{fig:noisetest_qso}
A quantitative test of the noise estimation fidelity in the spectra.
Each point shows the ratio of the pixel variance divided by the estimated noise
variance, averaged over the restframe $1460\,\ang < \lambrest < 1510\,\ang$ flat spectral region of
500 BOSS quasars within redshift bins of $\Delta \zq = 0.1$ and plotted as a function of the
corresponding observed wavelength of the flat spectral region. If there is no bias in the noise estimation, this ratio
should be unity. The black asterisks show this quantity estimated using the BOSS
pipeline co-added spectra and noise estimates, while the red triangles show 
the results from the MCMC co-addition and noise estimation procedure described in 
\S~\ref{sec:mcmc}. The MCMC method clearly provides a better noise estimation than the 
BOSS pipeline.
}
\end{figure}

The BOSS data reduction pipeline also delivers noise estimates in the form of
variance vectors, 
which are however known to be inaccurate \citep{mcdonald:2006,desjacques:2007,lee:2013,palanque-delabrouille:2013a}. 

To quantify the fidelity of the BOSS noise estimate, 
we used the so-called `side-band' method described in \citet{lee:2014} and \citet{palanque-delabrouille:2013a}, which
uses the variance in flat, absorption-free, regions of the quasar spectra to quantify the fidelity of the noise estimate.
First, we randomly selected 10,000 BOSS quasars (omitting BAL quasars) from the \citet{paris:2012} catalog
 in the redshift range $1.4 \leq \zq < 3.4$, evenly distributed into 20 redshift bins of width $\Delta \zq = 0.1$ (i.e., 500 objects
 per bin).
 We then consider the flat $1460\,\ang < \lambrest < 1510\,\ang$ spectral region in the quasar restframe, which is
 dominated by the smooth power-law continuum and relatively unaffected by broad emission lines 
 \citep[e.g.,][]{vanden-berk:2001,suzuki:2006} or absorption lines. The pixel variance in this flat portion of the spectrum should therefore 
 be dominated by spectral noise, allowing us to examine whether the noise estimate provided by the pipeline is accurate.  
We then evaluate the ratio of, $\sigma_{\rm side}$, the pixel flux RMS in the restframe $1460\,\ang < \lambrest < 1510\,\ang$ region divided
by the average pipeline noise estimate, $\sigma_\lambda$:
\beq
\left< \frac{\sigma_{\rm side}}{\sigma_\lambda} \right> = 
\frac{ \left[\sum f^2_\lambda - \bar{f}_\lambda^2 \right]^{1/2}}{\sum \sigma_\lambda},
\eeq
where the summations and average flux is evaluated in the quasar restframe $1460\,\ang < \lambrest < 1510\,\ang$.

In Figure~\ref{fig:noisetest_qso}, this quantity is averaged over the 500 individual quasars
per redshift bin and plotted as a function of the observed wavelength corresponding to $\lambda = (1+\langle \zq \rangle )1485\,\ang$.
With a perfect noise estimate, $\langle \sigma_{\rm side}/\sigma_\lambda \rangle$ should be unity at all wavelengths, but we see that
the BOSS pipeline underestimates the true noise in the spectra at $\lambda \lesssim 5000\;\ang$,
by up to $\sim 15\%$ at the blue end of the spectra, with an overall tilt that changes over to an
overestimate at $\lambda \gtrsim 4500\,\ang$.
\citet{lee:2013} and \citet{palanque-delabrouille:2013a} provide a set of correction vectors that can be applied to the pipeline
noise estimates to bring the latter to within several percent of the true noise level across the wavelength coverage
of the blue spectrograph. 

Unfortunately, these noise corrections are inadequate for our purposes, since we want to generate
realistic mock spectra that have different realizations of the \lya\ forest transmission field 
from the actual spectra, i.e., a different $\mathcal{F}_{\lambda}$. 
We therefore require a method that not only accurately estimates the noise in a given BOSS
spectrum, but also separates out the photon-counting and CCD terms in 
the variance, that results from applying the \citet{horne:1986} optimal spectral 
extraction algorithm:
\be
\sigma_{\lambda}^2 = S_{\lambda}\left(\mathcal{F}_{\lambda} + s_{\lambda}\right) + 
S_{\lambda}^2\sigma^2_{\rm RN} \label{eqn:hornenoise},  
\ee
where $\sigma_{\rm RN}$ is the CCD read-noise.

To resolve this issue, we apply our own novel statistical 
method to 
the individual BOSS exposures to generate co-added spectra while
simultaneously estimating the corresponding noise parameters for each
individual spectrum.  This procedure, which uses a Gibbs-sampled
Markov-Chain Monte Carlo (MCMC) algorithm, is described in detail in
the Appendix.  Initially, we attempted to model the noise with just a single constant
noise parameter which rescales the read-noise term of 
Equation~\ref{eqn:hornenoise}, but this was found to be inadequate.
This is likely because an optimal extraction algorithm weights by the
product of the S/N and object profile, causing the corresponding
variance to have a non-linear dependence on the flux and sky
level. Furthermore, systematic errors in the reduction, sky-subtraction
and calibration 
will result in additional noise contributions which could depend on
sky level, object flux, or wavelength, hence deviating from this
simple model.

After considerable trial-and-error to find a model that best minimizes the bias illustrated in Figure~\ref{fig:noisetest_qso}, 
we settled on the form:
\beq 
\sigma_{\lambda i}^2 = A_1 \hat{S}_{\lambda i}\left(\mathcal{F}_{\lambda}+ s_{\lambda_i}\right) + 
 A_2 \hat{S}_{\lambda i}^2\sigma^2_{\rm RN, eff} \sigma_\mathrm{disp}(\lambda) \label{eqn:mcmc_noise}
\eeq
where 
\beq 
\hat{S}_{\lambda i}=S_{\lambda i} \left(1 - \exp(-A_3 \lambda + A_4) \right),
\eeq
where the $A_{j}$ are free parameters in our noise model, while the
$\sigma_\mathrm{disp}(\lambda)$ factor in the 2nd term (the pixel dispersion) provides a rough approximation for the
{
wavelength-dependence of the spot-size (i.e. the size of the raw CCD image in the spatial direction).
Meanwhile, $\sigma_\mathrm{disp}=12$ is the average CCD read-noise per wavelength bin in the BOSS spectra
(D.J.\ Schlegel et al., in preparation).
}
The quantities $s_{\lambda,i}$, $S_{\lambda,i}$, and $\sigma_\mathrm{disp}(\lambda)$ 
(sky flux, calibration vector, and dispersion, respectively) are taken directly from the
BOSS pipeline.

In addition, we assume that the pixel noise can be modeled as a
Gaussian distribution with a variance given by
Equation~\ref{eqn:mcmc_noise}.  The first, photon counting, term in
the equation should formally be modeled as a Poisson distribution, but since
the BOSS spectrograph always receives $\gtrsim 30-40$
counts even at the blue end of the spectrograph where the counts are
the lowest, it is reasonable to use the Gaussian approximation
because even in the limit of low S/N (i.e.\ when the spectrum is
dominated by the sky flux), the moderate resolution ensures that
there are at least several dozen sky photons per pixel in each
exposure.

For each BOSS spectrum, we use the MCMC procedure described in the Appendix to combine the
multiple exposures while simultaneously
estimating the noise parameters $A_j$ and true observed spectrum, $\mathcal{F}_{\lambda}$.
With the optimal estimates of $A_j$ and $\mathcal{F}_{\lambda}$ for a given spectrum, the estimated noise
 variance is then simply Equation~\ref{eqn:mcmc_noise}.

An important advantage of the form in Equation~\ref{eqn:mcmc_noise} is that the
object photon noise $\propto \mathcal{F}_{\lambda}$ is explicitly separated out. 
This facilitates the construction of a mock spectrum with the same
noise characteristics as a true spectrum, \emph{but with a different
 spectral flux}. For example, a mock spectrum of the Ly$\alpha$ forest will
have a very different transmission field than the original data, 
and so the variance due to object photon counting noise can be added appropriately, in addition to
contributions from the known sky, and the read noise
term (Equation~\ref{eqn:mcmc_noise}). Our empirical determination of the parameters
govering this noise model for each individual spectrum form 
a crucial ingredient in our forward model, which we will describe in 
\S~\ref{sec:model}.

\begin{figure*}
\begin{center}
\includegraphics[height=0.3\textheight]{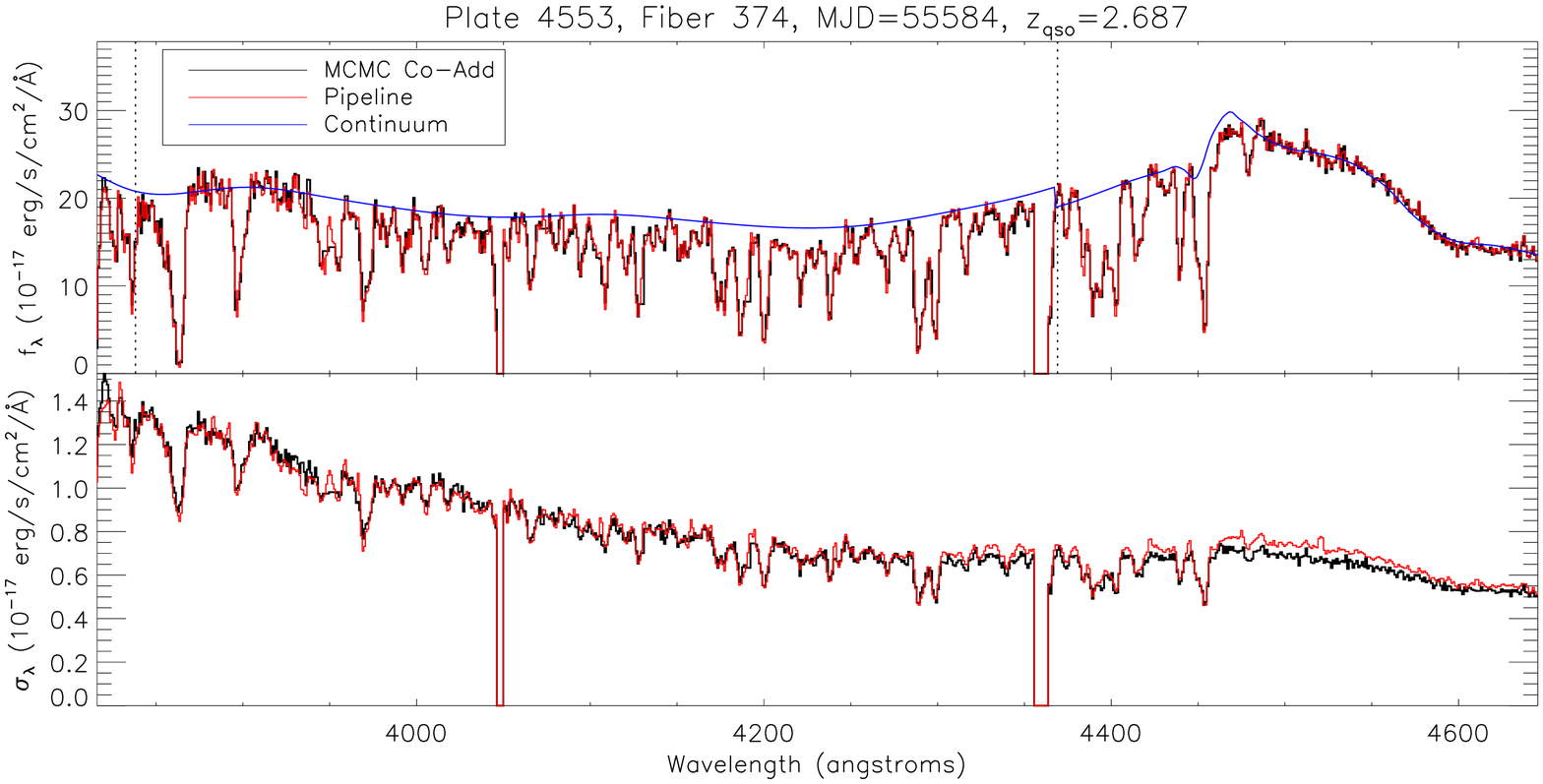} \\
\vspace{0.3cm}
\includegraphics[height=0.3\textheight]{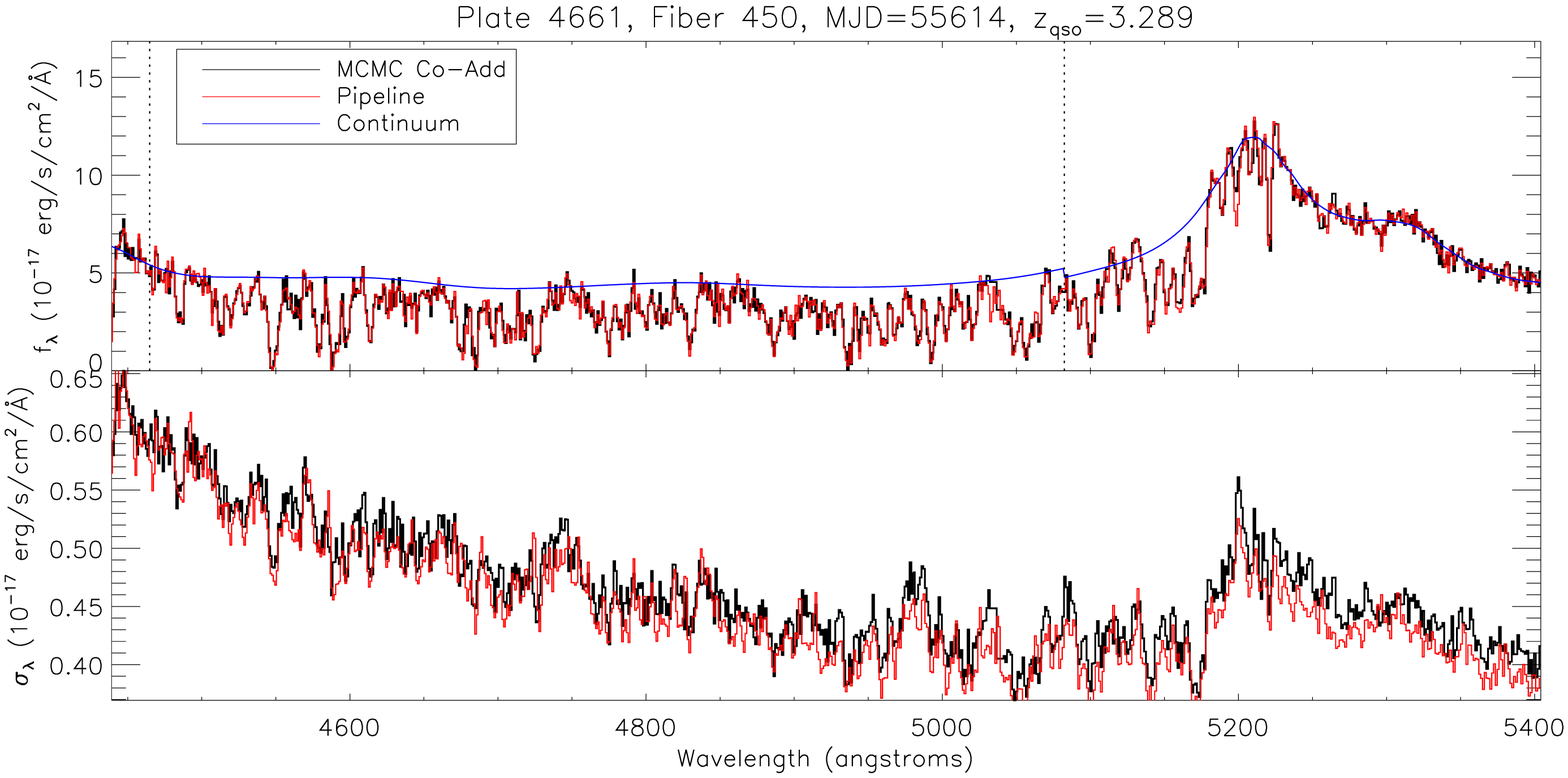}
\end{center}
\caption{\label{fig:plot_mcmc}
Examples of co-added BOSS spectra from the MCMC procedure described in \S~\ref{sec:mcmc}
(red), and from the BOSS pipeline (black) are shown in the upper panels, in the restframe interval
$1035-1260\,\ang$. The corresponding pixel noise estimates are shown in the upper panels.
The blue line shows the MF-PCA continuum used to extract the 
\lya\ forest transmitted flux, while the vertical dotted lines delineate the $1041-1185 \ang$ 
restframe interval which we define as the \lya\ forest. The continuum discontinuity at $\lambrest = 1185\ang$ 
is where we have applied the `mean-flux regulation' correction to the \lya\ forest. 
In the top figure, masked pixels have
had their flux and noise set to zero. 
The signal-to-noise ratios for the two spectra are $\snr \approx 11$ (top) and $\snr \approx 6$ (bottom)
within the \lya\ forest.
}
\end{figure*}

\begin{figure*}
\begin{center}
\includegraphics[height=0.3\textheight]{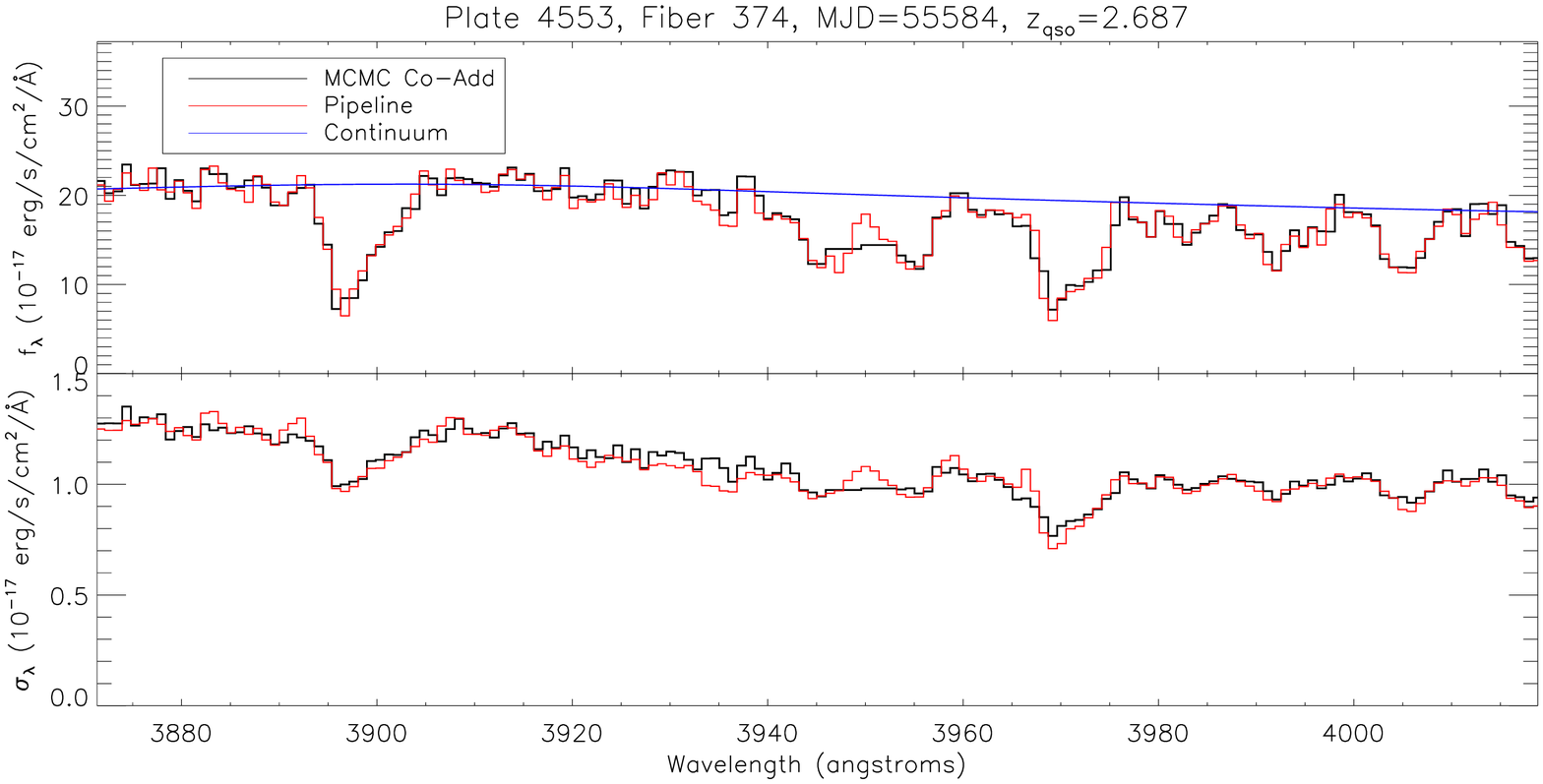} \\
\vspace{0.3cm}
\includegraphics[height=0.3\textheight]{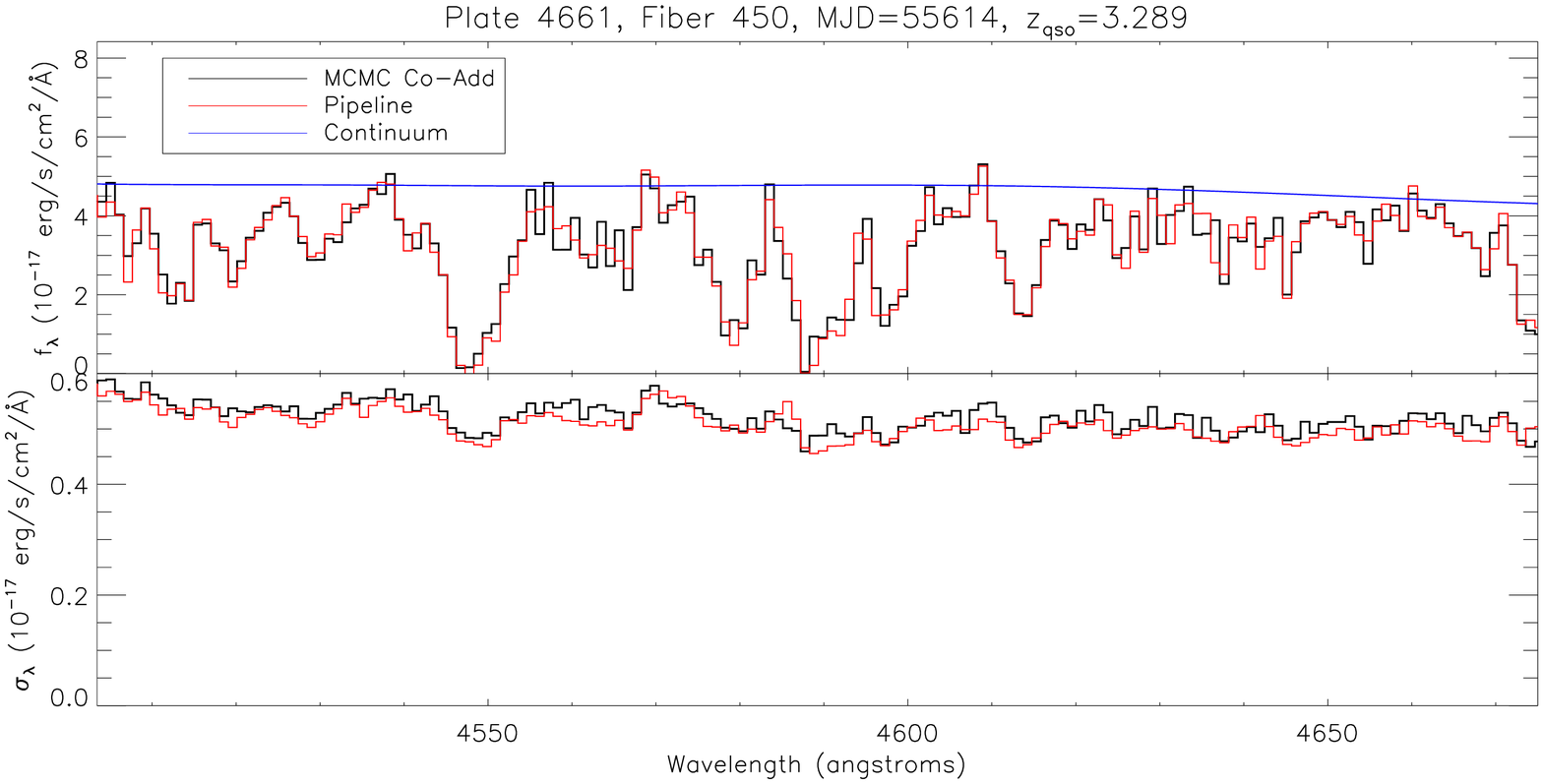}
\end{center}
\caption{\label{fig:plot_mcmc_zoom}
Same as Figure~\ref{fig:plot_mcmc}, but the $1050\,\ang < \lambrest < 1090\,\ang$ restframe region
is expanded to better illustrate the differences between the MCMC and pipeline co-added spectra.
}
\end{figure*}

Our MCMC procedure works for spectra from a single camera, either red
or blue; we have not yet generalized it to combine blue and red
spectra of each object.  However, the spectral range of the blue
camera alone ($\approx 3600-6400 \ang$) covers the \lya\ forest up to
$z \sim 5$, i.e., most practical redshifts for \lya\ forest analysis.
For the purposes of this paper, we restrict ourselves to spectra from
the blue camera alone.

In Figures~\ref{fig:plot_mcmc} and \ref{fig:plot_mcmc_zoom}, we show
examples of co-added BOSS quasar spectra, using both the MCMC
procedure and the standard BOSS pipeline. 
 In the upper panels, the
MCMC co-adds are not noticeably different from the BOSS pipeline,
although the numerical values are different.  In the lower panels, we
show the estimated noise from both methods --- the differences are
larger than in the fluxes but still difficult to distinguish by eye.

We therefore return to the statistical analysis by calculating $\langle \sigma_{\rm side}/\sigma_\lambda\rangle$, the
ratio of the pixel variance against the estimated noise from the flat $1460
\,\ang < \lambrest < 1510\,\ang$ region of BOSS quasars; this ratio, computed for our
MCMC coadds, is
plotted in Figure~\ref{fig:noisetest_qso}. {With these new co-adds, we
see that this ratio is within roughly $\pm 3\%$ of unity across} the entire $\lambda
\sim 3800 - 5000\,\ang$ wavelength range relevant to our subsequent analysis, with an overall bias of
$1\%$ (i.e. the noise is still underestimated by this level).
Crucially, we have removed the strong wavelength dependence of $\langle \sigma_{\rm side}/\sigma_\lambda\rangle$
that was present in the standard pipeline, and we suspect most of the scatter about unity is caused by the
limited number of quasars (500 per bin) available for this estimate, which will be mitigated by the larger number of quasars spectra available in the subsequent BOSS data releases. 
{In principle, we could correct the remaining $1\%$ noise bias, but since our selected spectra have $\snr>6$, this remaining
$1\%$ noise bias would smooth the forest transmission PDF by an amount roughly $1/25$ of the average PDF bin width
($\Delta F = 0.05$).  As we shall see, there are other systematic uncertainties in our modeling that have much larger effects than 
this, therefore we regard our noise estimates as adequate for the subsequent 
transmission PDF analysis, without requiring any further correction. }

\subsection{Mean-Flux Regulated Continuum Estimation} \label{sec:cont}

In order to obtain the transmitted flux $F$ of the \lya\ 
forest\footnote{Note that the ideal/model observed flux described in the noise modelling section, $\mathcal{F}_{\lambda}$, and
the \lya\ forest transmission $F$, are completely different quantities.}
we first need to divide the observed flux,
$\mathcal{F}_{\lambda}$, by an estimate for the quasar continuum, $c$.
We use the version of mean-flux regulated/principal component
analysis (MF-PCA) continuum fitting \citep{lee:2012a} described in \citet{lee:2013}. 
Initially, PCA fitting with 8 eigenvectors is performed on each quasar spectrum redwards
of the \lya\ line ($\lambrest = 1216-1600\ang$) in order to obtain a prediction for the continuum shape in 
the $\lambrest < 1216\ang$  \lya\ forest region \citep[e.g.,][]{suzuki:2005}. 
The slope and amplitude of this initial continuum estimate is then corrected to agree with 
the \lya\ forest mean transmission, $\fcont (z)$, at the corresponding absorber redshifts, using a linear correction function. 

The only difference in our continuum-fitting with that in \citet{lee:2013} is that here we use the latest
mean-flux measurements of \citet{becker:2013} to constrain our continua. 
{Their final result yielded the power-law redshift evolution of
the effective optical depth in the unshielded \lya\ forest, defined 
in their paper $\nhi \leq 10^{17.2}\,\persqcm$ (although they only removed contributions from $\nhi \geq 10^{19}\,\persqcm$ absorbers).
This is given by}
\beq \label{eq:fmean}
\tau_\mathrm{Ly\alpha,B13}(z) \equiv -\ln(\fmean(z)) = \tau_0\left(\frac{1+z}{1+z_0}\right)^{\beta} + C,
\eeq
with best-fit values of $[\tau_0, \beta, C] = [0.751, 2.90, -0.132]$ at $z_0 = 3.5$. 

{
However, the actual raw measurement made by \citet{becker:2013} is the effective total absorption within the \lya\ forest region
of their quasars, which also contain contributions from metals and optically-thick systems:
\beq \label{eq:tau_cont}
\tau_\mathrm{eff}(z) \equiv \tau_\mathrm{Ly\alpha,B13}(z)  + \tau_{\rm{metals}} + \tau_{\rm{LLS}}(z),
\eeq
where $\tau_{\rm{metals}}$ and $\tau_{\rm{LLS}}(z)$ denote the IGM optical depth contributions
from metals and Lyman-limit systems, respectively.
For the purposes of our continuum-fitting, the quantity we require is $\tau_\mathrm{eff}(z)$, since
the $\tau_{\rm{metals}}$ and $\tau_{\rm{LLS}}(z)$ contributions are also present in our BOSS spectra.
\citet{becker:2013} did not publish their raw $\tau_\mathrm{eff}(z)$, therefore we must now `uncorrect' the metal
and LLS contributions from the published $\tau_\mathrm{Ly\alpha,B13}(z)$. The discussion below therefore 
attempts to retrace their footsteps and does not necessarily reflect our own beliefs regarding the actual
level of these contributions.
}

We find $\tau_{\rm{metals}}=0.02525$ by simply averaging over the \citet{schaye:2003} metal correction 
tabulated by \citet{faucher-giguere:2008a} (i.e., the $2.2\leq z \leq 2.5$ values in $\Delta z = 0.1$ bins from their Table~4), 
that were used by \citet{becker:2013} to normalize their relative mean-flux measurements.
{Note that there is no redshift dependence on $\tau_{\rm{metals}}$ in this context, because \citet{becker:2013}
argued that the metal contribution does not vary significantly over their redshift range. Whether or not this is really
true is unimportant to us at the moment, since we are merely `uncorrecting' their measurement.}

The LLS contribution to the optical depth is re-introduced by integrating over $f(\nhi, b, z) $, the column-density distribution
of neutral hydrogen absorbers:
\begin{eqnarray}
\tau_\mathrm{LLS}(z) &\approx& \frac{1+z}{\lambda_\mathrm{Ly\alpha}} \int^{N_{max}}_{N_{min}} \mathrm{d} \nhi \int \mathrm{d} b \nonumber \\
 & & \hphantom{yoloooo} \times f(\nhi, b, z) W_0(\nhi,b),
\end{eqnarray}
where $b$ is the Doppler parameter and $W_0(\nhi,b)$ is the rest-frame equivalent width (we use the analytic approximation given 
by \citealt{draine:2011}, valid in the saturated regime). 

Following \citet{becker:2013}, we adopted a fixed 
value of $b=20\;\kms$ and 
assumed that $f(\nhi,z) = f(\nhi) \mathrm{d}n/\mathrm{d}z$, 
where $f(\nhi)$ is given by the $z=3.7$ broken
power-law column density distribution of \citet{prochaska:2010} and $\mathrm{d}n/\mathrm{d}z \propto (1+z)^2$.
\citet{becker:2013} had corrected for super-LLSs and DLAs in the column-density 
range $[N_{min}, N_{max}] = [10^{19}\;\persqcm, 10^{22}\;\persqcm]$, 
but as discussed above we have discarded all sightlines that include $\nhi \geq 10^{20.3} \persqcm$ DLAs, therefore we 
 reintroduce the optical depth contribution for super-LLSs, i.e., $[N_{min}, N_{max}] = [10^{19}\;\persqcm, 10^{20.3}\;\persqcm]$.
 We find $\tau_\mathrm{LLS}(z)  = 0.0022\times [(1+z)/3]^3$. This is a small correction, giving rise to only a $0.5\%$ change in \fmean\ 
 at $z=3.0$.

This estimate of the raw absorption, $\feff (z) = \exp[-\tau_\mathrm{eff}(z)]$, 
is now the constraint used
to fit the continua of the BOSS quasars, i.e. we set $\fcont = \langle F \rangle_{\rm eff} (z)$. 
Note that in our subsequent modelling of the data, 
we will use the same $\fcont(z)$
to fit the mock spectra to ensure an equal treatment between data and mocks.
Since $\fcont(z)$ includes a contribution from $\nhi < 10^{20.3}\,\persqcm$ optically-thick systems,
our mock spectra will need to account for these systems as we shall describe in $\S\ref{sec:lls_fid}$. 

The MF-PCA technique requires spectral coverage in the quasar restframe interval 
$1000-1600 \ang$. However, as noted in the previous section, we work with co-added 
BOSS spectra from only the blue cameras covering $\lambda \lesssim 6400\ang$; this covers 
the full $1000-1600\ang$ interval required for the PCA fitting 
only for $z \lesssim 3$ quasars.
However, the differences in the fluxes between our MCMC co-adds and the BOSS pipeline co-adds 
are relatively small, and we do not expect the
relative shape of the quasar spectrum to vary significantly.
We can thus carry out PCA fitting on the BOSS pipeline co-adds, 
which cover the full observed range ($3700-10000\ang$), to predict the overall quasar continuum shape.
This initial prediction is then used to perform mean-flux regulation using the MCMC co-adds and noise estimates, to
fine-tune the amplitude of the continuum fits. 
 
The observed flux, $f_{\lambda}$, is divided by the continuum estimate, $c$, 
 to derive the \lya\ forest transmission, $F = f_{\lambda}/c$. 
For each quasar, we define the \lya\ forest as the rest wavelength interval
$1041-1185 \ang$. This wavelength range conservatively avoids
the quasar's \lyb/\ion{O}{6} emission line blend by $\Delta v \sim 3000\,\kms$ on the blue end,
as well as the proximity zone close to the quasar redshift by staying $\Delta v \sim 10,000\,\kms$ from
the nominal quasar systemic redshift.
We are now in a position to measure the transmission PDF,
which is simply the histogram of pixel transmissions $F \equiv \exp(-\tau)$. 

\subsection{Observed transmission PDF from BOSS}\label{sec:bosspdf}

\begin{figure*}
\begin{center}
\includegraphics[width=0.32\textwidth]{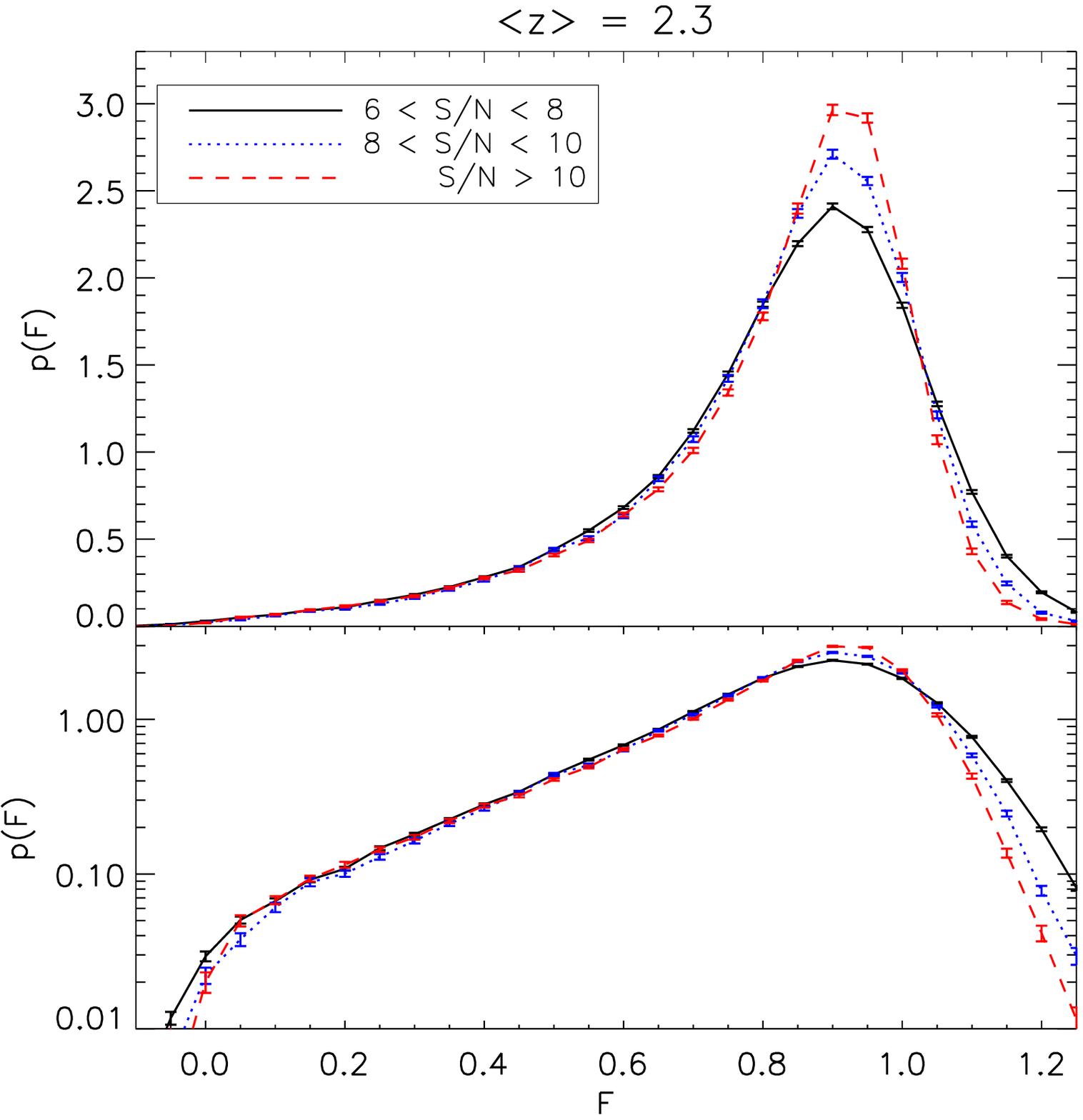}
\includegraphics[width=0.32\textwidth]{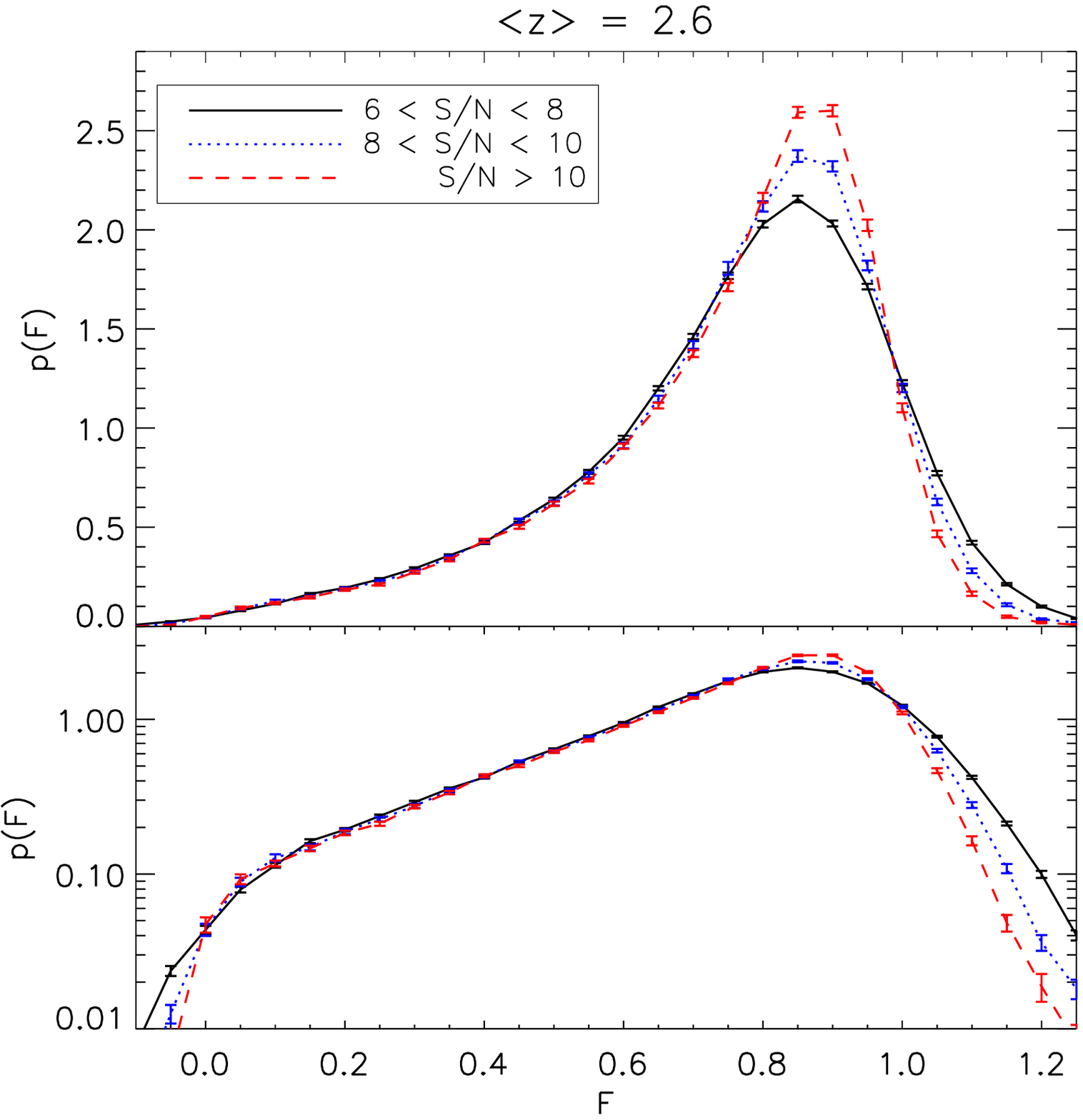}
\includegraphics[width=0.32\textwidth]{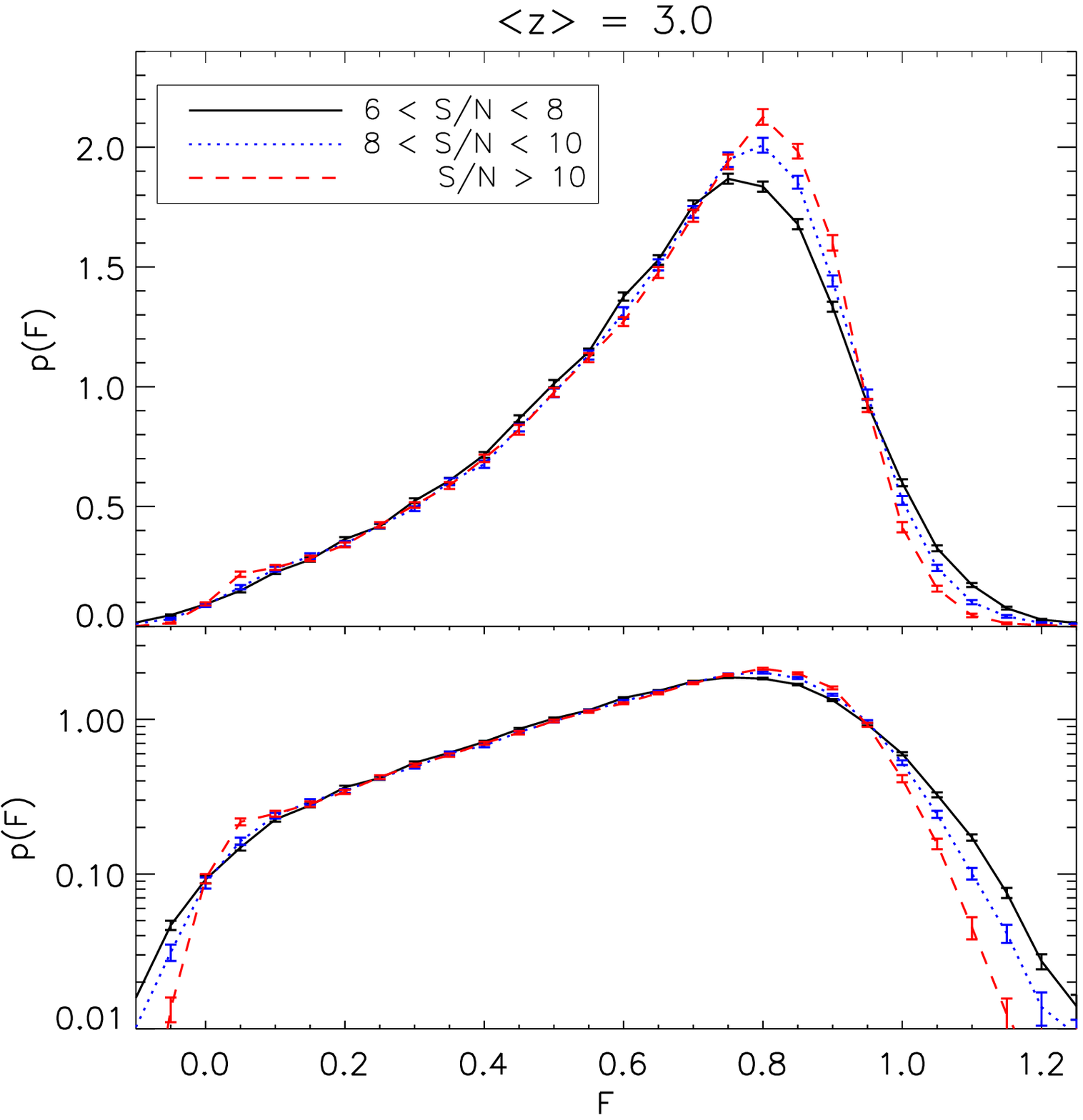}
\end{center}
\caption{\label{fig:pdfboss}
\lya\ forest transmission probability distribution functions, $p(F)$, measured from different subsamples of our BOSS sample, 
at various redshift (with $\Delta z = 0.3$) and S/N. 
Both the upper- and lower-panels show the PDF, but with linear and logarithmic ordinate axes, respectively.
The different colors and line-styles denote our different S/N subsamples at each redshift.
The error bars are estimated from bootstrap resampling over $\Delta v = \scien{2}{4}\,\kms$ segments 
from the contributing spectra. Table~\ref{tab:pdfbins} summarizes the number of spectra and pixels
which contribute to each bin.}
\end{figure*}

\begin{figure}
\includegraphics[width=0.48\textwidth]{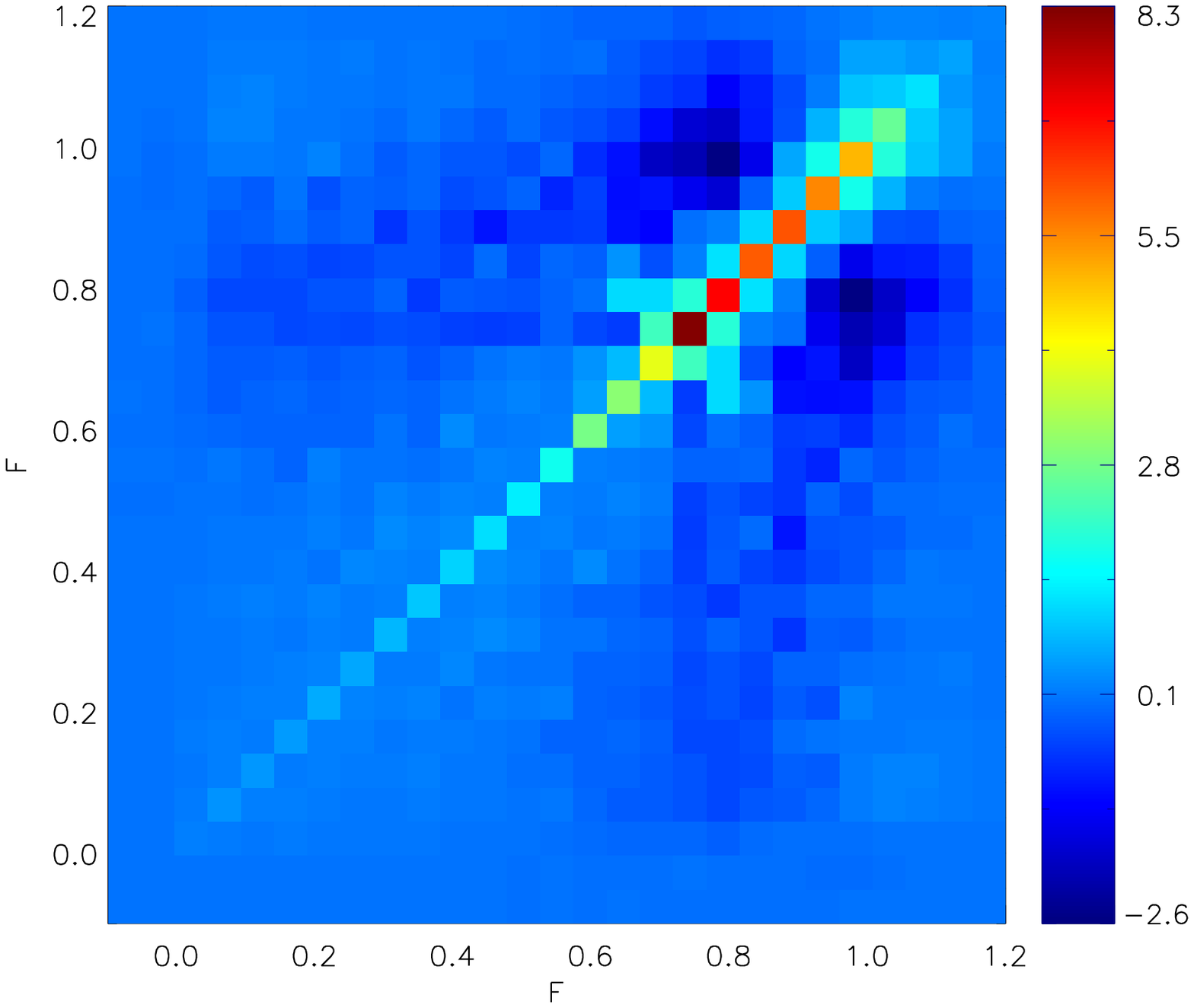}
\includegraphics[width=0.48\textwidth]{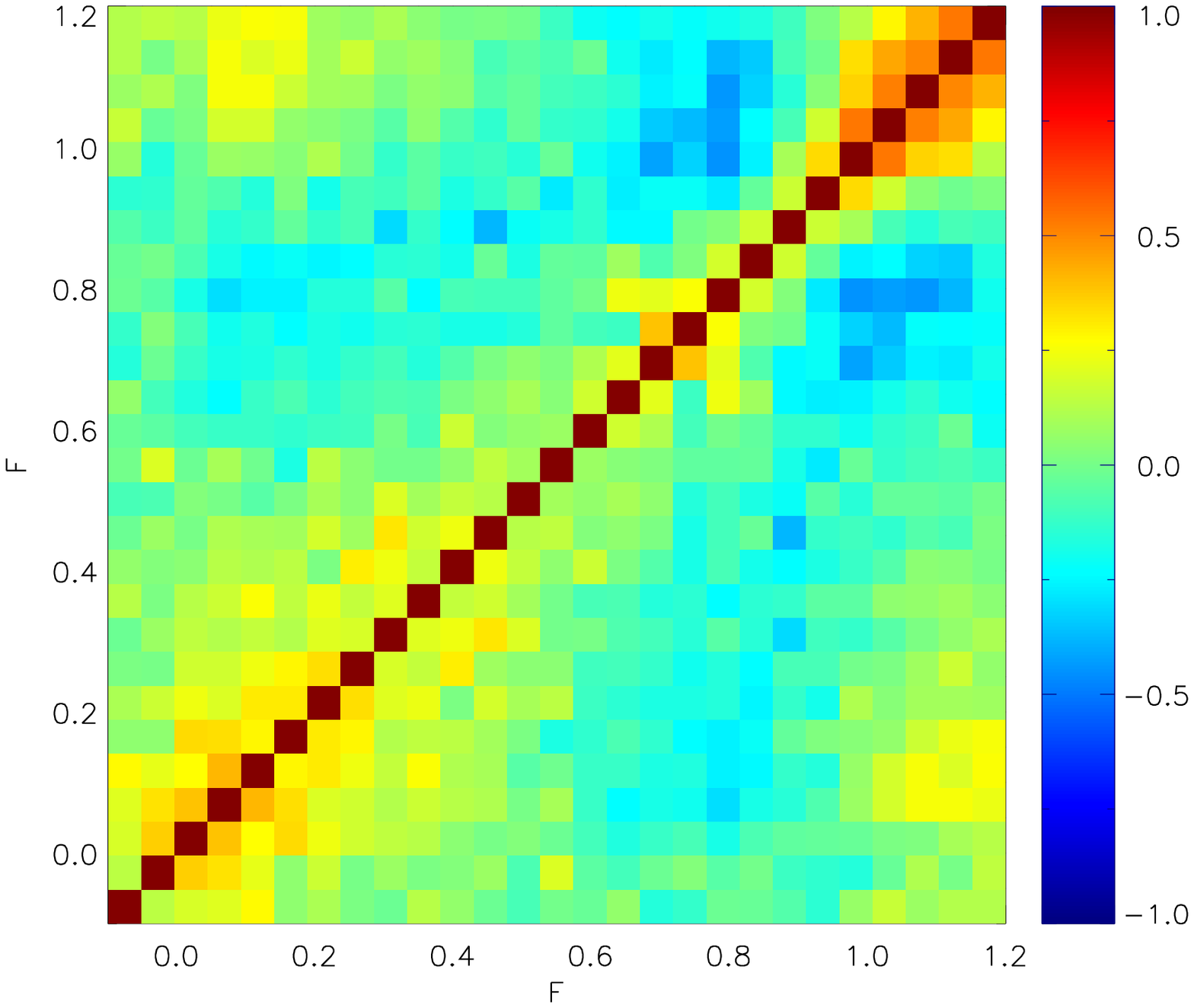}
\caption{\label{fig:covar_boot}
(Top) 2D density plot of the error covariance matrix for the \lya\ forest transmission PDF from the $\zav = 2.6$, S/N=$8-10$ BOSS subsample
as a function of transmission bins, along with (bottom) the corresponding correlation function. The covariance matrix was estimated
through bootstrap resampling, and the values been multiplied by $10^4$ for clarity. The covariances are largely diagonal, except for some 
cross-correlations between neighboring bins.
}
\end{figure}

\begin{figure}
\includegraphics[width=0.48\textwidth]{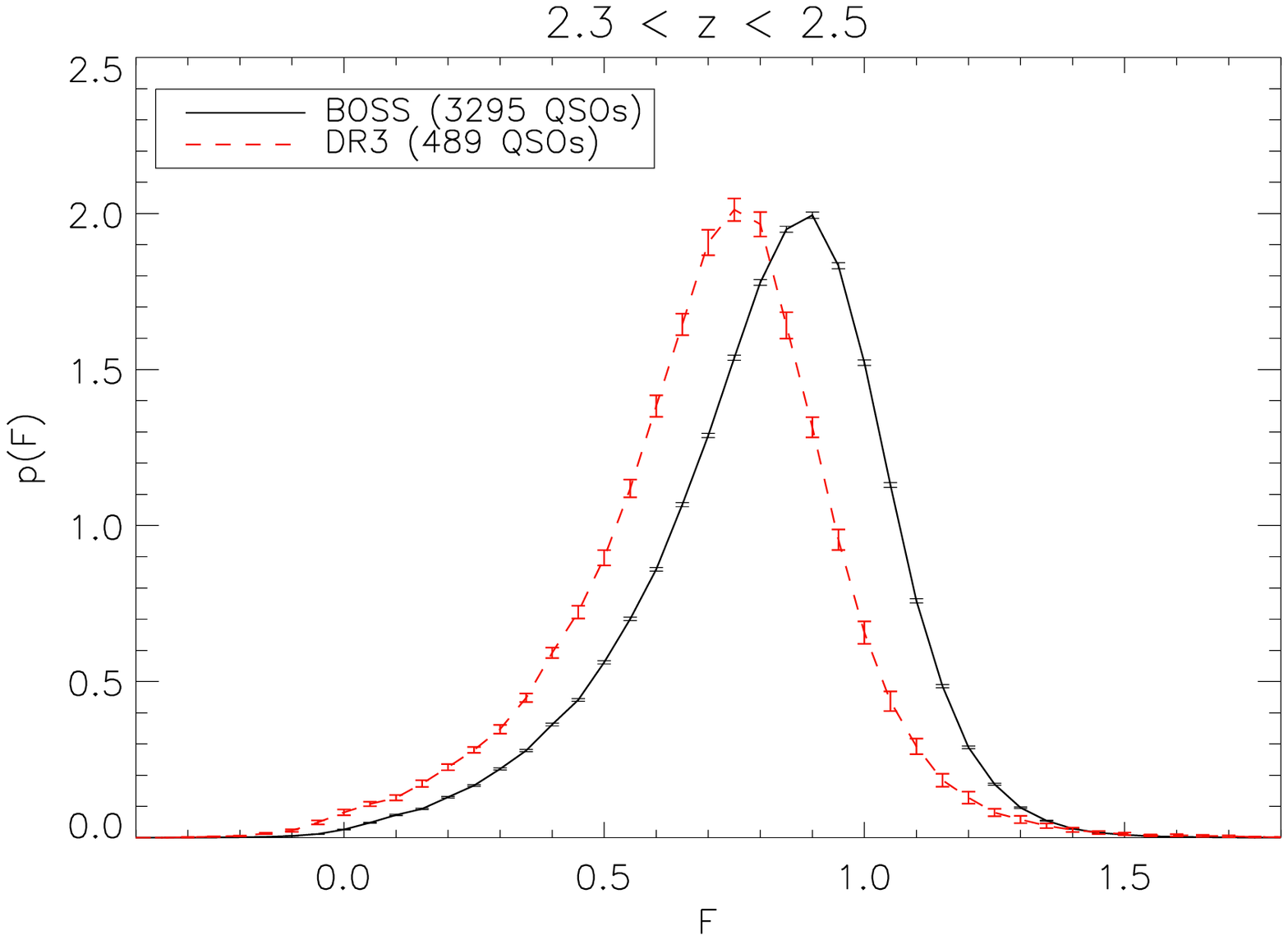}\\
\includegraphics[width=0.48\textwidth]{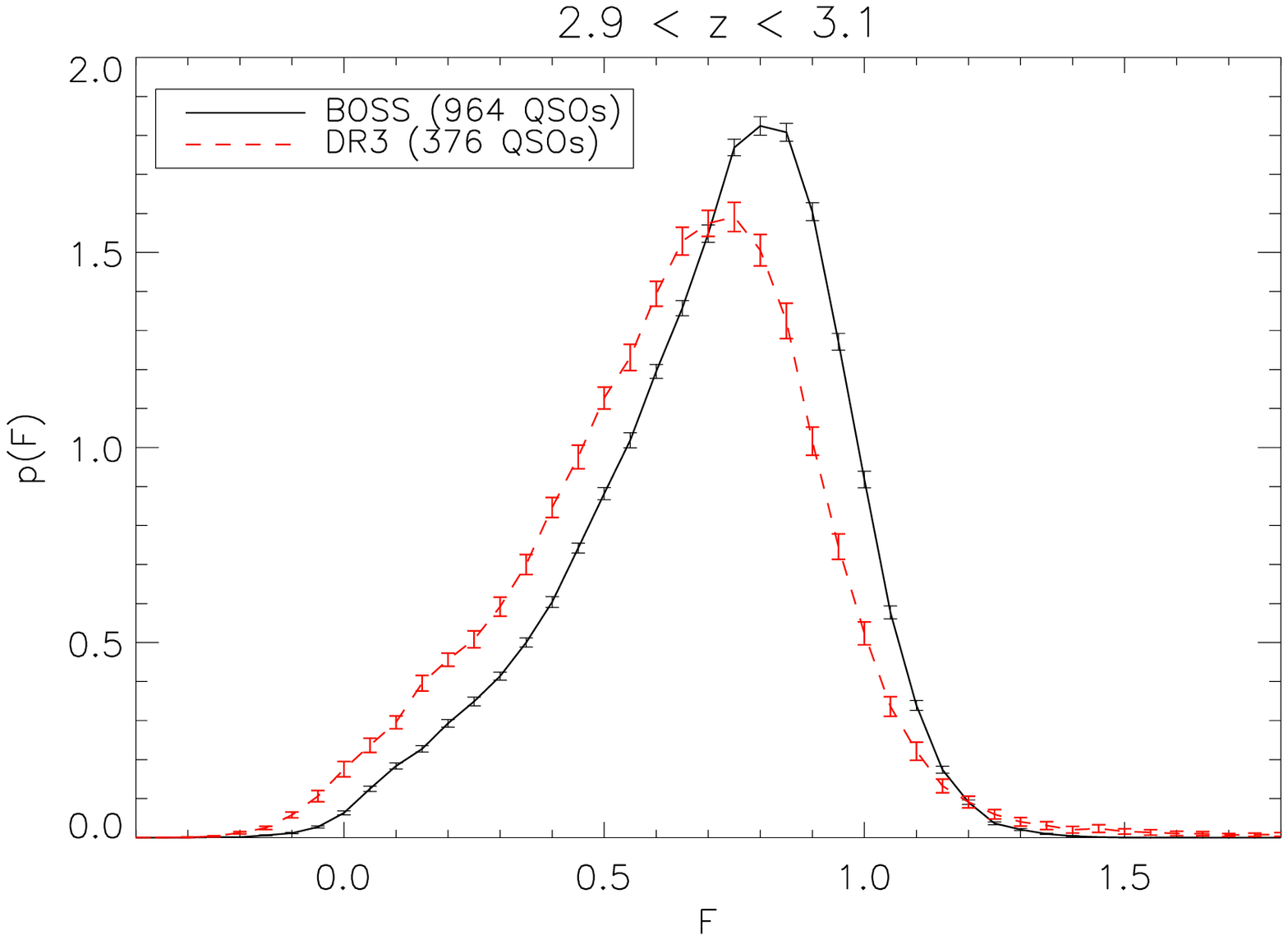}
\caption{\label{fig:pdf_desj}
A comparison between the \lya\ forest transmission PDFs measured from our BOSS DR9
sample (black solid lines), and the SDSS DR3 sample from \citet{desjacques:2007}
(red dashed-lines).  
Only sightlines with $\snr > 4$ were used in evaluating these PDFs.
The lower average transmission of the DR3 PDFs is because \citet{desjacques:2007}
had directly extrapolated a power-law from $\lambrest > 1216\ang$ for 
continuum estimates, which does not take into account a 
flattening of the quasar continuum that occurs at $\lambrest \sim 1200\ang$;
our BOSS spectra, in contrast, have been normalized to mean-transmission
values in agreement with latest measurements and takes this effect into account.
}
\end{figure}

Since the \lya\ forest evolves as a function of redshift, 
we measure the BOSS \lya\ forest transmission PDF in three bins with mean redshifts 
of $\langle z \rangle = 2.3$, 
$\langle z \rangle = 2.6$, and $\langle z \rangle =3.0$, 
and bin sizes of $\Delta z = 0.3$.
These redshifts bins were chosen to match the simulations outputs (\S~\ref{sec:sims}) that we 
will later use to make mock spectra to compare with the observed PDF; this choice
of binning leads to the gap at $2.75 < z < 2.85$ as seen in Figure~\ref{fig:hist_zabs}.
In this paper, we restrict ourselves to $z \lesssim 3$ since the primary purpose is to 
develop the machinery to model the BOSS spectra.
In subsequent papers, we will apply these techniques to 
analyze the transmission PDF in the full $2 \lesssim z \lesssim 4$ range
 using the larger samples of subsequent BOSS data releases \citep[DR10,][]{ahn:2014}.

Another consideration is that the transmission PDF is strongly affected
by the noise in the data. While we will model this effect in detail (\S~\ref{sec:model}), there is a large
distribution of S/N within our subsample ranging from $\snr =6$ per pixel to $\snr \sim 20$ per pixel. 
We therefore further divide the sample into three bins depending on the median
S/N per pixel within the \lya\ forest: $6 < \snr < 8$, $8 < \snr < 10$, $\snr > 10$.
The consistency of our results across the S/N bins will act as an important check for the
robustness of our noise model (\S~\ref{sec:mcmc}).

We now have nine redshift and S/N bins in which we evaluate
the transmission PDF from BOSS; the sample sizes are summarized in Table~\ref{tab:pdfbins}. 
For each bin, we have selected quasars that have at least 30 \lya\ forest pixels 
within the required redshift range, and which occupy the quasar
restframe interval $1041-1185\ang$. 
The co-added spectrum is divided with its MF-PCA continuum estimate 
(described in the previous section) to obtain the transmitted flux, $F$, in the desired pixels. 
We then compute the transmission PDF from these pixels.

Physically, the possible values of the \lya\ forest transmission 
range from $F=0$ (full absorption) to $F=1$ (no absorption).
However, the noise in the BOSS \lya\ forest pixels, as well as continuum
fitting errors, lead to pixels with $F < 0$ and $F>1$.
We therefore measure the transmission PDF in the range $-0.2 < F < 1.5$, 
in 35 bins with width $\Delta(F) = 0.05$, and normalized such that the area under the curve is unity.
The statistical errors on the transmission PDF are estimated by the following method:
we concatenate all the individual \lya\ forest segments that contribute
to each PDF, 
and then carry out bootstrap resampling over $\Delta v = \scien{2}{4}\,\kms$ 
segments with 200 iterations. 
This choice of $\Delta v$ corresponds to $\sim 250 - 300\,\ang$ in the observed frame at $z \sim 2-3$ ---
according to \citet{rollinde:2013}, this choice of $\Delta v$ and number of iterations should be sufficient for the errors to converge 
\citep[see also Appendix B in][]{mcdonald:2000}.

In Figure~\ref{fig:pdfboss}, we show the \lya\ forest transmission PDF 
measured from the various redshift- and S/N subsamples in our 
BOSS sample. At fixed redshift, the PDFs from the lower S/N data 
have a broader shape as expected from increased noise variance. 
With increasing redshift, there are more  
absorbed pixels, causing the transmission PDFs to shift
towards lower $F$ values.
As discussed previously, there is a significant portion of  
$F > 1$ pixels due to a combination of pixel noise and continuum errors, with 
a greater proportion of $F>1$ pixels in the lower-S/N subsamples as expected.
Unlike the high-resolution transmission PDF, at $\zav \lesssim 3$ there are few pixels
that reach $F=0$. This effect is due to the resolution of the BOSS spectrograph, 
which smooths over the observed \lya\ forest such that even saturated \lya\ forest absorbers 
with $\nhi \gtrsim 10^{14}-10^{16}\,\persqcm$
rarely reach transmission values of $F \lesssim 0.3$.  
The pixels with $F \lesssim 0.3$ are usually contributed either by blends of absorbers 
or optically thick LLSs \citep[see also][]{pieri:2014}. 

An advantage of our large sample size is that also able to directly estimate the
error covariances, $\mathbf{C}_{\mathrm{boot}}$, via bootstrap
resampling--- an example is shown in Figure~\ref{fig:covar_boot}.  In
contrast to the \lya\ forest transmission PDF from high-resolution data which
have significant off-diagonal covariances \citep{bolton:2008}, 
the error covariance from
the BOSS transmission PDF is nearly diagonal with just some small correlations
between neighboring bins, although we also see some anti-correlation
between transmission bins at $F \sim 0.8$ and $F \sim 1$. 

It is interesting to compare the transmission PDF from our data with 
that measured by \citet{desjacques:2007} from SDSS DR3. This comparison is shown in 
Figure~\ref{fig:pdf_desj}, in which the transmission PDFs 
calculated from SDSS DR3 \lya\ forest spectra with $\snr > 4$ (kindly provided by 
Dr.\ V.\ Desjacques) are shown for two redshift
bins, juxtaposed with the BOSS transmission PDFs calculated from 
spectra with the same redshift and S/N cuts.  

 While there is some resemblance between the two PDFs, 
 the most immediate difference 
 is that the \citet{desjacques:2007} PDFs are shifted to lower transmission values, i.e., the mean
 transmission, $\langle F \rangle$, is considerably smaller than that from our BOSS data:
 $\langle F \rangle (z=2.4) = 0.73$ and $\langle F \rangle (z=3.0)= 0.64 $ from their
 measurement, whereas the BOSS PDFs have $\langle F \rangle (z=2.4)= 0.80$
 and $\langle F\rangle (z=3.0) = 0.70 $. 
 This difference arises because the \citet{desjacques:2007} used a power-law
 continuum (albeit with corrections for the weak emission lines in the quasar continuum) 
 extrapolated from $\lambrest > 1216\ang$ in the quasar restframe;
 this does not take into account the power-law break that appears to occur in low-redshift quasar spectra
 at $\lambrest \approx 1200\ang$ \citep{telfer:2002,suzuki:2006}. 
 Later in their paper, \citet{desjacques:2007} 
 indeed conclude that this must be the case in order to be consistent with other $\fmean(z)$ 
 measurements.
 Our continua, in contrast, have been constrained to match existing measurements
 of $\langle F \rangle (z)$, for which there is good agreement between different authors at 
 $ z \lesssim 3$ \citep[e.g.,][]{faucher-giguere:2008a, becker:2013}.
 
Another point of interest in Figure~\ref{fig:pdf_desj} is that the error bars of the
 BOSS sample are considerably smaller than those of the earlier measurement.
This difference is largely due to the significantly larger sample size of BOSS.
The proportion of pixels with $F \lesssim 0$ appears to be smaller in the BOSS
PDFs compared with the older data set, but this is because \citet{desjacques:2007}
did not remove DLAs from their data.

We next describe the creation of mock \lya\ absorption spectra designed to match the
properties of the BOSS data.

\section{Modeling of the BOSS Transmission PDF}\label{sec:model}
In this section, we will describe simulated \lya\ forest mock spectra 
designed, through a `forward-modelling' process, to have the same characteristics as the BOSS spectra, 
for comparison with the observed transmission PDFs described in the previous section.
For each BOSS spectrum which had contributed to our transmission PDFs in the previous section, 
we will take the \lya\ absorption from randomly selected simulation sightlines, then introduce the characteristics
of the observed spectrum using auxiliary information returned by our pipeline.  

Starting with simulated spectra from a set of detailed hydrodynamical IGM simulations, 
we carry out the following steps, which we will describe in turn in the subsequent subsections:
\begin{enumerate}
\item Introduce LLS absorbers 
\item Smooth the spectrum to BOSS resolution
\item Add metal absorption via an empirical method using lower-redshift SDSS/BOSS quasars
\item Add pixel noise, based on the noise properties of the real BOSS spectrum using parameters
estimated by our MCMC noise estimation technique
\item Simulate continuum errors by refitting the noisy mock spectrum
\end{enumerate}

\begin{figure}
 \epsscale{1.15}
 \plotone{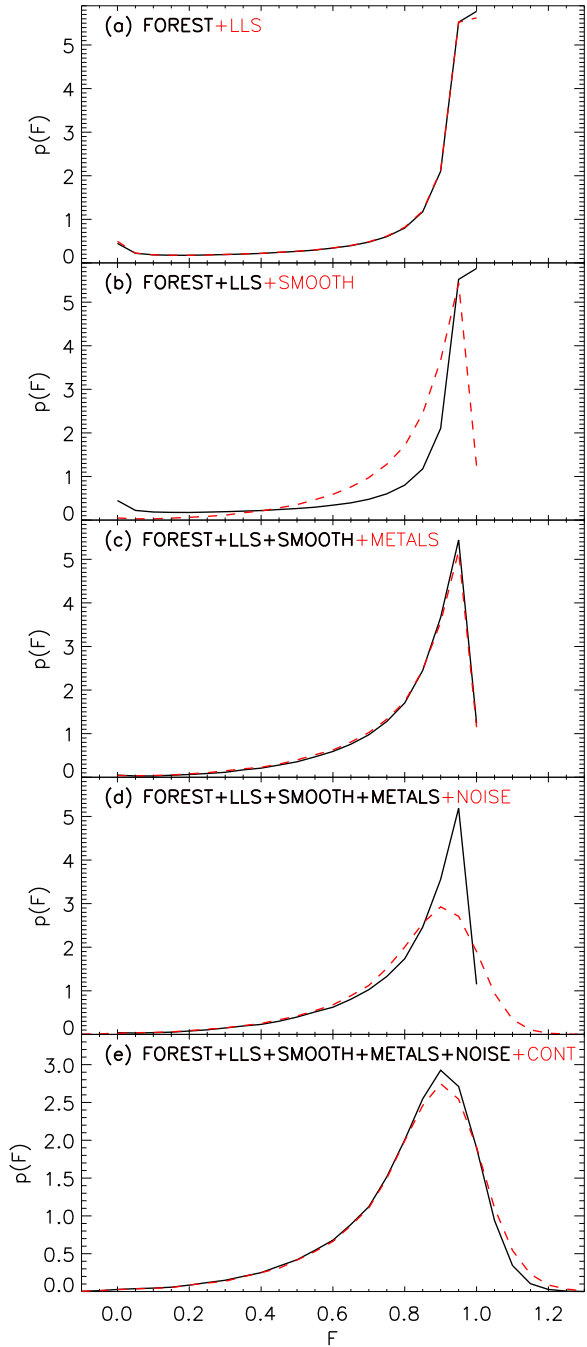}
 \caption{\label{fig:pdf_steps}
Cumulative effect of various aspects of our forward model that attempts to reproduce
the \lya\ forest transmission PDF from BOSS. Starting with the `raw' transmission PDF from the simulations (top),
 the black curve in each panel shows the PDF from the prior panel, while
the red curve shows the effect from:  (a) the addition of LLS; (b) smoothing from the finite spectrograph
resolution; (c) contamination from lower-redshift metals; (c) pixel noise; (e) continuum fitting errors. The transmission PDF modeled
in this figure is from the $\zav = 2.3$, $8 < \snr < 10$ bin.
 }
 \end{figure}

In the subsequent subsections, we will describe each step in detail. The effect
of each step in on the observed transmission PDF is illlustrated in Figure~\ref{fig:pdf_steps}.

\subsection{Hydrodynamical Simulations} \label{sec:sims}

As the basis for our mock spectra, we use hydrodynamic simulations 
run with a modification of the
publicly available {\small{GADGET-2}} code.  This code implements a
simplified star formation criterion \citep{springel:2005} that converts
all gas particles that have an overdensity above 1000 and a
temperature below $10^5$ K into star particles (see \citealt{viel:2004}).
The simulations used are described in detail in \citet{becker:2011} and in 
\citet{viel:2013}.

The reference model that we use is a box of
length 20 $h^{-1}$ comoving Mpc with $2\times512^3$ gas and cold DM
particles (with a gravitational softening length of 1.3 $h^{-1}$ kpc)
in a flat $\Lambda$CDM universe with cosmological parameters
$\Omega_{\rm m}=0.274$, $\Omega_{\rm b}=0.0457$, $n_{\rm
  s}=0.968,H_0=70.2\rm\,km\,s^{-1}\,Mpc^{-1}$ and $\sigma_8=0.816$, in
agreement both with WMAP-9yr  \citep{komatsu:2011} and Planck data 
\citep{planck-collaboration:2013}.  The
initial condition power spectra are generated with {\small
  CAMB} \citep{lewis:2000}. For the boxes considered in this work, we have verified that the transmission PDF
  has converged in terms of box size and resolution.

We explore the impact of different thermal histories on the \lya\
forest by modifying the ultraviolet (UV) background photo-heating
rates in the simulations as done in e.g., \citet{bolton:2008}.  A power-law
temperature-density relation, $T=T_{0} \Delta^{\gamma-1}$, arises
in the low density IGM ($\Delta <10$) as a natural consequence of
the interplay between photo-heating and adiabatic cooling
\citep{hui:1997, gnedin:1998}.  
The value of $\gamma$ within a simulation can be modified by varying a density-dependent
heating term \citep[see, e.g.,][]{bolton:2008}.
We consider a range of values for the temperature
at mean density, $T_{0}$, and the power-law index of the
temperature-density relation, $\gamma$, based on the observational
measurements presented recently by \citet{becker:2011}.  These
consist of a set of three different indices for the temperature-density
relation, $\gamma(z=2.5)\sim 1.0,1.3,1.6$, that are kept roughly
constant over the redshift range $z=[2-6]$ and three different
temperatures at mean density, $T_0(z=2.5)\sim [11000, 16000, 21500]\,$K,
which evolve with redshift, yielding a total of nine different thermal
histories. Between $z=2$ and
$z=3$ there is some temperature evolution and the IGM becomes hotter at
low redshift; at $z=2.3$, the models have $T_0 \sim [13000, 18000, 23000]\,$K.
We refer to the intermediate temperature model as our `reference' model, or \tref{}, 
while the hot and cold models are referred to as \thot{} and \tcold{}, respectively.
The values of $T_0$ of our simulations at the various redshifts are summarized in 
Table~\ref{tab:T0}.

\begin{deluxetable}{l  ccc}
\tablecolumns{5}
\tablewidth{0.45\textwidth}
\tablecaption{\label{tab:T0} Evolution of $T_0$ in Hydrodynamical Simulations} 
\tablehead{\colhead{$\langle z \rangle$} & \colhead{\tcold{}} & \colhead{\tref{}} & \colhead{\thot{}} } 
\startdata
2.3  & $13000$K & $18000$K & $23000$K \\
2.6 & $11000$K  &  $16000$K & $21500$K \\
3.0 & $9000$K & $14000$K  & $19000$K
\enddata
\end{deluxetable}

Approximately 4000 core hours were required for each simulation
run to reach $z=2$.  The physical properties of the \lya\
forest obtained from the {\sc TreePM/SPH} code {\sc GADGET-2} are in
 agreement at the percent level with those inferred from the
moving-mesh code \textsc{AREPO} \citep{bird:2013} and with the Eulerian code
\textsc{ENZO} \citep{oshea:2004}.

For this study, the simulation outputs were saved at $z = [2.3, 2.6, 3.0]$, 
from which we extract 5000 optical depth sightlines binned to 2048 pixels each.
To convert these to transmission spectra, the optical depths were rescaled such that the skewers
collectively yielded a desired mean-transmission, $\langle F \rangle_\mathrm{Ly\alpha} \equiv \exp(-\tau_\mathrm{Ly\alpha})$.
For our fiducial models, we would like to use the mean-transmission values estimated byestimated by \citet{becker:2013}, which we denote as for $\langle F \rangle_\mathrm{Ly\alpha,B13} \equiv  \exp(-\tau_\mathrm{Ly\alpha,B13})$.
However, their estimates assume certain corrections from optically-thick systems and metal absorption.
We therefore add back in the corrections they made (see discussion in \S\ref{sec:cont}) 
to get their `raw' measurement for 
$\fmean$ that now includes all
optically thick systems and metals, and then remove these contributions assuming \emph{our} own 
LLS and metal absorption models (see below).

{
 Later in the paper, we will argue that our PDF analysis in fact places independent constraints on \fmeanlya. }
 
  \subsection{Lyman-limit systems} \label{sec:lls_fid}
  In principle, all optically-thick \lya\ absorbers such as Lyman-limit systems (LLSs) and
  damped \lya\ absorbers (DLAs) should be discarded from \lya\ forest analyses, 
  since they do not trace the underlying matter density field in the same way 
  as the optically-thin forest (Equation~\ref{eq:fgpa}), and require radiative transfer simulations to accurately capture
  their properties \citep[e.g.,][]{mcquinn:2011c,rahmati:2013}.
  
 While DLAs are straightforward to identify through their saturated
absorption and broad damping wings even in noisy BOSS data \citep[see, e.g.,][]{noterdaeme:2012}, the
detection completeness of optically-thick systems through their
\lya\ absorption drops rapidly at $\nhi \lesssim 10^{20} \,\persqcm$.
Even in high-S/N, high-resolution spectra, optically thick systems can only be reliably
detected through their \lya\ absorption at $\nhi \gtrsim 10^{19}\,\persqcm$ (``super-LLS'').
Below these column densities, optically-thick systems can be identified either through
their restframe $912\,\ang$ Lyman-limit (albeit only one per spectrum) or using higher-order Lyman-series lines 
\citep[e.g.,][]{rudie:2013}. Neither of these approaches have been applied in previous \lya\ forest transmission
PDF analyses \citep{mcdonald:2000,kim:2007,calura:2012, rollinde:2013}, 
so arguably all these analyses are contaminated by LLSs.

Instead of attempting to remove LLSs from our observed spectra, we instead incorporate them into our
mock spectra through the following procedure.
For each PDF bin, we evaluate the total
 redshift pathlength of the contributing BOSS spectra (and
 corresponding mocks) --- this quantity is summarized in
 Table~\ref{tab:pdfbins}.  This is multiplied by $l_\mathrm{LLS}(z)$,
 the number of LLS per unit redshift, to give the total number of LLS
 expected within our sample. We used the published estimates of this
 quantity by \citet{ribaudo:2011}\footnote{Note that the value $l_{z0}=0.30$ given
 in Table~6 of \citet{ribaudo:2011} is actually erroneous, and the
 correct normalization is in fact $l_{z0} = 0.1157$, consistent with
 the data in their paper, which is used in Equation~\ref{eq:lls_lz}.
 Dr.\ J.\ Ribaudo, in private communication, has concurred with this
 conclusion.} which is valid over $0.24 < z <
 4.9$: \beq \label{eq:lls_lz} l_\mathrm{LLS}(z) = l_{z0} (1 +
 z)^{\gamma_\mathrm{LLS}}, \eeq where $l_{z0} = 0.1157$ and
 $\gamma_\mathrm{LLS}= 1.83$.  

 \begin{figure*}
 \plotone{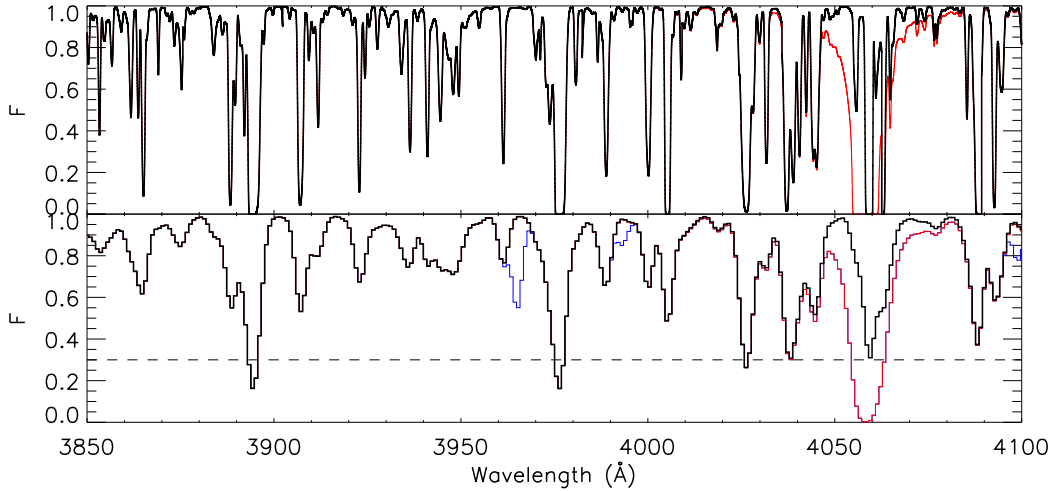}
 \caption{\label{fig:steps1} Simulated $\zav=2.3$ \lya\ forest skewer
   from our hydrodynamical simulations, without smoothing (top panel)
   and smoothed to BOSS resolution (bottom panel). The black curve is
   the simulated transmission directly extracted from the simulations,
   while the red curve is the same transmission field but with a LLS
   added at $\lambda = 4057\,\ang$ or $ z= 2.337$. The blue curve in
   the bottom panel shows the effect of the metal absorbers added
   using our empirical method. For illustrative purposes, we have
   specifically chosen to this simulated sightline to have significant
   LLS and metal absorption; it is possible for a sightline to have
   neither. The dashed-horizontal line denotes $F = 0.3$, below which
   our fiducial transmission PDF model disagrees with BOSS (see
   \S~\ref{sec:results}).  }
 \end{figure*}

After estimating the total number of LLSs in our mock spectra, $l_\mathrm{LLS}(z) \Delta z$, we add them at random points
 within our set of simulated optical depth skewers. 
 We also experimented with adding LLSs such that they are correlated with regions that already have high column
 density \citep[e.g.,][]{font-ribera:2012b}, but we found little significant changes to the transmission PDF and therefore 
 stick to the less computationally-intensive random LLSs.

For each model LLS, we then draw a column density using the published LLS column density distribution, $f(\nhi)$,
from \citet{prochaska:2010}. This distribution is measured at $z\approx 3.7$, so we make the assumption that
$f(\nhi)$ does not evolve with redshift between $2 \lesssim z \lesssim 3.7$. 
For our column densities of interest, this distribution is represented by the broken power-laws:
\beq \label{eq:lls_plaw}
f(\nhi) = \begin{cases}
   k_1 \nhi^{-0.8} & \mathrm{if}\; 10^{17.5} < \nhi < 10^{19.0} \\
    k_2 \nhi^{-1.2}  & \mathrm{if}\;  10^{19.0} < \nhi < 10^{20.3} \end{cases}.
\eeq
For the normalizations $k_1$ and $k_2$, we demand that
\beq \label{eq:lls_norm}
\int^{10^{19.0}}_{10^{17.5}} k_1 \nhi^{-0.8}\, \mathrm{d}\nhi + \int^{10^{20.3}}_{10^{19.0}} k_2 \nhi^{-1.2}\, \mathrm{d}\nhi = 1,
\eeq
and require both power-laws to be continuous at $\nhi = 10^{19.0}\,\persqcm$. These constraints 
produce $k_1 = 10^{-4.505}$ and $k_2 =  10^{3.095}$.
After drawing a random value for the column density of each LLS, 
we add the corresponding Voigt profile to the optical depth in the simulated skewer.

In addition to the LLS with column densities of $10^{17.5}\,\persqcm <
\nhi < 10^{20.3}\,\persqcm$ that are defined to have $\tau_\mathrm{HI}
\geq 2$, there is also a population of partial Lyman-limit systems
(pLLSs) that are not well-captured in our hydrodynamical simulations
since they have column densities ($10^{16.5}\,\persqcm \lesssim \nhi <
10^{17.5}\,\persqcm$) at which radiative transfer effects become
significant ($\tau_\mathrm{HI} \gtrsim 0.1$).  However, the incidence
rates and column-density distribution of pLLSs are ill-constrained
since they are difficult to detect in normal LLS searches.  We
therefore account for the pLLS by extrapolating the low-end of the
power-law distribution in Equation~\ref{eq:lls_plaw} down to $\nhi =
10^{16.5}\,\persqcm$, i.e.\ 
\beq f(10^{16.5}\,\persqcm < \nhi <
10^{17.5}\,\persqcm) = k_1 \nhi^{-0.8}.  
\eeq
This simple extrapolation does not take into account constraints from the mean-free path 
of ionizing photons \citep[e.g.,][]{prochaska:2010} which predicts a steeper slope for the pLLS
distribution, but we will explore this later in \S\ref{sec:mod_lls}.

 Comparing the integral
of this extrapolated pLLS distribution with Equation~\ref{eq:lls_norm}
leads us to conclude that \beq l_\mathrm{pLLS}(z) = 0.197\;
l_\mathrm{LLS}(z), \eeq and we proceed to randomly add pLLSs to our mock
spectra in the same way as LLSs.

{The other free parameter in our LLS model is their effective $b$-parameter distribution. 
However, due to the observational difficulty in identifying $\nhi\lesssim 18.5\,\persqcm$ 
LLSs the $b$-parameter distribution of this distribution has, to our knowledge, never been quantified.
Due to this lack of knowledge, it is common to simply adopt
a single $b$-value when attemping to model LLSs \citep[e.g.,][]{font-ribera:2012b,becker:2013}.
We therefore assume that all our pLLSs and
LLSs have a $b$-parameter of $b = 70\,\kms$ similar to DLAs \citep{prochaska:1997},
 an `effective' value meant to capture the blending of multiple \lya\ components.
However, the $b$-parameter for this population of absorbers is a highly uncertain quantity 
and as we shall see, it will need to be modified to provide a satisfactory fit to the data although it will
turn out to not strongly affect our conclusions regarding the IGM temperature-density relationship.}

 \subsection{Spectral Resolution}
 
The spectral resolution of SDSS/BOSS spectra is 
$R \equiv \lambda/\Delta \lambda \approx 1500-2500$ \citep{smee:2013}.
 The exact value varies significantly both as a function of
 wavelength, and across different fibers and plates
 depending on observing conditions (Figure~\ref{fig:disp}).
 
For each spectrum, the BOSS pipeline provides an estimate of the
$1\sigma$ wavelength dispersion at each pixel, $\sigma_{\rm disp}$, in
units of the co-added wavelength grid size ($\Delta \log_{10} \lambda
= 10^{-4}$). The spectral resolution at that pixel can then be
obtained from the dispersion,
through the following conversion: $R \approx ( 2.35
\times \scien{1}{-4} \ln 10\, \sigma_{\rm disp})^{-1}$.
Figure~\ref{fig:disp} shows the pixel dispersions from 236
randomly-selected BOSS quasar as a function of wavelength at the blue
end of the spectrograph.  Even at fixed wavelength, there is a
considerable spread in the dispersion, e.g., ranging from $\sigma_{\rm
  disp} \approx 0.9 - 1.8$ at $3700 \ang$.  The value of $\sigma_{\rm
  disp}$ typically decreases with wavelength (i.e., the resolution
increases).
 
 In their analysis of the \lya\ forest 1D transmission power spectrum, \citet{palanque-delabrouille:2013a} made their own
 study of the BOSS spectral resolution by directly analysing the line profiles of the mercury and cadmium arc lamps used
 in the wavelength calibration. They found that the pipeline underestimates the spectral resolution as a function of fiber position 
 (i.e.\ CCD row) and wavelength: the discrepancy is $<1\%$ at blue wavelengths and near the CCD edges, but increases to as much 
 as 10\% at $\lambda \sim 6000\,\ang$ near the center of the blue CCD (c.f.\ Figure~4 in \citealt{palanque-delabrouille:2013a}). 
 Our analysis is limited to $\lambda \leq 5045\,\ang$, i.e. $z \leq 3.15$, where the discrepancy is under 4\%.  
 Nevertheless, we implement these corrections to the BOSS resolution estimate
  to ensure that we model the spectral resolution to an accuracy of $< 1\%$.

For each BOSS \lya\ forest segment that contributes to the observed 
transmission PDFs discussed in \S~\ref{sec:bosspdf}, we concatenate randomly-selected
 transmission skewers from the simulations described in 
the previous section. This is because the simulation box size of $L=20\,\mpc$ 
($\Delta v \sim 2,000\,\kms$) is significantly shorter than the path length of
our redshift bins ($\Delta z = 0.3$, or $\Delta v \approx 27,000\,\kms$).
This ensures that each BOSS spectrum in our sample has a mock spectrum that is
exactly matched in pathlength.

We then directly convolve the simulated skewers by a Gaussian kernel with a standard deviation that varies with wavelength, 
using the estimated resolution from the
real spectrum, multiplied by the \citet{palanque-delabrouille:2013a} resolution corrections. 
The effect of smoothing on the transmission PDF is illustrated by the 
red-dashed curve in Figure~\ref{fig:pdf_steps}b. 
Smoothing decreases the proportion of pixels
with high-transmission ($F \approx 1$) and with 
high-absorption ($F\approx 0$), and increases
the number of pixels with intermediate transmission values.

\subsection{Metal Contamination} \label{sec:metalmodel}

 \begin{figure}
 \includegraphics[width=0.47\textwidth,clip=true, trim=15 0 4 00]{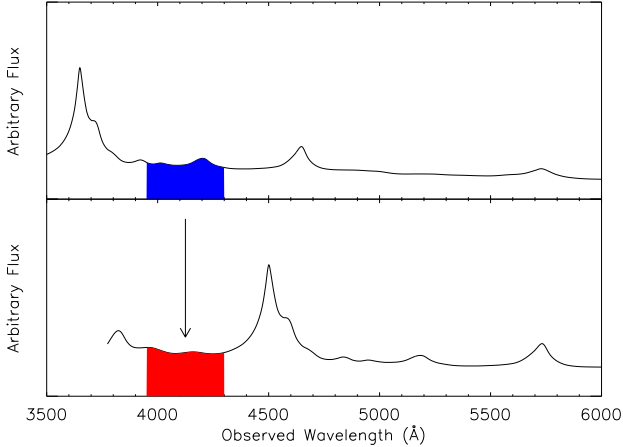}
 \caption{\label{fig:metalcartoon} An illustration of our empirical
   `sideband' model of metal contamination in our mock
   \lya\ forest spectra.  The lower panel shows the $\zq=2.7$ quasar
   along with its \lya\ forest region (red) which we wish to model.
   To its corresponding mock spectrum, we add metals observed in the
   $\lambrest \approx 1260-1390\,\ang$ region of a lower-redshift
   ($\zq = 2.0$) quasar (blue region in top panel).  }
 \end{figure}

 \begin{figure*}
 \begin{center} \includegraphics[width=0.7\textwidth, clip=true, trim= 10 0 21 0 ]{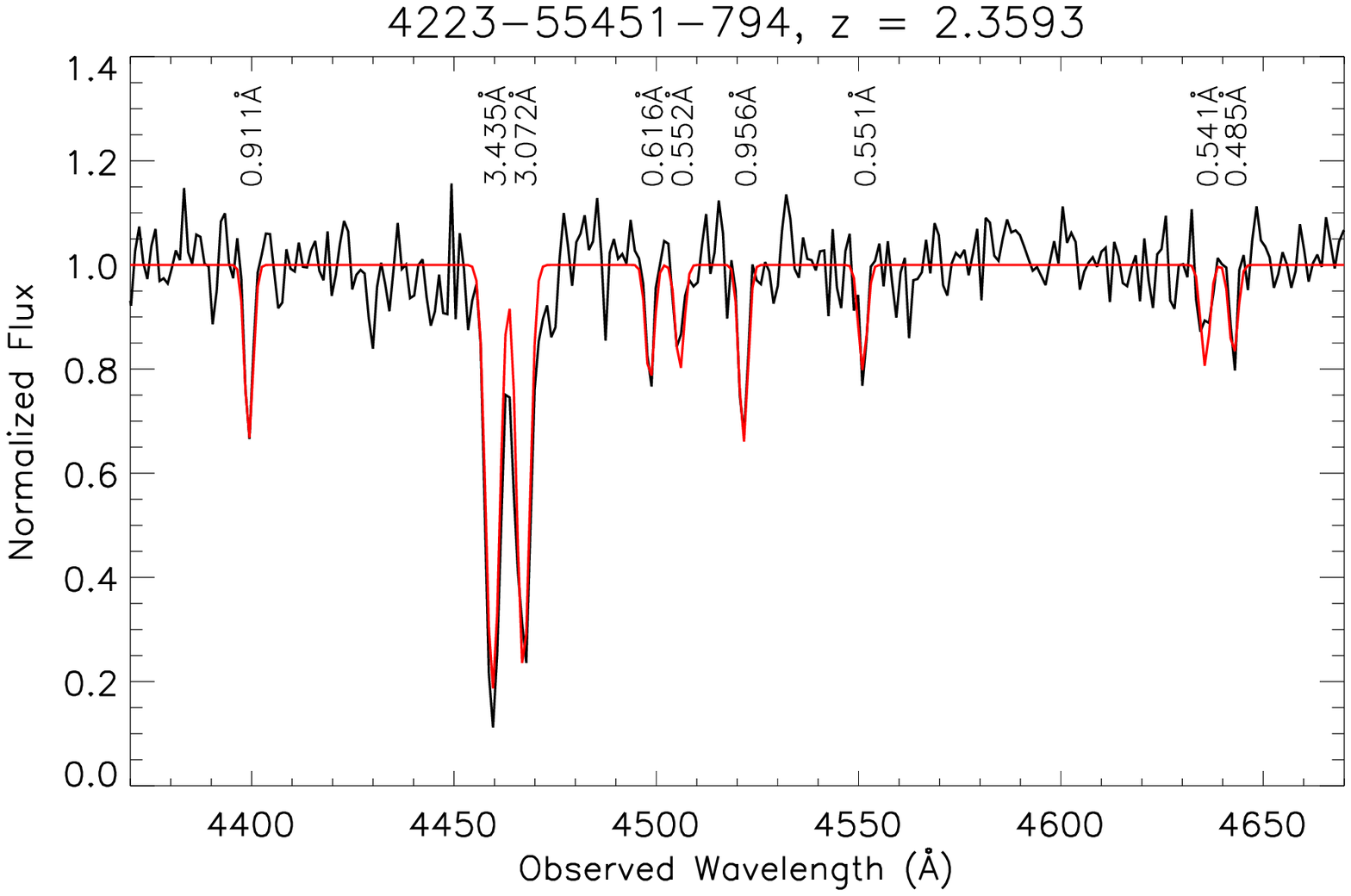}  \end{center}
 \caption{\label{fig:metalfit} A continuum-normalized spectrum of a
   BOSS quasar showing the metal absorbers in the $1300\,\ang <
   \lambrest < 1390\,\ang$ `sideband' region, which would be used to
   add metals to $\zav = 2.6$ mock \lya\ forest spectra.  The red
   curve shows our metal model for this spectrum, generated from the
   observed wavelengths and equivalent widths in the absorber catalog
   generated by the automatic algorithm of \citet{lundgren:2009}.  We
   also assume that the absorbers all lie on the saturated portion of
   the curve-of-growth and have $\tau_0 = 3$, with the equivalent
   width (labeled above each absorption line) proportional to the
   $b$-parameter.  The model absorption profiles represented by the
   red curve would be added to our mock \lya\ forest spectra. We have
   chosen to plot this particular `sideband' because it has more
   absorbers than average --- the typical spectrum has less metal
   absorption than this.  }
 \end{figure*}

Metal absorption along our observed \lya\ forest sightlines acts as a
contaminant since their presence alters the observed statistics of the
\lya\ forest.  In high-resolution data, this contamination is usually
treated by directly identifying and masking the metal absorbers, 
although in the presence of line blending it is unclear how thorough this approach can be.  

With the lower S/N and
moderate resolution of the BOSS data, direct metal identification and
masking is not a viable approach. Furthermore, most of the weak metal
absorbers seen in high-resolution spectra are not resolved in the BOSS
data.

{Rather than removing metals from the BOSS \lya\ forest spectra, 
we instead add metals as observed in lower-redshift quasar spectra.}
In other words, we add absorbers observed in the restframe $\lambrest \approx 1260-1390\,\ang$ region of lower-redshift quasars
with $1 + \zq \approx (1216\,\ang/1300\,\ang)(1 + \zav)$, 
such that the observed wavelengths are matched to the \lya\ forest segment with average redshift $\zav$.
Figure~\ref{fig:metalcartoon} is a cartoon that illustrates this concept.
This method makes no assumption about the nature of the metal absorption in the \lya\ forest, and
includes all resolved metal absorption spanning the whole range of redshifts down to $z \sim 0$.
The disadvantage of this method is that it does not include metals with intrinsic wavelengths 
$\lambda \lesssim 1300\,\ang$, but the relative contribution of such metal species towards the transmission PDF should
be small\footnote{\ion{Si}{3} an obvious exception, but we will later account for this omission in our error bars (\S\ref{sec:systematics}).}
 since most of the metal contamination comes from low-redshift ($z \lesssim 2$) \ion{C}{4} and \ion{Mg}{2}. 

We use a metal catalogue generated by B.\ Lundgren \etal\ \citep[in
  prep; see also][]{lundgren:2009}, which lists absorbers in SDSS
\citep{schneider:2010} and BOSS quasar spectra \citep{paris:2012} ---
the SDSS spectra were included in order to increase the number of $\zq
\approx 1.9-2.0$ quasars needed to introduce metals into the $\zav =
2.3$ \lya\ forest mock spectra, which are not well sampled by the BOSS
target selection \citep{ross:2012}.  We emphasize that we work with
the `raw' absorber catalog, i.e.\ the individual absorption lines have not been identified in terms
of metal species or redshift.  For each quasar, the catalog provides a
line list with the observed wavelength, equivalent width (EW, $W_r$),
full-width at half-maximum (FWHM), and detection S/N,
$W_r/\sigma_{W_r}$.  {To ensure a clean catalog, we use only
$W_r/\sigma_{W_r} \geq 3.5$ absorbers in the catalog that were
identified from quasar spectra with $\snr > 15$ per angstrom redwards
of \lya. The latter criterion ensures that even relatively weak lines (with $\mathrm{EW} \gtrsim 0.5\,\ang$)
 are accounted for in our catalog.}
Figure~\ref{fig:metalfit} shows an example of the lower-redshift quasar
spectra that we use for the metal modelling.

However, we want to add a smooth model of the metal-line absorption to add to our mock spectra, rather
than adding in a noisy spectrum. We therefore use a simple model as follows:
For each \lya\ forest segment we wish to model at redshift $\zav$, we
select an absorber line-list from a random quasar with $1 + \zq
\approx (1216\,\ang/1300\,\ang)(1 + \zav)$.  
We next assume that all resolved metals in the
SDSS/BOSS spectra are saturated and thus in the flat regime of the
curve-of-growth.  The equivalent width is then given by
\beq \label{eq:metal_ew} W_r \approx \left( \frac{2b}{c} \right)
\sqrt{\ln(\tau_0/ \ln 2)}, \eeq where $\tau_0$ is the optical depth at
line center, $b$ is the velocity width and $c$ is the speed of light.
In the saturated regime, $W_r$ is mostly sensitive to changes in $b$
while being highly insensitive to changes in $\tau_0$. 
We can thus adopt $\tau_0$ as a global constant and solve for $b$, given the $W_r$
of each listed absorber in the selected 'sideband' quasar. We have
found that $\tau_0 = 3$ provides a good fit for most of the absorbers.

We then add the Gaussian profile into our simulated optical depth skewers:
\beq
\tau = \tau_0 \exp\left[-\left(\frac{c}{b}\right) \left(\frac{\Delta \lambda}{\lambda}\right)^2 \right] 
\eeq
centered at the same observed wavelength, $\lambda$, as the real absorber.
The red curve in Figure~\ref{fig:metalfit} shows our model for the observed absorbers, using just the 
observed wavelength, $\lambda$, and equivalent width, $W_r$, from the absorber catalog.

{
Our method for incorporating metals is somewhat crude since one should, in principle, first deconvolve the
spectrograph resolution from the input absorbers, and then add the metal absorbers into our mock spectra prior to
convolving with the BOSS spectral resolution. 
In contrast, we fit $b$-parameters to the absorber catalog without spectral deconvolution, therefore these $b$-parameters
can be thought of as combinations of the true absorber width, $b_\mathrm{abs}$ and the spectral dispersion, $\sigma_\mathrm{disp}$, i.e.\
$b^2 \sim b^2_\mathrm{abs} + \sigma^2_\mathrm{disp}$.
While technically incorrect, this seems reasonable since the template quasar spectra and forest spectra that we are
attempting to model both have approximately the same resolution, and in practical terms this
 \textit{ad hoc} approach does seem to be able to reproduce the observed metals in 
the lower-redshift quasar spectra (Figure~\ref{fig:metalfit}). 
The other possible criticism of our approach is that it does not incorporate weak metal absorbers,
although we attempted to mitigate this by setting a very high S/N threshold on the template
quasars for the metals. However, we have checked that such weak metals do not significantly change
the forest PDF (and indeed metals in general do not seriously affect the PDF, c.f.\ Figure~\ref{fig:pdf_steps}c).
}

We also tried adding metals with similar redshifts to --- and correlated with --- forest absorbers 
(e.g., absorption by \ion{Si}{2} and \ion{Si}{3}) 
measured in \citet{pieri:2010} and \citet{pieri:2014} using a method described in the appendix of \citet{slosar:2011}. 
We found a negligible impact on the transmission PDF owing mainly to the fact that these correlated metals contribute only
$\sim 0.3\%$ to the overall flux decrement, so we neglect this contribution in our subsequent analysis.

\subsection{Pixel Noise} \label{sec:noisemodel}
It is non-trivial to introduce the correct noise to a simulated \lya\ forest spectrum:
given a noise estimate from the observed spectrum, one needs to first
ensure that the mock spectrum has approximately the same flux normalization
as the data. This is challenging, as the \lya\ forest transmission at any given pixel,
which ranges from 0 to 1, will vary considerably between the simulated spectrum
and the real data.

 \begin{figure*}
 \epsscale{0.9}
 \plotone{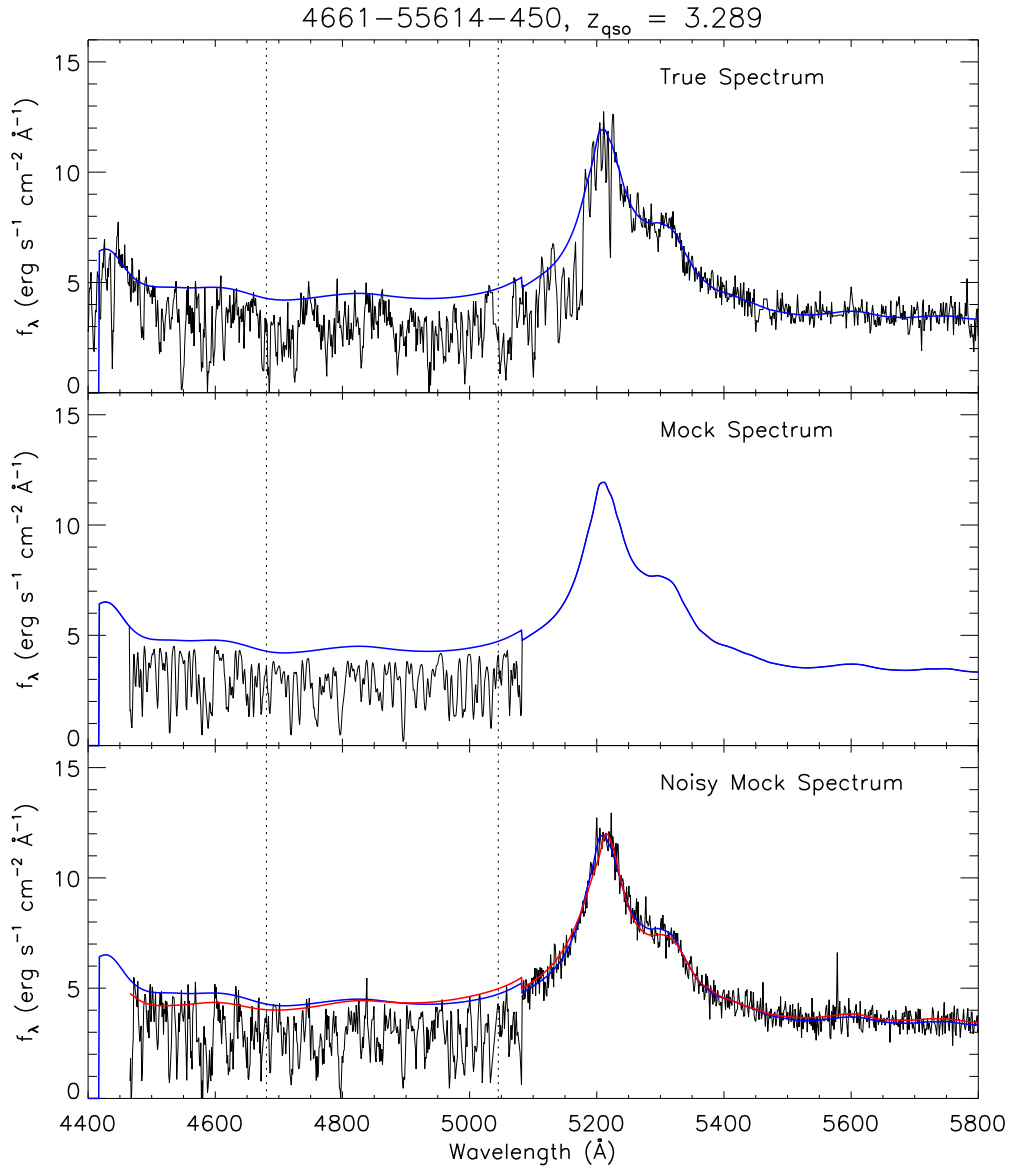}
 \caption{\label{fig:steps2}
Simulating the noise properties and continuum errors of a BOSS quasar. The top panel shows the
observed spectrum of a BOSS quasar, and its associated continuum fit, $c$, in blue. The middle panel
shows the simulated transmission spectra (after adding LLS, smoothing and adding metals) {multiplied by the quasar continuum fitted to the true spectrum}. In the lower panel, we have added noise to the mock spectrum
using the noise parameters estimated from the true spectrum (see \S~\ref{sec:mcmc}).
A new continuum, $c'$, (red) is re-fitted to the noisy mock spectrum. The difference between new continuum $c'$
and `true' continuum, $c$, of the mock (blue) introduces continuum errors to our model.
The vertical dotted lines indicates the range of pixels that contribute to the $\zav=3.0$ subsample in our transmission PDF;
a small segment between $(1+\zq)1040\,\ang = 4461\,\ang$ and $(1+2.75)1216\,\ang = 4560\,\ang$ also contributes to the $\zav = 2.6$ bin.
 }
 \end{figure*}

The simplest method of adding noise to a mock spectrum
is simply to introduce Gaussian deviates using the pipeline noise
estimate for each spectrum--- this was essentially the method used by \citet{desjacques:2007} and
the BOSS mocks described in \citet{font-ribera:2012a}.
However, with the MCMC co-addition procedure described in \S~\ref{sec:mcmc}, 
we are in a position to model the noise in a more robust and self-consistent fashion. 

Recall that the MCMC procedure returns posterior probabilities for two 
quantities: the true underlying spectral flux density, $\mathcal{F}_{\lambda}$, and the four free parameters $A_{j}$, 
which parametrize the noise in each spectrum.
This estimate of the $A_j$ from each quasar spectrum 
allows us to accurately model the pixel noise
using Equation~\ref{eqn:mcmc_noise}.

The MF-PCA method (\S~\ref{sec:cont}) produces an estimate of the
 quasar continuum, $c$, providing approximately the correct {flux level} 
at each point in the spectrum. 
We can now multiply $c$ with the simulated \lya\ forest
transmission spectra, $F$, which had already been smoothed to the same dispersion 
as its real counterpart (the estimated quasar continuum is already at approximately
the correct smoothing, since it was fitted to the observed spectrum). 

This procedure produces a noiseless mock spectrum with the correct
flux normalization and smoothing.  
{We can now generate noisy spectra corresponding to a given BOSS quasar, using our MCMC noise estimation described in 
Section~\ref{sec:mcmc}.
First, we substitute our mock spectrum as $\mathcal{F}_{\lambda}$ into Equation~\ref{eqn:mcmc_noise}, 
and then combine the $A_j$ noise parameters (estimated through our MCMC procedure) as well as the calibration vectors
$S_{\lambda,i}$ and sky estimates $s_\mathrm{\lambda,i}$. This lets us generate
self-consistent noise vectors corresponding to each individual exposure that make up the mock quasar spectrum, 
 $\sigma_{\lambda i}$. The noise vectors are then used to draw random Gaussian deviates that are added to the mock spectrum, 
 on a per-pixel basis, to create the mock spectral flux density,  $f_{\lambda i}$.
 Finally, we combine these individual mock exposures into the optimal spectral flux density for the mock spectrum, through
 the expression (see Appendix):}
\be f_{{\rm opt},\lambda} \equiv
\frac{1}{\sigma^2_{{\rm opt},\lambda}} \sum_{i} \frac{f_{\lambda
    i}}{\sigma_{\lambda i}^2}, \ee where \be \frac{1}{\sigma_{{\rm
      opt},\lambda}^2} \equiv \sum_i \frac{1}{\sigma_{\lambda
    i}^2}. \label{eqn:opt} \ee

Figure~\ref{fig:pdf_steps}c illustrates the effect of adding pixel noise
to the smoothed \lya\ forest transmission PDF. As expected, this scatters a significant fraction of pixels 
 to $F > 1$, and also to $F < 0$ to a smaller extent. 

\subsection{Continuum Errors}\label{sec:contmodel}
With the noisy mock spectrum in hand (see, e.g., bottom panel of Figure~\ref{fig:steps2}), 
we can self-consistently include the effect of continuum errors into our model transmission PDFs 
by simply carrying out our MF-PCA continuum-fitting procedure on the individual noisy mock spectra.
Dividing out the mock spectra with the new continuum fits then incorporates an estimate of the continuum
errors (estimated by \citealt{lee:2012a} to be at the $\sim 4-5\%$ RMS level) into the evaluated model transmission PDF.
This estimated error includes uncertainties stemming from the estimation of the quasar
continuum shape due to pixel noise, as well as the random variance in the mean \lya\ forest
absorption in individual lines-of-sight. 

Note that regardless of the overall mean-absorption in the mock spectra (i.e. inclusive of our models
for metals, LLSs, and mean forest absorption --- see \S~\ref{sec:mffit}), we always use $\fcont(z)$, the same input 
mean-transmission derived from \citet{becker:2013} (described in \S~\ref{sec:cont}) to fit the continua in both the
data and mock spectra. While the overall absorption in our fiducial model is consistent with that
from \citet{becker:2013}, as we shall see later, the shape of the transmission PDF retains information 
on the true underlying mean-transmission even if fitted with a mean-flux regulated continuum with a wrong input $\fmean(z)$. 

The effect of continuum errors on the transmission PDF is shown in Figure~\ref{fig:pdf_steps}e:
like pixel noise, it degrades the peak of the PDF, but only near $F \sim 1$.

\section{Model Refinement} \label{sec:results}
In an ideal world, one would like to do a blind analysis by generating the transmission PDF
model (\S\ref{sec:model}) in isolation from the data, before `unblinding' to compare with 
data --- this would then in principle yield results free from psychological bias in the model building.
However, as we shall see in \S\ref{sec:compare_init}, this does not give acceptable fits to the data
so we have to instead modify our model to yield a better agreement, in particular our LLS model (\S\ref{sec:mod_lls})
and assumed mean-transmission (\S\ref{sec:mffit}).

\subsection{Initial Comparison with \tref{} Models} \label{sec:compare_init}
\begin{figure*}
\begin{center}
\includegraphics[width=0.32\textwidth]{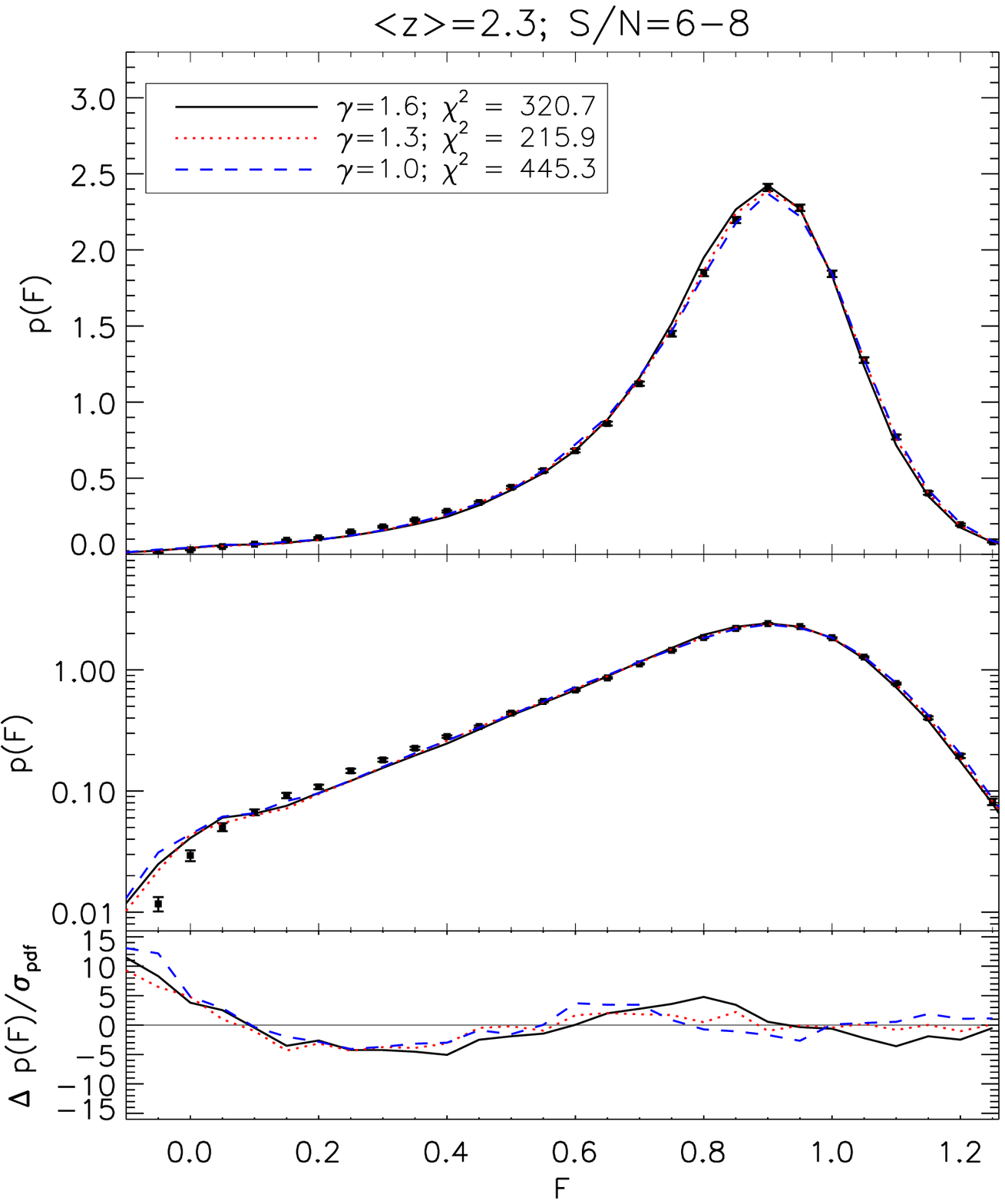}
\includegraphics[width=0.32\textwidth]{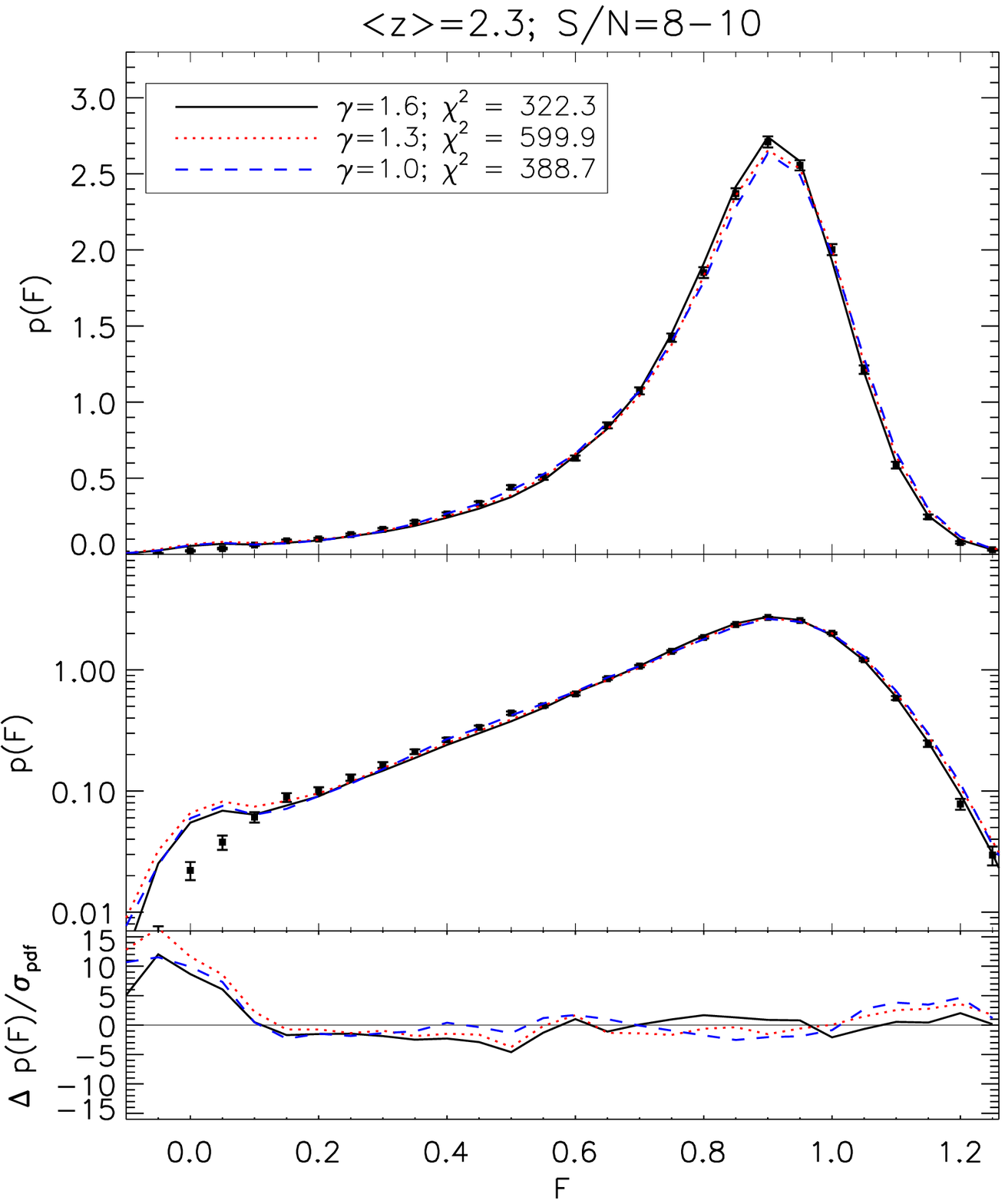}
\includegraphics[width=0.32\textwidth]{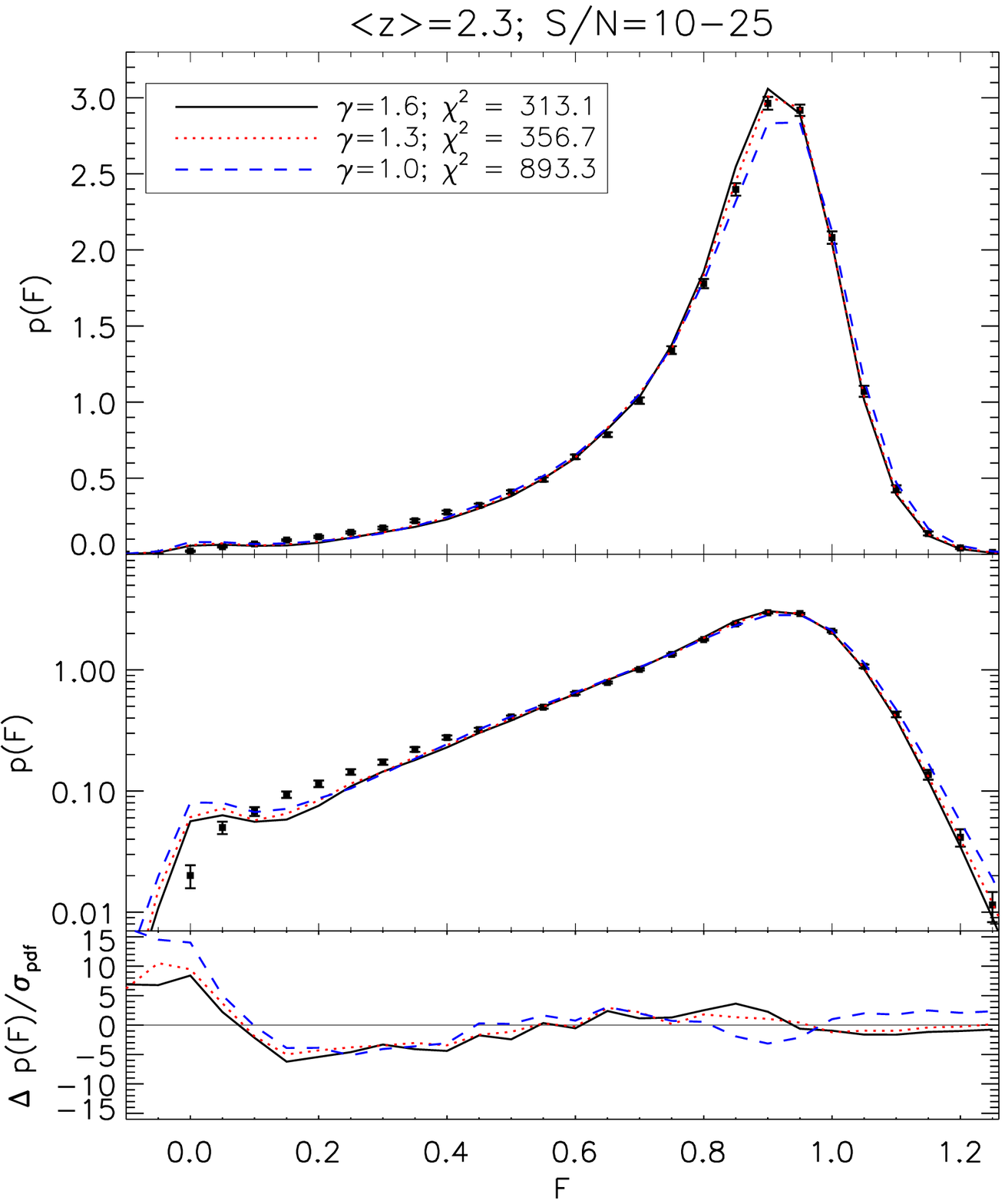} \\ 
\vspace{1em}
\includegraphics[width=0.32\textwidth]{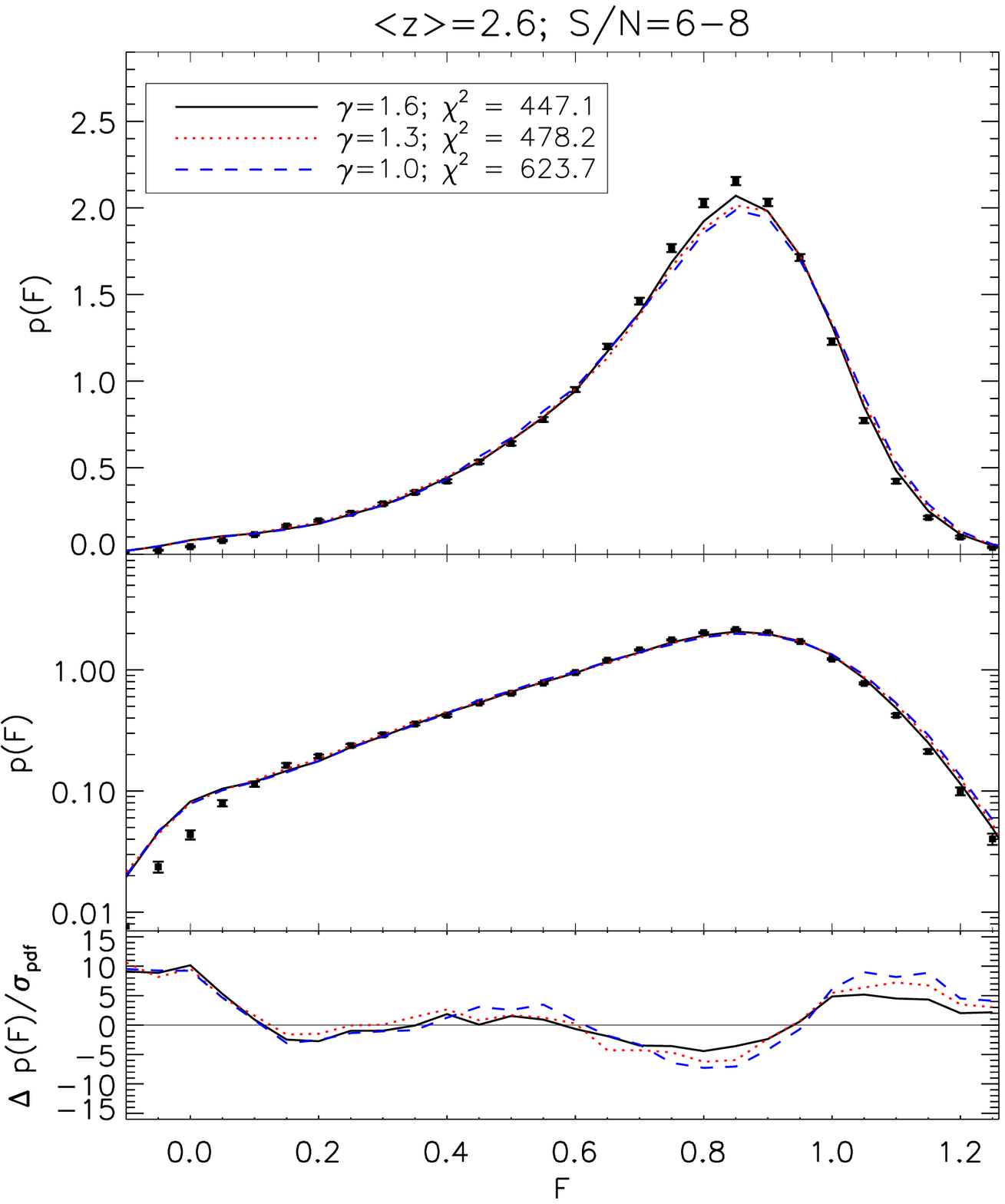}
\includegraphics[width=0.32\textwidth]{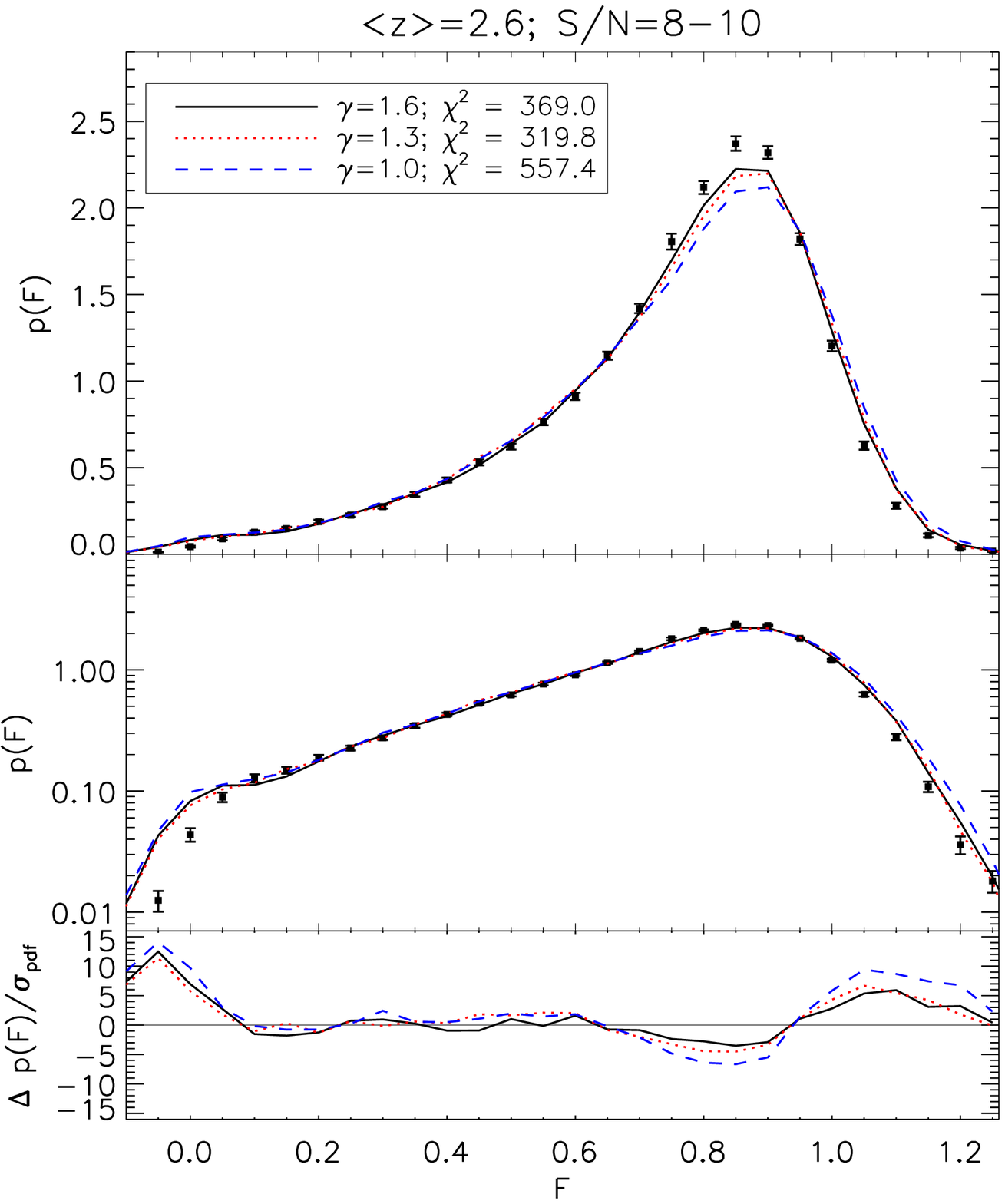}
\includegraphics[width=0.32\textwidth]{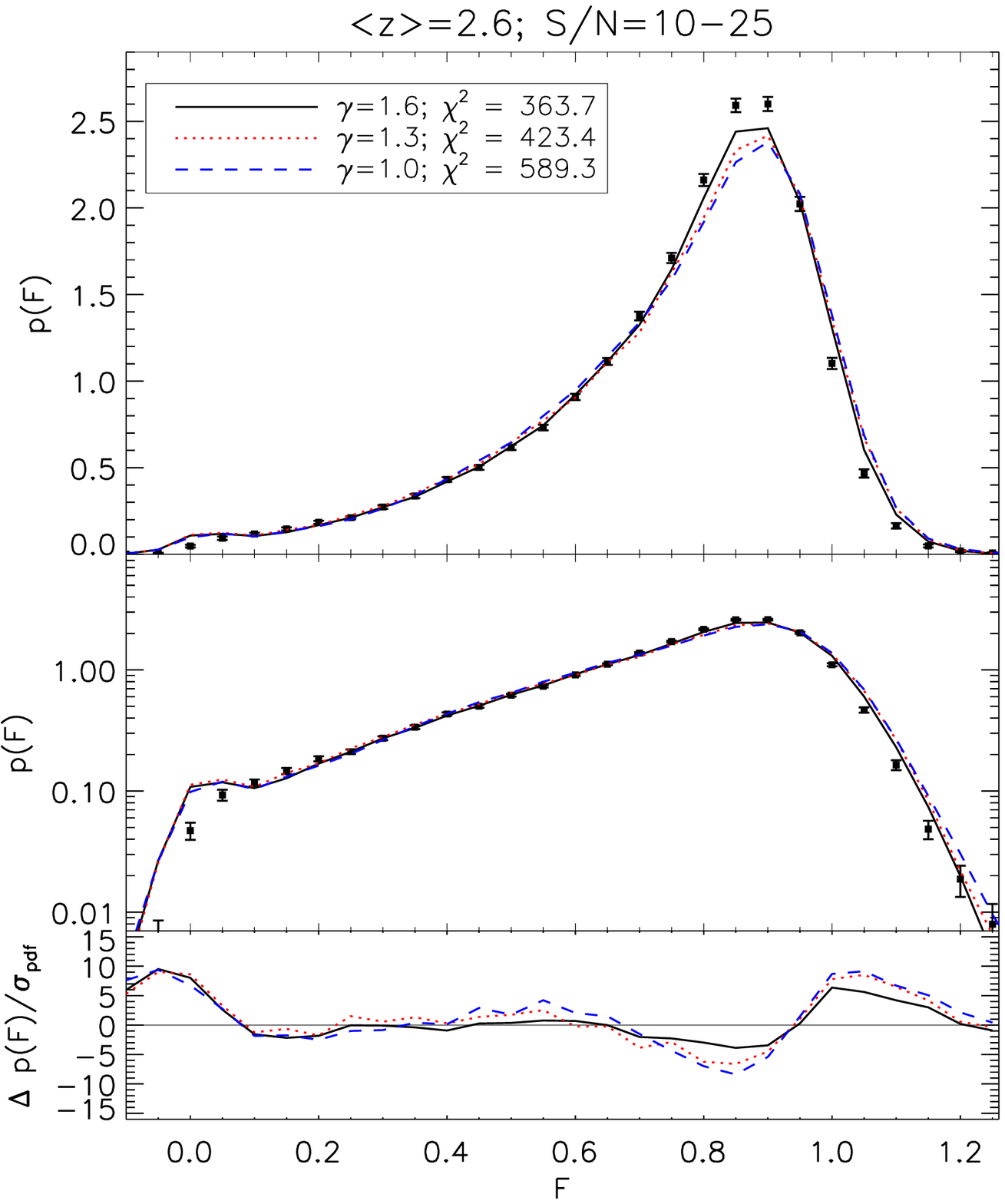} \\
\vspace{1em}
\includegraphics[width=0.32\textwidth]{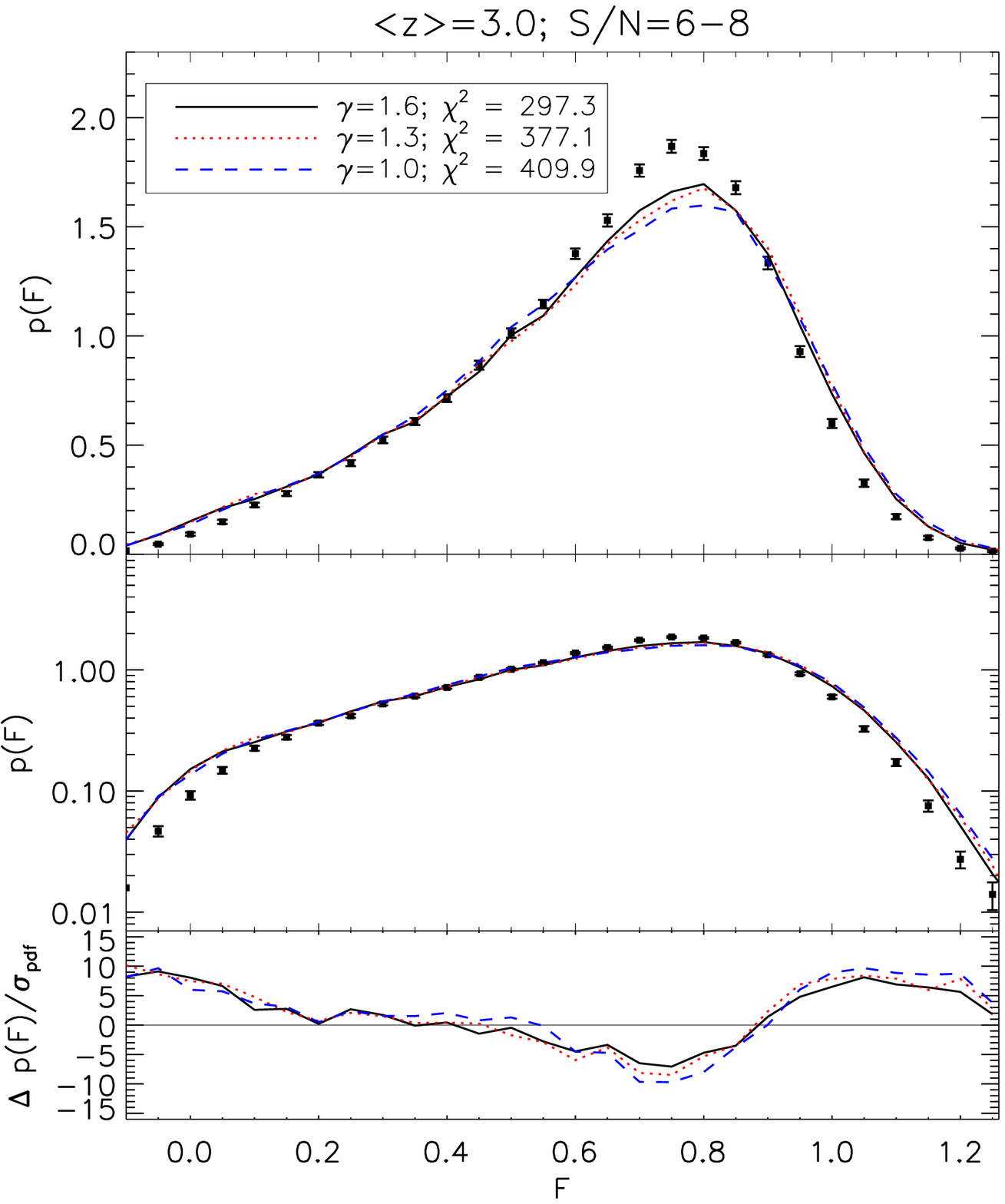}
\includegraphics[width=0.32\textwidth]{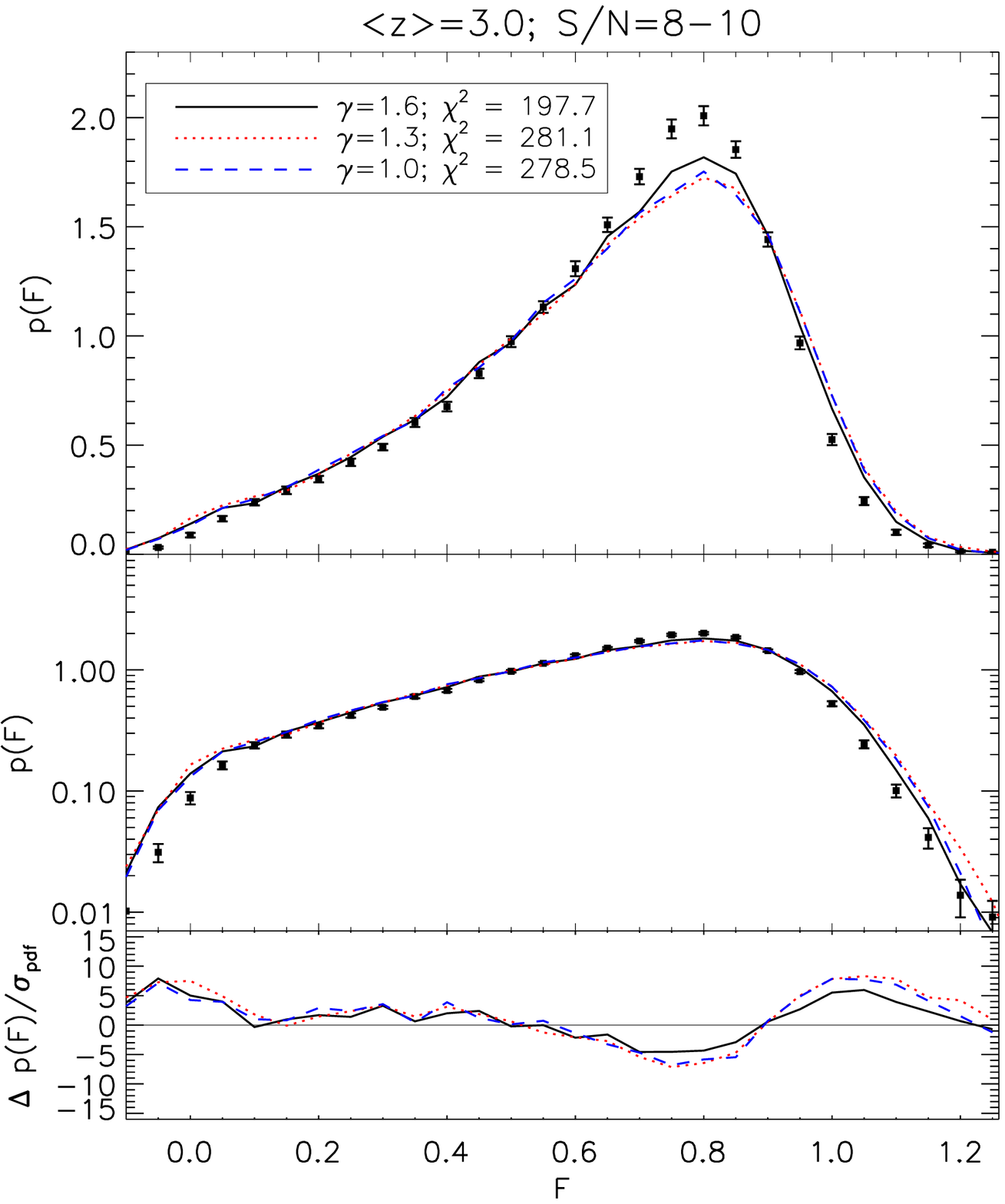}
\includegraphics[width=0.32\textwidth]{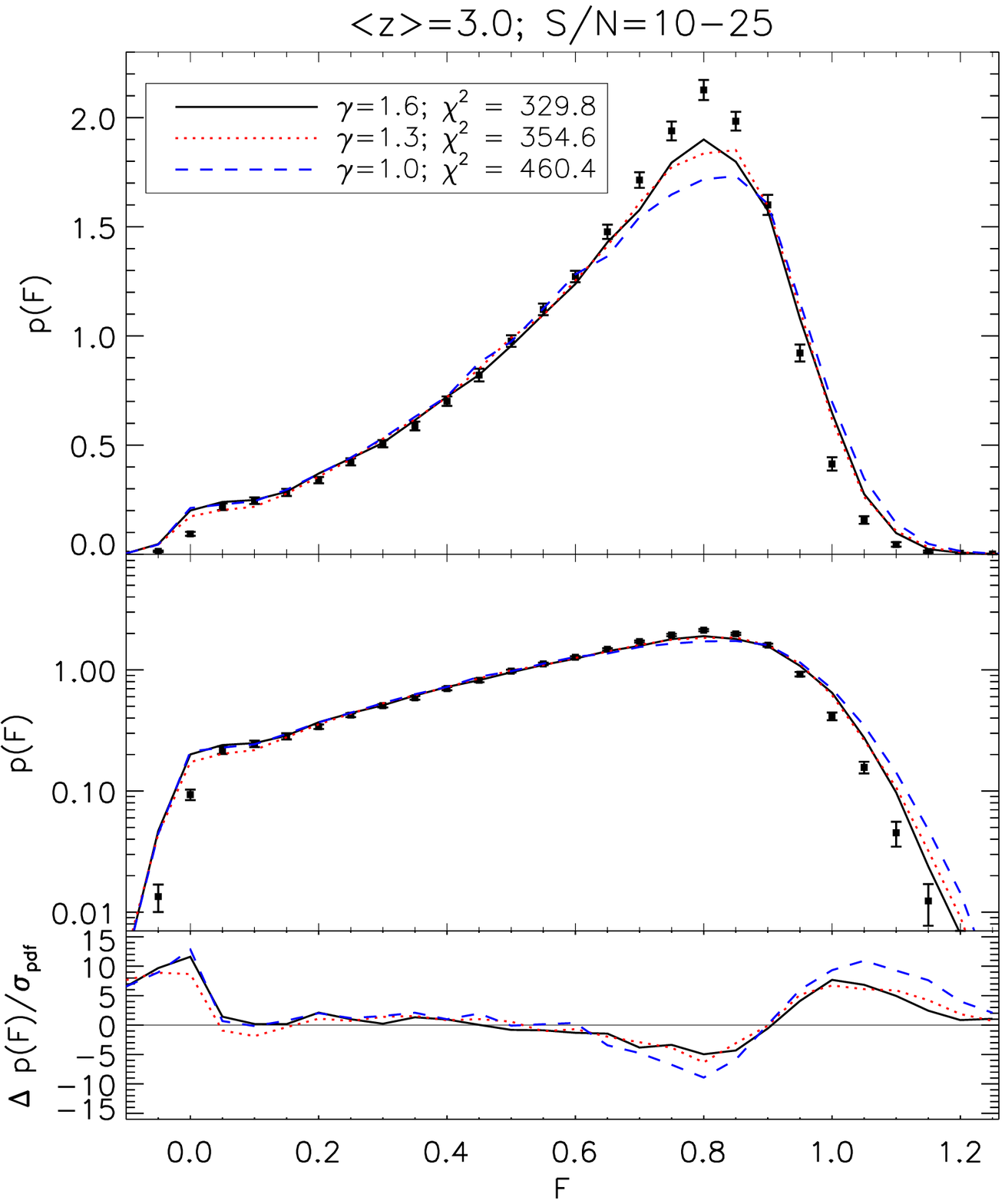} 
\end{center}
\caption{\label{fig:pdfgamma_fid}
An initial comparison between the transmission PDFs observed from BOSS \lya\ forest data (error bars) and simulated
PDFs generated from the \tref{} hydrodynamical simulations (curves) with the method described in \S~\ref{sec:model};
each row is at the same redshift, while the different columns display the different S/N cut.
The points with the error bars are the PDFs measured from the BOSS data (estimated from bootstrap resampling,
 while the black, dotted-red and dashed-blue
curves denote simulated PDFs with $\gamma=[1.5,1.3,1.0]$ respectively.
The top and middle panels show the transmission PDFs with linear and logarithmic axes, while the 
lower panels show the pull, i.e.\ residuals between the simulated PDF and the data PDF, divided by the error.
The $\chi^2$ values indicated in these plots are for 24 d.o.f., and clearly indicate unacceptable fits
to the data --- modifications to the model are required.
}
\end{figure*}

For each of our 9 hydrodynamical simulations (sampling 3 points each in
$T_0$ and $\gamma$), we determine the transmission PDF from the \lya\ forest
mock spectra that include the effects described in the previous
section, for the various redshift \& S/N subsamples in which we had
measured the PDF in BOSS (\S\ref{sec:bosspdf}).  In
Figure~\ref{fig:pdfgamma_fid}, we show the transmission PDFs for all our
redshift and S/N subsamples in BOSS, compared with the corresponding
simulated transmission PDFs from the \tref{} simulation with $\gamma=[1.0,1.3,1.6]$. 
Note that the error bars shown
are the diagonal elements of the covariance matrix estimated through bootstrap
resampling on the data.  

At first glance, the model transmission PDFs seem
to be a reasonably match for the data, especially considering we have
carried out purely forward modelling without fitting for any
parameters.  However, when comparing the `pull', $(p_{\mathrm{data},i}
- p_{\mathrm{model},i})/\sigma_{p,i}$, between the data and model (bottom panels of 
Figure~\ref{fig:pdfgamma_fid}), we
see significant discrepancies in part due to the extremely small bootstrap error bars.  
Nevertheless, it is gratifying to see
that the shape of the residuals are relatively consistent across the
different S/N subsamples at fixed redshift and $\gamma$, since this
indicates that our spectral noise model is robust.

We proceed to quantify the differences between the simulated transmission
PDFs, $p_{\mathrm{model}}$, and observed transmission PDFs, $p_\mathrm{data}$,
with the $\chi^2$ statistic: \beq \label{eq:chisq} \chi^2 =
\sum_{ij} (p_{\mathrm{model},i}-p_{\mathrm{data},i})^T
C^{-1}_{ij}(p_{\mathrm{model},j}-p_{\mathrm{data},j}), \eeq where
we use the bootstrap error covariance matrix, $\mathbf{C}_\mathrm{boot}$.
Note that we also include a bootstrap error term that accounts for the sample variance in the model
transmission PDFs, since our pipeline for generating mock spectra is too computationally expensive to include sufficiently large amounts of
skewers to fully beat down the sample variance in the models\footnote{We aim for $3-4\times$ more 
mock spectra than in the corresponding data sample, 
but later when we have to compute large model grids we are limited to models with the same size
as the data.}.

{We limit our model comparison to the range $-0.1 \leq F \leq 1.2$, i.e. 27 transmission
bins with bin width $\Delta(F) = 0.05$.  
This range covers pixels that have been scattered to `unphysical' values of
$F<0$ or $F>1$ due to pixel noise, as is expected from the low-S/N of our BOSS data, 
and also captures $>99.8\%$ of the pixels within each of our data subsets.
In particular, it is important to retain the bins with $F > 1$ because the $F \sim 1$ transmission bins are highly
sensitive to $\gamma$ \citep{lee:2012} and therefore we want to fully
sample that region of the PDF even if it will require careful modeling of
pixel noise and continuum errors.}

There are two constraints on all our transmission
PDFs: the normalization convention 
\beq \label{eq:pdf_norm1} \int
p(F)\; \mathrm{d}F = 1 \eeq and the imposition of the same mean
transmission due to the mean-flux regulated continuum-fitting
\beq \label{eq:pdf_norm2} \int F\, p(F)\; \mathrm{d}F = \langle F
\rangle_\mathrm{cont} \eeq such that all the mock spectra have the
same absorption, $\fcont(z)$.  This is because the mock spectra have
been continuum-fitted (\S\ref{sec:contmodel}) in exactly the same way
as the BOSS spectra, which assumes the same mean \lya\ transmission
inferred from the \citet{becker:2013} measurements (\S\ref{sec:cont}).
The `true' optically-thin mean-transmission, $\fmeanlya$, imposed on the simulation skewers
is in principle a different quantity from $\fcont$, since the latter
includes contribution from metal contamination and optically-thick
LLSs.  

This leaves us with $\nu = 27 -1 -2=24$ degrees of freedom
(d.o.f.) in our $\chi^2$ comparison.  
The $\chi^2$ for all the models
shown in Figure~\ref{fig:pdfgamma_fid} are shown in the corresponding
figure legends.

In this initial comparison, the $\chi^2$ values for the models in Figure~\ref{fig:pdfgamma_fid} are clearly unacceptable:
we find $\chi^2 \gtrsim 200$ for 24 d.o.f. in all cases. However, it is interesting to note that the 
$\gamma = 1.6$ or $\gamma=1.3$ models are preferred at all redshifts and S/N cuts.
Note that the S/N=$8-10$ subsamples (middle column in Figure~\ref{fig:pdfgamma_fid}) tends to have a slightly better agreement 
between model and data compared to the other
S/N cuts at the same redshift: this simply reflects the smaller quantity of data of the subsample (c.f.\ Table~\ref{tab:pdfbins})
and hence larger bootstrap errors.

A closer inspection of the residuals in Figure~\ref{fig:pdfgamma_fid}
indicate that there are two major sources of discrepancy between the
models and data: firstly, at the low-transmission end, we underproduce pixels at $0.1 \lesssim F
\lesssim 0.4$ while simultaneously over-producing $F\lesssim 0.1$ pixels, especially at $\zav = 2.3$ and $\zav = 2.6$.  
This seems to affect all $\gamma$ models equally.
\citet{pieri:2014} found that at BOSS resolution, pixels with $F\lesssim 0.3$ come predominantly
from saturated \lya\ absorption from LLS. 
We therefore investigate possible modifications to our LLS model in \S\ref{sec:mod_lls}. 

The other discrepancy in the model transmission PDFs manifests
at the higher-transmission end in the $\zav = 2.6$ and $\zav = 3$ subsamples, where we see a
sinusoidal shape in the residuals at $F > 0.6$ that appears consistent across different S/N.  
This portion of the transmission PDF depends on both $\gamma$ and, as we shall see, on the assumed 
mean-transmission $\fmean(z)$, which we shall discuss in more detail in \S\ref{sec:mffit}.

Finally, our transmission PDF model includes various uncertainties in the modelling of metals, 
LLSs, and continuum-fitting which have not yet been taken into account. 
In \S\ref{sec:systematics}, we will estimate the contribution of these uncertainties,
by means of a Monte-Carlo method, in our error covariances.

\subsection{Modifying the LLS Column Density Distribution} \label{sec:mod_lls}
With the moderate spectral resolution of BOSS, there are few individual pixels in the optically-thin \lya\ forest 
that reach transmission values of $F \lesssim 0.4$. 
Such low-transmission pixels are typically due to either the blending of multiple absorbers \citep[see, e.g., Figure~2 in][]{pieri:2014}, 
or optically-thick systems (see Figure~\ref{fig:steps1} in this paper).

As we have seen in Figure~\ref{fig:pdfgamma_fid}, at low-transmission
values the discrepancy between data and model has a distinct shape,
which is particularly clear at $\zav = 2.3$: the models underproduce
pixels at $0.1 \lesssim F \lesssim 0.4$ while at the same time
overproducing saturated pixels with $F \approx 0$.

To resolve this particular discrepancy would
therefore require either drastically increasing the amount of clustering in the \lya\ forest, or modifying our assumptions on the LLSs
in our mock spectra.
The first possibility seems rather unlikely since the \lya\ forest power on relevant scales are well-constrained 
\citep{palanque-delabrouille:2013a}, and would in any case require new simulation suites to address ---  beyond the scope 
of this paper.

\begin{figure}
\includegraphics[width=0.5\textwidth, clip=true, trim = 10 0 0 0]{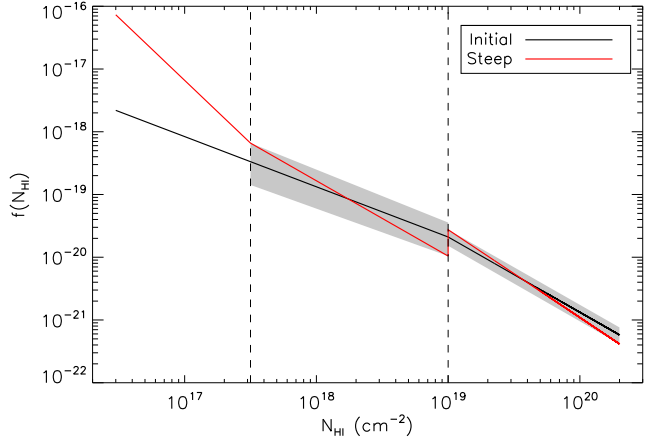}
\caption{\label{fig:lls_dist}
LLS and pLLS column-density power-law distributions used in our initial model (black; \S\ref{sec:lls_fid}) 
and steeper modification (red; \S\ref{sec:mod_lls}).
The distributions are normalized assuming the overall LLS incidence rate at $z=2.25$ (c.f.\ Eq.~\ref{eq:lls_lz}). 
The vertical dashed-lines denotes the $\nhi = 10^{17.5}\,\persqcm$ boundary between pLLS and LLS, and $\nhi = 10^{19}\,\persqcm$
boundary between LLS and super-LLS. The shaded regions show the range of possible distributions as determined by
\citet{prochaska:2010}, but there are few robust constraints in the $10^{16.5}\,\persqcm \leq \nhi \leq 10^{17.5}\,\persqcm$ pLLS regime.
The `initial' distribution was used in the preliminary data comparisons in \S\ref{sec:compare_init},
but all subsequent analysis (after \S\ref{sec:mod_lls}) assumes the `steep' distribution.}
\end{figure}

On the other hand, it is not altogether surprising that our fiducial column density distribution (\S~\ref{sec:lls_fid}) ---
 which was measured at $z \approx 3.7$ \citep{prochaska:2010} --- do not reproduce the BOSS data at $\zav = 2.3-2.6$.
 We therefore search for a LLS model that better describes the low-transmission end of the BOSS \lya\ forest. 
Looking at the $\zav=2.3$ PDFs in Figure~\ref{fig:pdfgamma_fid}, we see that our fiducial model
 \emph{over}-produces pixels at $F=0$, yet is deficient at slightly higher $F$. This suggests that our model
 is over-producing super-LLS ($\nhi > 10^{19}\,\persqcm$) that contribute large absorption troughs with $F=0$, while 
 not providing sufficient lower-column density absorbers that can individually reach minima of $0.1 \lesssim F \lesssim 0.4$
 when smoothed to BOSS resolution. In other words, our fiducial model appears to have an excessively `top-heavy' LLS column density
 distribution. 
 
 \begin{figure}
 \includegraphics[width=0.5\textwidth]{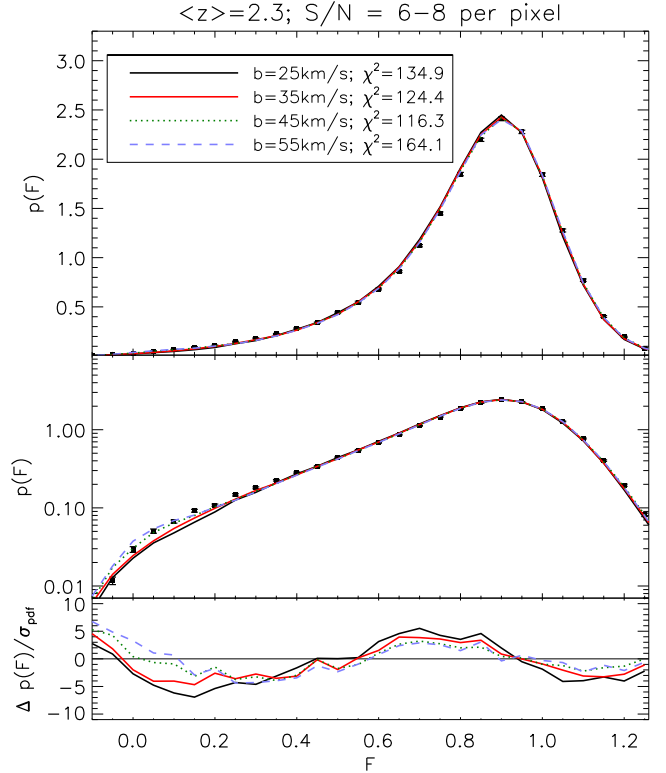}
 \caption{\label{fig:pdf_bpar}
Variation of the transmission PDF as a function of LLS $b$-parameter.
All model transmission PDFs here are computed from the \tref{}, $\gamma=1.6$ model assuming 
the revised pLLS/LLS distribution described in 
\S\ref{sec:mod_lls} (curves), compared with the S/N = $6-8$ BOSS transmission PDF at $\zav = 2.3$ (error bars).
The quoted $\chi^2$ values are for 24 d.o.f., and evaluated using only bootstrap error covariances.
We find that $b = 45\,\kms$ gives the best fit to the data. 
 }
 \end{figure}

For a change, we will try a LLS column density distribution with a more ample bottom-end, using the steepest power-laws 
within the $1\sigma$ limits estimated by \citet{prochaska:2010}:
\beq \label{eq:lls_plawsteep}
f(\nhi) = \begin{cases}
   k_1 \nhi^{-1.2} & \mathrm{if}\; 10^{17.5} < \nhi < 10^{19.0} \\
    k_2 \nhi^{-1.4}  & \mathrm{if}\;  10^{19.0} < \nhi < 10^{20.3} \end{cases}.
\eeq
We use the same $ l_\mathrm{LLS}(z)$ as before, and obey the integral
constraints from \citet{prochaska:2010} that demand that the ratios of
$\int f(\nhi)\; \mathrm{d}\nhi$ between the two column-density regimes
be fixed.  This gives us $k_1 = 10^{2.819}$ and $k_2 = 10^{7.039}$,
although the new distribution is no longer continuous at $\nhi =
10^{19} \,\persqcm$.
This new distribution is illustrated by the red power-laws in Figure~\ref{fig:lls_dist}.

Another change we have made is to the partial LLS model, which was possibly too conservative in the fiducial model.
Instead of extrapolating from the LLS distribution, we now adopt the pLLS power-law slope of $\beta_\mathrm{pLLS} = -2.0$ 
inferred from the total mean-free path to ionizing photons by \citet{prochaska:2010}. This dramatically
increases the incidence of pLLS in our spectra relative to LLS: we now have $l_\mathrm{pLLS} = 1.8\, l_\mathrm{LLS}$, where 
$l_\mathrm{LLS}$ is the same value we used previously (Equation~\ref{eq:lls_lz}).
This increase, while large, is not unreasonable in light of the large uncertainties in direct measurements
on the \ion{H}{1} column-density distribution from direct \lya\ line-profile fitting \citep[e.g.,][]{janknecht:2006, rudie:2013}.
Note also that even this increased pLLS incidence only amounts to, on average, 
less than one pLLS per quasar ($\Delta(z) \sim 0.3-0.4$ per quasar at our redshifts).

\begin{figure}
\includegraphics[width=0.5\textwidth]{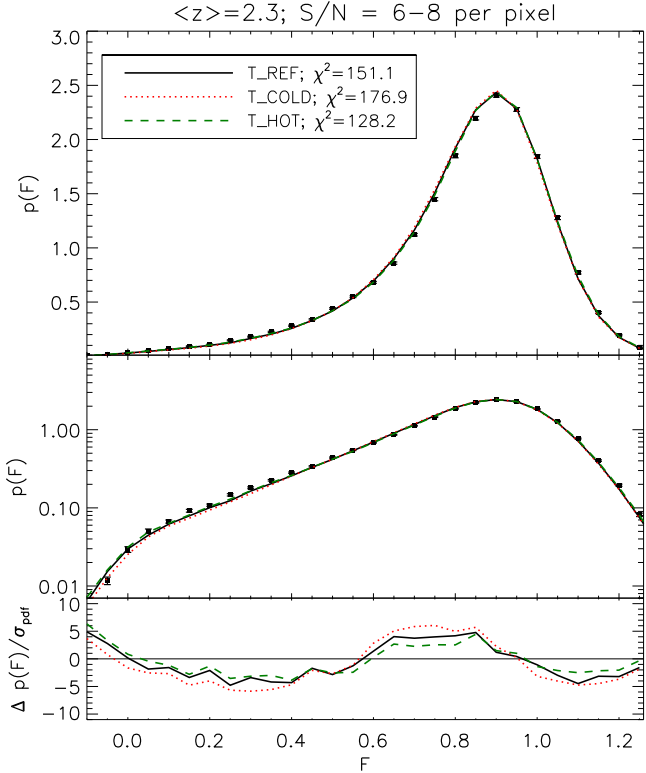}
\caption{\label{fig:pdf_temperature}
Variation of the transmission PDF as a function of the IGM temperature at mean-density, $T_0$. 
All model transmission PDFs here have the same temperature-density relationship, $\gamma=1.6$, and
are compared with the S/N = 6-8 BOSS transmission PDF at $\zav = 2.3$ (error bars). 
The quoted $\chi^2$ values are for 23 degrees of freedom.
Note that in these models we have already implemented the improved LLS/pLLS model decribed in \S\ref{sec:mod_lls},
hence the much improved $\chi^2$ values compared those quoted in Fig.~\ref{fig:pdfgamma_fid}.
}
\end{figure}

We found that while increasing the number of pLLS relieves
the tension between data and model at $0.1 \lesssim F \lesssim 0.4$, it
does not resolve the excess at the fully absorbed $F \approx 0$ pixels
in the models.  However, changing the $b$-parameter of
the LLS and pLLS from our original fiducial value of $b = 70\,\kms$
modifies the PDF in a way that improves the agreement.  This is a
reasonable step, since the effective $b$-parameter is otherwise observationally
ill-constrained for the LLS and pLLS populations.  
This is because LLSs are typically complexes of multiple systems separated in velocity space, 
and while there have been analyses of the $b$-parameter in
these individual components, the `effective' 
$b$-parameter for complete LLS systems has never been quantified to our knowledge.  

We therefore search
for the best-fit $b$-parameter with respect to the \tref{},
$\gamma=1.3$ model at $\zav = 2.3$, focusing primarily on the
agreement in the $0 \leq F \leq 0.4$ bins (Figure~\ref{fig:pdf_bpar}).
Our choice of model for this purpose should not significantly affect
our subsequent conclusions regarding the IGM temperature-density slope, since
there is little sensitivity towards the latter in the relevant
low-transmission bins (c.f.\ Figure~\ref{fig:pdfgamma_fid}). 
However, there will be some degeneracy between the LLS $b$-parameter and
$T_0$ (Figure~\ref{fig:pdf_temperature}) since changing the latter does somewhat change the 
low-transmission portion of the PDF --- we will come back to this point in \S\ref{sec:discussion}.

As shown in
Figure~\ref{fig:pdf_bpar}, a value $b = 45\,\kms$ gives 
the best agreement with the data at $0 \leq F
\leq 0.4$.  
This yields $\chi^2 = 116$ for 24 d.o.f., which is dramatically improved over those
quoted in Figure~\ref{fig:pdfgamma_fid}, but still not quite a good fit.
In the subsequent results, we will adopt this steeper
pLLS/LLS model and $b$-parameter as the fiducial model in our analysis, and will
correspondingly decrease the degrees of freedom in our $\chi^2$
analysis to account for the fitting of $b$.

{Note that while significantly improving the PDF fit, this new $b$-parameter still
does not give a perfect fit to the low-transmission ($F<0.4$) end. 
This is probably due to the simplified nature of our LLS model, which neglects the
finite distribution of $b$-parameters and internal velocity dispersion of individual components.
These properties are currently not well-known, and it seems likely that an improved model would
allow a better fit to the low-transmission end of the PDF.
}

\subsection{Estimation of Systematic Uncertainties} \label{sec:systematics}

\begin{figure}
\includegraphics[width=0.5\textwidth, clip=true, trim= 0 1 0 0 ]{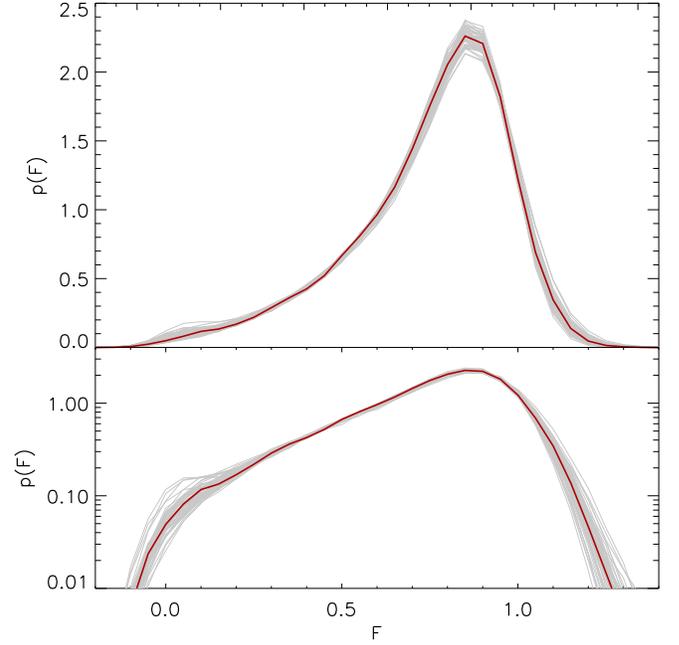}
\caption{\label{fig:pdfs_sys}
Grey curves show 50 model transmission PDFs with a random sampling of different LLS incidence rates, 
metal absorption, and continuum scatter, 
evaluated for the $\zav = 2.3$, S/N=$8-10$ BOSS subsample and using the \tref{} simulation with $\gamma = 1.6$. 
The red curve shows the
transmission PDF at our fiducial level of LLS incidence, metal absorption, and continuum scatter. 
The top panel is has a linear abscissa, while the lower panel has a logarithmic abscissa.
}
\end{figure}

\begin{figure}
\includegraphics[width=0.48\textwidth]{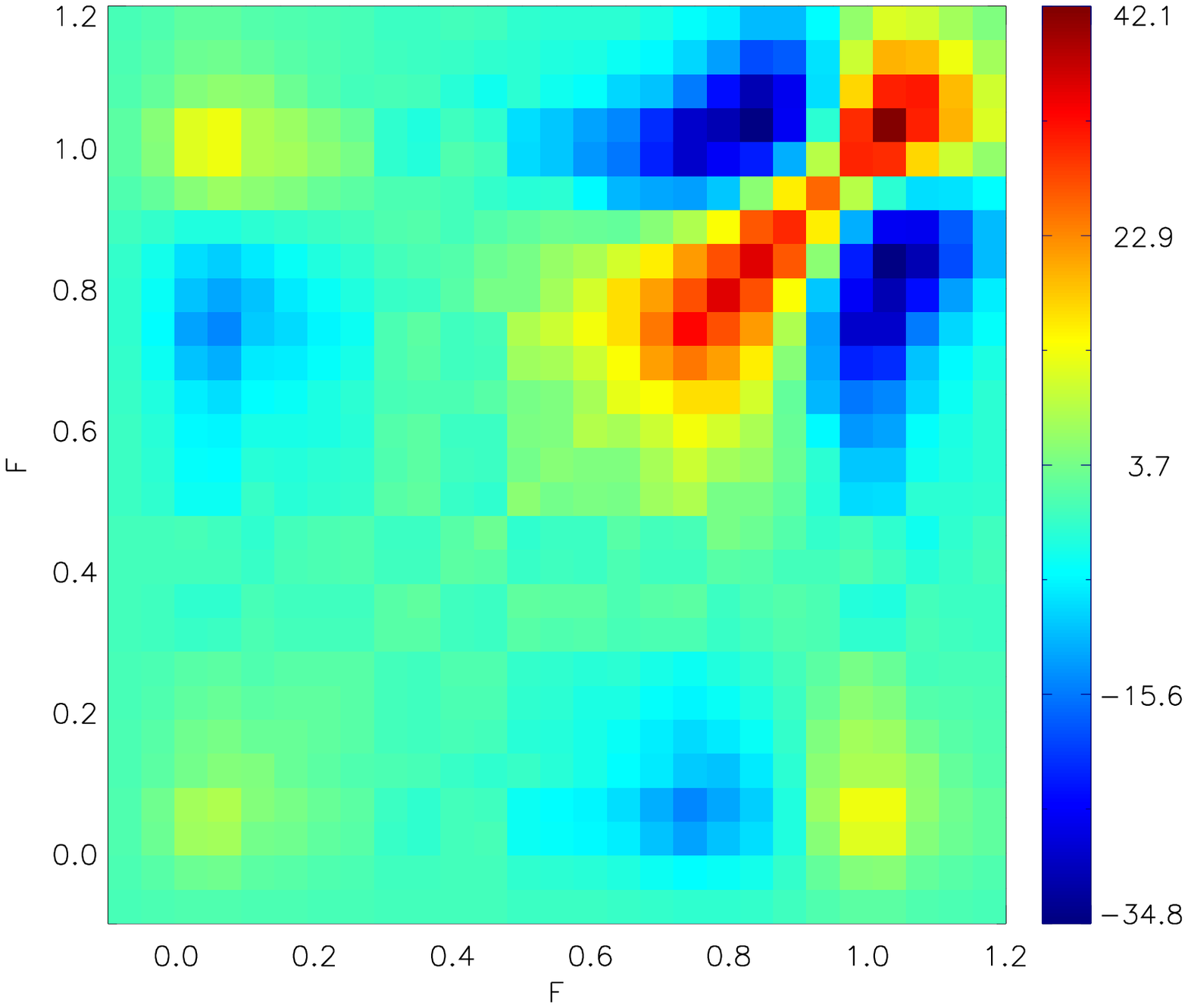}
\includegraphics[width=0.48\textwidth]{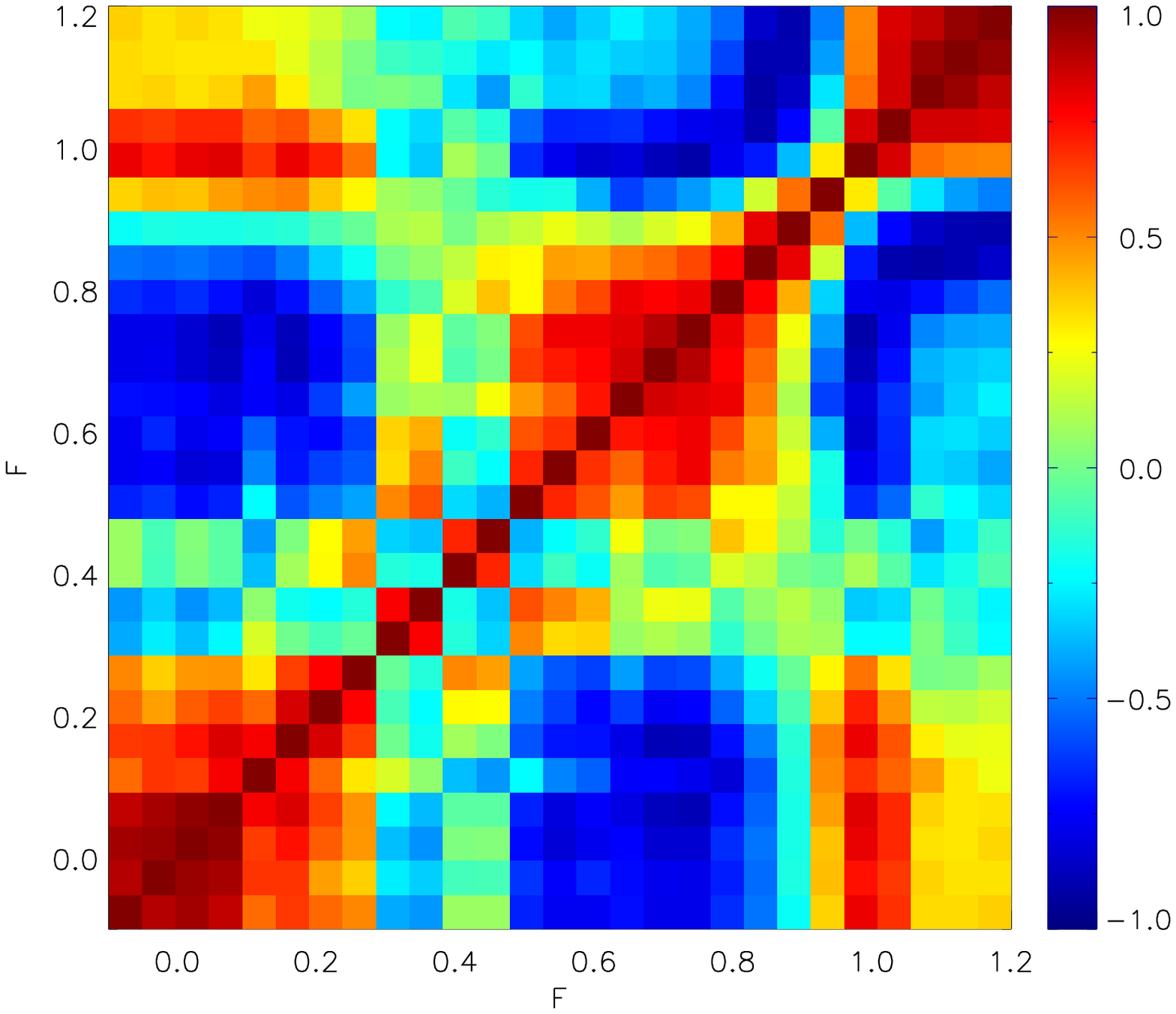}
\caption{\label{fig:covar_sys}
(Top) 2D density plot of the error covariance matrix representing our systematic uncertainties in the LLS incidence rate, pLLS 
column-density distribution, LLS $b$-parameter, metal absorption, and continuum scatter, 
as estimated through the Monte Carlo method described in \S~\ref{sec:systematics}. 
The bottom plot shows the corresponding correlation function. This particular covariance matrix was estimated for
the $\zav = 2.6$, S/N = $8-10$ subsample, and the values in the covariance have been multiplied by $10^4$ for clarity. 
}
\end{figure}

While we have estimated the sample variance of our BOSS transmission PDFs by bootstrap resampling on the spectra, 
there are significant uncertainties associated with each component of our transmission PDF model as described above, 
e.g.,\ the LLS incidence rate and level of continuum error. 
These uncertainties can be incorporated into a systematics covariance matrix,  
$\mathbf{C}_{\mathrm{sys}}$ that can then be added to the bootstrap covariance, $\mathbf{C}_\mathrm{boot}$,
 when computing the model
likelihoods. This requires assuming that $\mathbf{C}_{\mathrm{sys}}$ and $\mathbf{C}_\mathrm{boot}$ are
uncorrelated, and that the errors are Gaussian distributed. 

We adopt a Monte Carlo approach to estimate $\mathbf{C}_{\mathrm{sys}}$ 
by generating 200 model transmission PDFs that randomly vary the systematics.
We then evaluate the covariance of the transmission PDFs, 
$p_{i}$, relative to the fiducial model, $p_{\mathrm{ref},i}$  at each transmission bin
$i$.  This allows us to construct a covariance matrix with the elements
\beq
C_{\mathrm{sys},ij} = \langle (p_i - p_{\mathrm{ref},i})(p_j - p_{\mathrm{ref},j}) \rangle
\eeq
that encompasses the errors from the
uncertainties in the LLS model, metal absorption, and continuum scatter.  
{Note that estimation of systematic uncertainties is typically a subjective process, and 
for most of these contributions we can only make educated guesses as to their uncertainty.}

Our Monte Carlo iterations sample the various components of our model as follows:

\begin{description} 
\item[LLS Incidence] We sample the uncertainty in the power-law exponent $\gamma_\mathrm{LLS}$
 of the redshift evolution in LLS incidence rate
(Equation~\ref{eq:lls_lz}), which is $\sigma_{\gamma_\mathrm{LLS}}\pm 0.21$ as reported by \citet{ribaudo:2011}. 
We assume this uncertainty is Gaussian and draw $l_\mathrm{LLS}(z)$ accordingly. This primarily affects the low-flux regions
$-0.1 \lesssim F \lesssim 0.3$ of the PDF.
\item[partial-LLS Slope] Our choice of slope for the distribution of partial LLS ($\nhi <10^{17.5}\,\persqcm$ absorbers
 is from an indirect constraint with significant uncertainty \citep{prochaska:2010}. We therefore vary the pLLS slope around the 
fiducial $\beta_\mathrm{pLLS}=-2.0$ by $\pm 0.5$ assuming a flat prior in this range, which primarily alters the $0 \lesssim F \lesssim 0.4$
portion of the PDF since pLLS typically do not saturate at BOSS resolution.
\item[LLS $b$-parameters] Also in the previous section, we found that a global $b$-parameter of $b=45\,\kms$
gives the best agreement with the data, but this is an \emph{ad hoc} approach with significant uncertainties. 
In our Monte Carlo Sampling we therefore adopt a conservative $b = 45\,\kms \pm 20\,\kms$ with a uniform prior. 
This primarily affects the PDF at $-0.1 \leq F \leq 0.4$ as can be seen in Figure~\ref{fig:pdf_bpar}.
\item[Intervening Metals] Although we used an empirical method to model intervening metals (\S~\ref{sec:metalmodel}),
we may have missed metals with rest wavelengths $\lambda \lesssim 1300 \,\ang$. 
Furthermore, we have a relatively small set ($\sim 300-400$) of 
`template' quasars from which our metal model is derived,
which may contribute some sampling variance.
We therefore guess at an Gaussian error of $\pm 30\%$ for the metal incidence rate. This modulates the extent to which metals
pulls the overall PDF towards lower $F$-values (c.f.\ Figure~\ref{fig:pdf_steps}c).
\item[Continuum Errors] The overall r.m.s. scatter in our continuum estimation also 
affect the flux PDF (Figure~\ref{fig:pdf_steps}e).  This can be varied in our model by 
rescaling the quantity $c'(\lambda)/c(\lambda)-1$, where $c$ is the `true' continuum
used to generate the mock spectrum, while $c'$ is the model continuum which we subsequently fit (Figure~\ref{fig:steps2}).
  For each iteration in our Monte Carlo systematics estimation,
we dilate or reduce $c'(\lambda)/c(\lambda)-1$ by a
Gaussian deviate assuming $\pm 20\%$ scatter. 
This primarily affects the high-transmission ($F > 0.8$) end of the PDF.
\end{description}

For these Monte Carlo iterations, we used the identical thermal model ($\gamma = 1.6$, \tref{})
as well as fixed the same random number seeds used for the selection of simulation skewers and generation of noise
vectors in our spectra, in order to ensure that the only variation between the different iterations are from the randomly-sampled systematics.
Figure~\ref{fig:pdfs_sys} shows 50 of these Monte Carlo iterations on the transmission PDF for the $\zav = 2.3$, S/N=$8-10$ subsample.

Figure~\ref{fig:covar_sys} shows an example of the
systematic contribution to the covariance matrix.  The overall amplitude of
the systematic contribution is considerably higher than that
estimated from the bootstrap resampling
(c.f.\ Figure~\ref{fig:covar_boot}), indicating that we are in the
systematics-limited regime.  We also see significant anti-correlations
at almost the same level as the positive correlations, which are due
mostly to correlations between transmission bins on either side of `pivot
points' as the transmission PDF varies from the systematics --- these
anti-correlations will somewhat counteract the increased size of the
diagonal components. In the
subsequent analysis, we will use an error covariance matrix, 
$\mathbf{C} = \mathbf{C}_\mathrm{boot} + \mathbf{C}_\mathrm{sys}$, in which 
the systematics covariance
matrix estimated in this sub-section is added to the bootstrap
covariance matrix (described in \S~\ref{sec:bosspdf}) estimated from
the BOSS transmission PDFs.

We have at this point yet to address one more parameter which can 
significantly change the shape of our model transmission PDFs, namely
the \lya\ forest mean-transmission assumed in the mock spectra, $\fmean_\mathrm{Ly\alpha}$. 
However, this is an important astrophysical parameter which we did not want to treat as a `systematic', so the
next sub-section will describe our treatment of $\fmean_\mathrm{Ly\alpha}$.

\subsection{Modifying the Mean-transmission} \label{sec:mffit}

\begin{figure}
\includegraphics[width=0.48\textwidth]{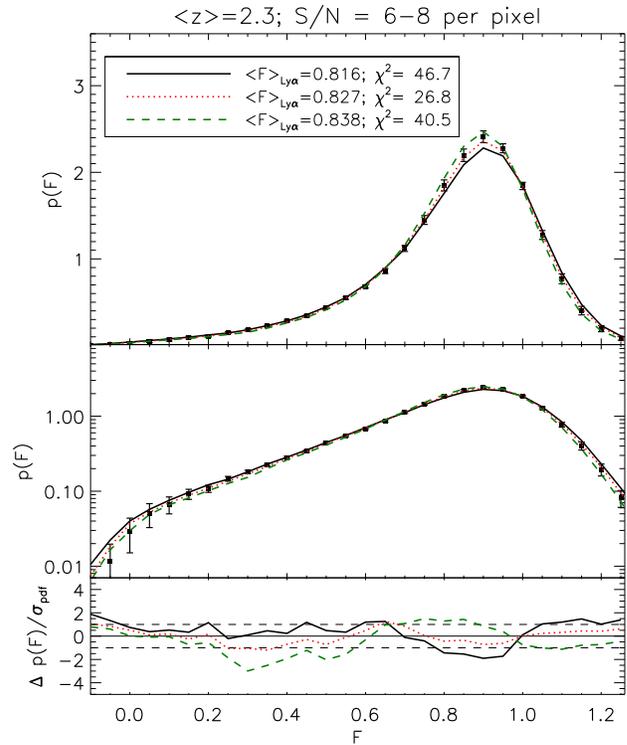}
\caption{\label{fig:pdf_fmeanlya}
Variation of the model transmission PDFs (curves) with respect to changing the mean-transmission, $\fmeanlya$, of the \lya\ forest
simulations. The model PDFs were generated from the $\gamma=1.6$, \tref{} model, while the error bars 
show the corresponding transmission PDFs from BOSS data.
In the bottom panel, the dashed horizontal lines indicate $\pm 1\sigma$ discrepancies between models and data, 
although we caution against `chi-by-eye' due to the significantly non-diagonal covariances in the errors.
The central $\fmeanlya$ value shown here corresponds to that estimated by \citet{becker:2013}, while
the other two are evaluated at $\pm 1\sigma$ of their reported errors.
The mean-transmission value, $\fcont$, assumed in the mean-flux regulated continuum fitting is constant in all cases.
Note that the $\chi^2$ values, which are for 23 d.o.f., are much improved over the previous data comparisons, 
since they now include the improved LLS/pLLS model as well as the full covariance matrix including systematic uncertainties.}
\end{figure}

\begin{figure}
\begin{center}
\includegraphics[width=0.48\textwidth]{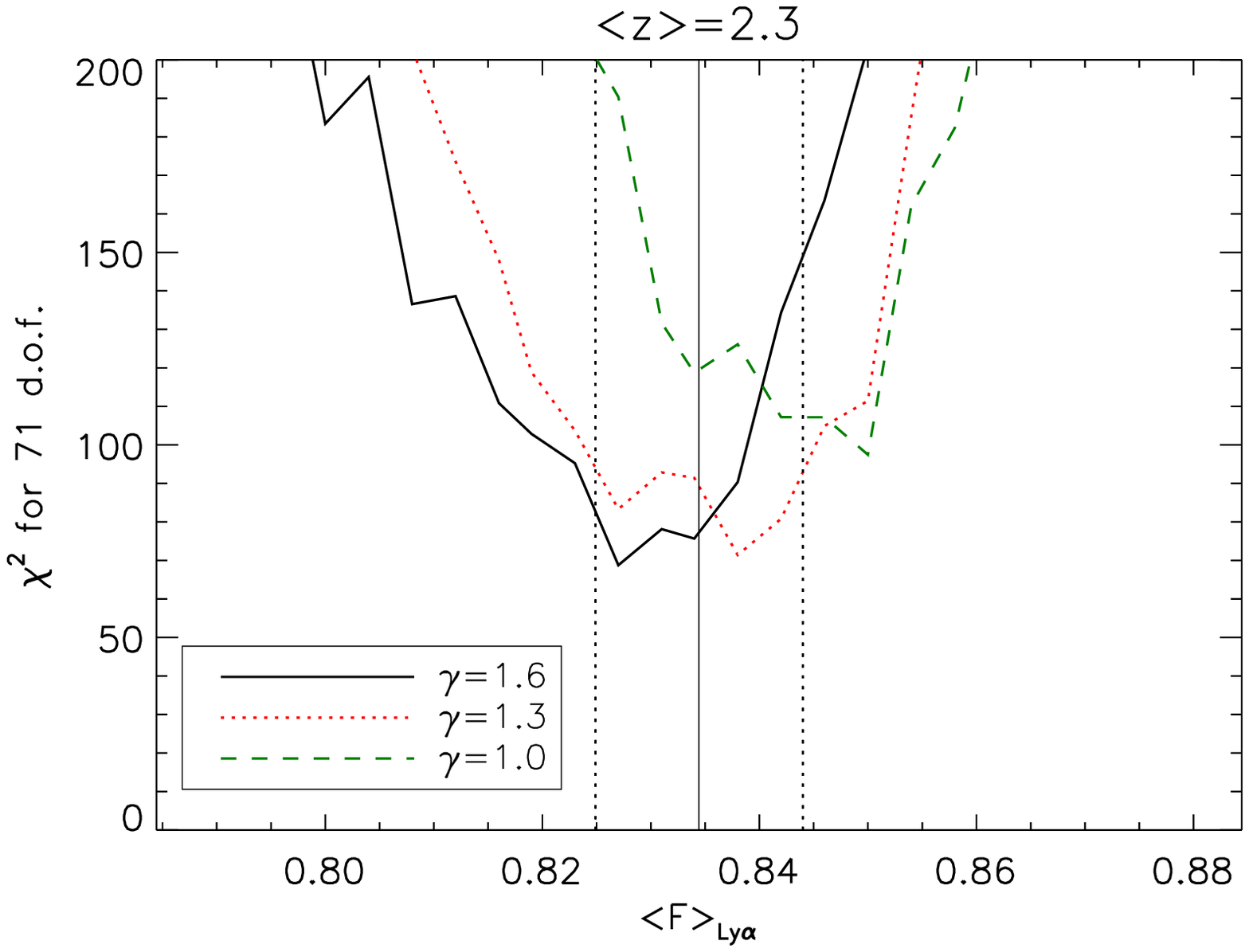} \\
\includegraphics[width=0.48\textwidth]{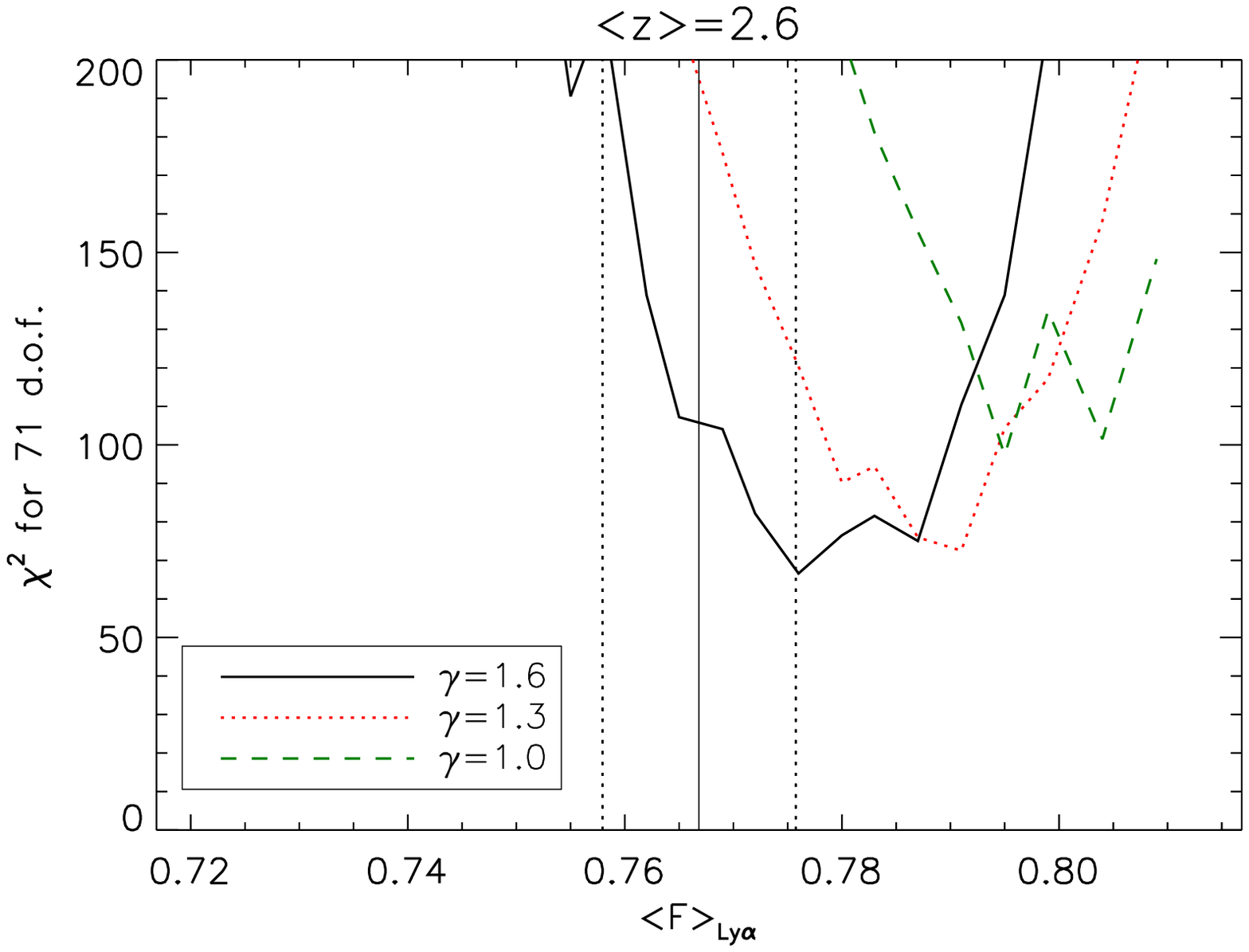} \\
\includegraphics[width=0.48\textwidth]{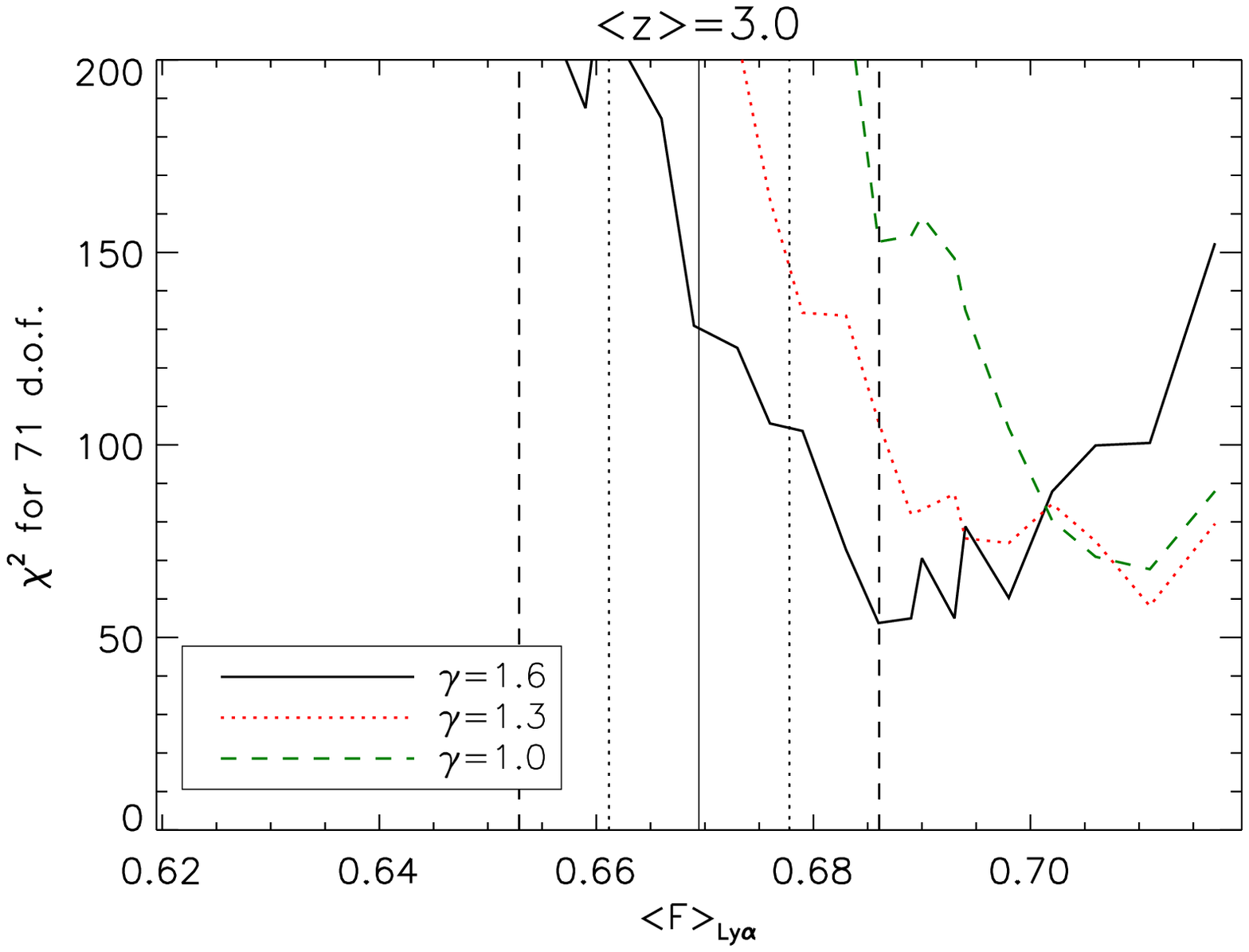}
\end{center}
\caption{\label{fig:chisq_vs_fmean}
$\chi^2$ values for the \tref{} models (with different $\gamma$) plotted as a function of \lya\ forest mean-transmission 
values, \fmeanlya, used to normalize the simulation skewers.
The quoted $\chi^2$ values (with $\nu=71$ d.o.f.) were obtained by summing over the $\chi^2$
for the different S/N subsamples at each redshift.
The fiducial transmission values inferred from \citet{becker:2013} is shown as the solid vertical lines, 
while the dot-dashed vertical lines denote
their $1\sigma$ errors. The dashed lines in the $\zav = 3$ panel denote the inflated error bars we use to account for the quasar selection bias
shown in Fig.~\ref{fig:igmtrans}. In \S\ref{sec:margmf} we will marginalize over the uncertainties in $\fmeanlya$ to obtain our final results.
}
\end{figure}

\begin{figure*}
\begin{center}
\includegraphics[width=0.33\textwidth]{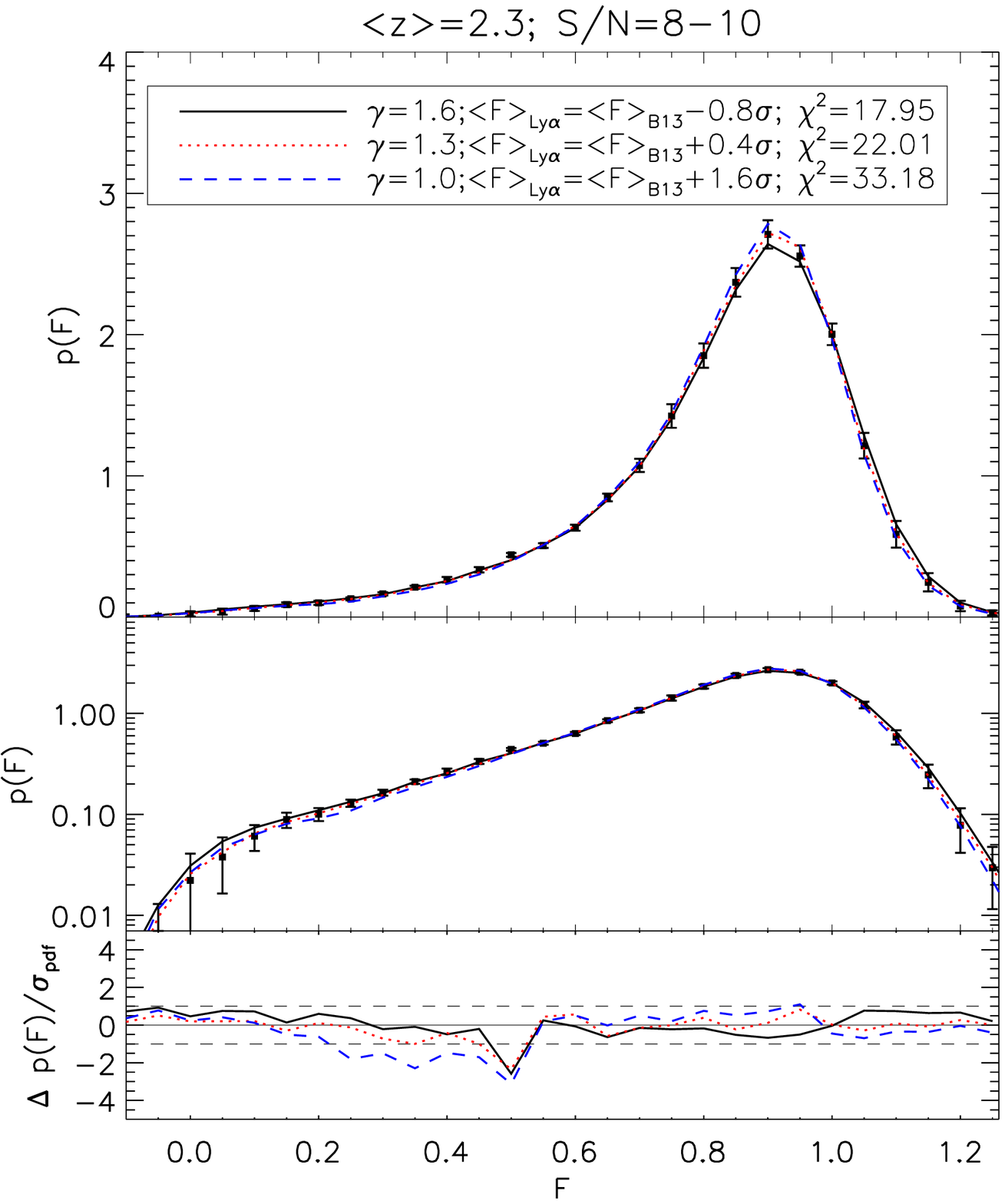}
\includegraphics[width=0.33\textwidth]{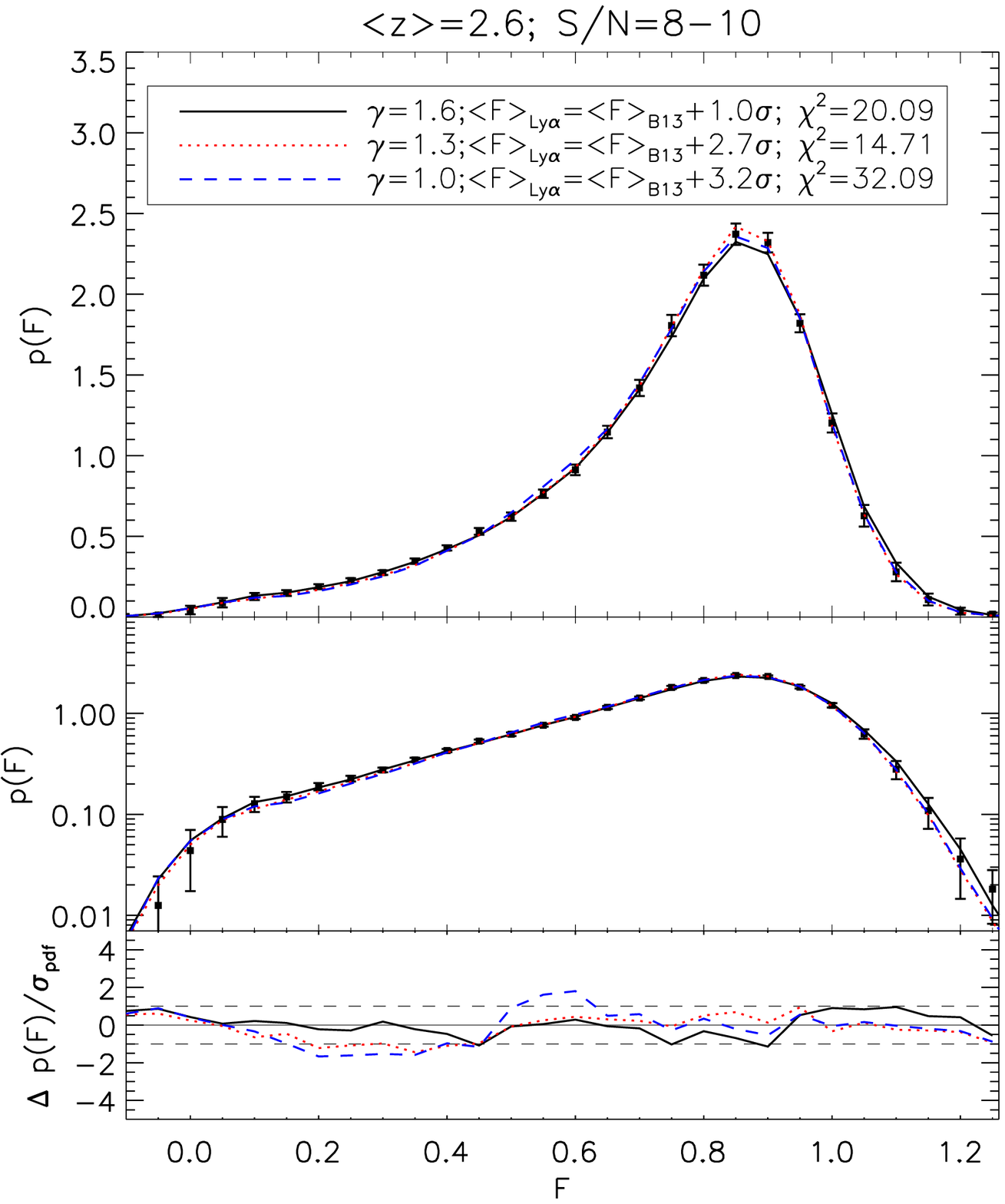}
\includegraphics[width=0.33\textwidth]{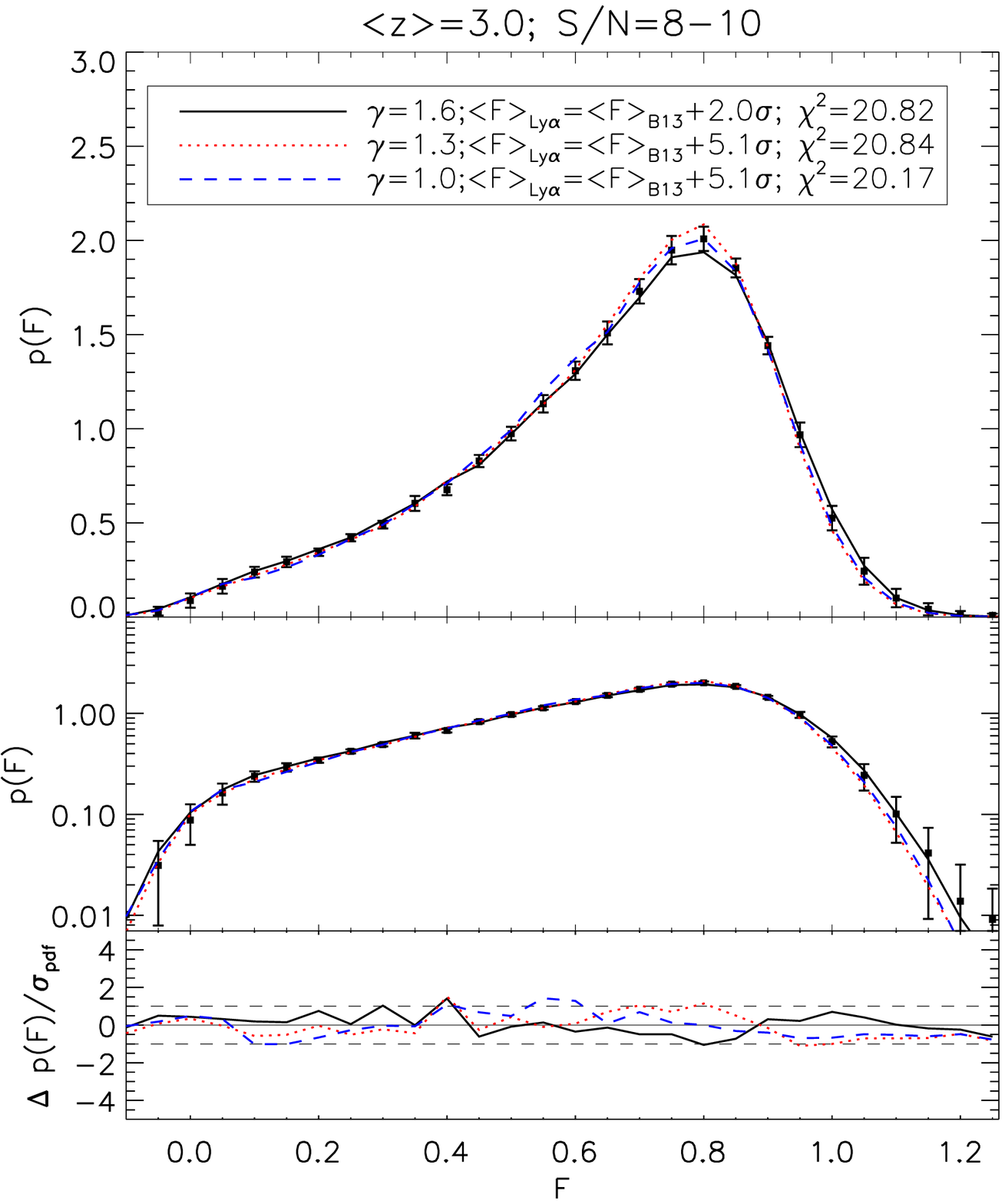}
\end{center}
\caption{\label{fig:pdfgamma_bestfit}
Model transmission PDFs (curves) with the best-fit \lya\ forest mean-transmission $\fmeanlya$ for different $\gamma$ values
from the \tref{} family of models (using the improved LLS/pLLS model).
These are for the S/N=8-10 subsample and compare with the corresponding BOSS data transmission PDFs (error bars).
The upper two panels
in each plot show the transmission PDFs in linear and logarithmic ordinate axes, respectively, while the bottom panels
show residuals divided by the errors, with dashed horizontal lines indicating the $\pm 1\sigma$ region relative
to the data. The best-fitting $\fmeanlya$ values correspond to the minima
in Fig.~\ref{fig:chisq_vs_fmean}, but here we have labeled them relative to $\fmean_\mathrm{Ly\alpha,B13}$, 
the fiducial \citet{becker:2013} values and errors. The $\chi^2$ values quoted are for 23 d.o.f. 
(taking into account the fitting of the LLS $b$-parameter), and were computed using the full error
covariances including both bootstrap and systematic terms.  
}
\end{figure*}

{
In the initial comparison of the model transmission PDFs shown in Figure~\ref{fig:pdfgamma_fid}, 
the models show a discrepancy with the data at higher transmission bins
$F \gtrsim 0.6$. 
Such differences can be alleviated by varying the mean-transmission of the pure \lya\ forest, 
$\fmeanlya \equiv \exp(-\tau_{\rm Ly\alpha})$, i.e. ignoring the contribution from metals and LLS.
This quantity can be varied directly in the simulation skewers (Section~\ref{sec:sims}).
When we vary \fmeanlya\ in the simulations, the quantity \fcont, 
which is used to normalize the continuum level of the mock quasar spectrum, is always kept
fixed to $\feff(z) = \exp[-(\tau_{\rm Ly\alpha}+\tau_{\rm metals}+\tau_{\rm LLS})]$ as 
derived from \citet{becker:2013} (see Section~\ref{sec:cont}). However, since we are applying the same \fcont\ to both 
the real and mock spectra, \fcont\ can be best thought of as a normalization that does not actually need to match 
\feff.
Once both the real and mock spectra have been normalized by \fcont, the transmission PDF retains information
on the respective contributions from the \lya\ forest, metals and LLSs regardless of the assumed \fcont, 
because these contributions affect the shape of the
PDF in different ways. In principle, it is possible to vary these all components to infer their relative contributions, but due
to the crudeness of our metal and LLS models, we choose have only \fmeanlya\ as a free parameter while keeping 
$\langle F \rangle_{\rm metals} = \exp(-\tau_{\rm metals})$ and $\langle F \rangle_{\rm LLS} = \exp(-\tau_{\rm LLS})$ fixed.
The possible variation of these latter two components are instead incorporated into the systematic uncertainties determined in 
Section~\ref{sec:systematics}. The effect of varying \fmeanlya\ is illustrated in Figure~\ref{fig:pdf_fmeanlya},
where we plot the same IGM model with different underlying values of \fmeanlya\ in the simulation skewers whilst
 keeping fixed the contribution from metals, LLSs etc.
}

We therefore explore a range of \fmeanlya\ around the vicinity of that estimated by \citet{becker:2013}, 
$\fmean_\mathrm{Ly\alpha,B13}$, and at each value of $\fmeanlya$ evaluate the 
 $\chi^2$ summed over
all the S/N subsamples for each $\zav$ and $\gamma$ combination. 
In addition, we now adopt the updated LLS/pLLS model described in \S\ref{sec:mod_lls}, while
the $\chi^2$ evaluation now uses the full covariance matrix including both the bootstrap and
systematics (\S\ref{sec:systematics}) uncertainties to compare with the transmission PDFs measured from 
the BOSS data.

{The models are compared with the BOSS data as we vary $\fmeanlya$, and for each 
\fmeanlya\ we compute the 
total chi-squared summed over all three S/N subsamples, where each subsample 
contributes $27-1-2=24$ d.o.f. (c.f. Equations~\ref{eq:pdf_norm1} and \ref{eq:pdf_norm2})
along with a further reduction of one d.o.f. since we have effectively
fitted for the LLS $b$-parameters in \S~\ref{sec:mod_lls}, for a total of $\nu=71$ d.o.f.
The result of this exercise is shown in
Figure~\ref{fig:chisq_vs_fmean} which shows the $\chi^2$ values for the \tref{} models with different 
$\gamma$ ---
we only vary $\gamma$ and not $T_0$ because the $F\gtrsim 0.6$ portions
of the transmission PDF that change the most with \fmeanlya\ do not vary as much with respect to changes in $T_0$
 (c.f. Figure~\ref{fig:pdf_temperature}). 
 Examples
of the corresponding best-fit model PDFs in one S/N subsample are
shown in Figure~\ref{fig:pdfgamma_bestfit}, where we see that varying
$\fmeanlya$ can indeed change the shape of the $F \gtrsim 0.6$
portion of the transmission PDF sufficiently, improving the fits in those transmission ranges
compared to the fiducial models (Figure~\ref{fig:pdfgamma_fid}).
}

In all our redshift bins, the best-fitting models seen in Figure~\ref{fig:chisq_vs_fmean} are $\gamma = 1.6$
with $\chi^2 = [69,67,54]$ for 70 d.o.f.\footnote{In this particular section, when we quote the $\chi^2$ for the 
best-fitting $\fmeanlya$ the d.o.f. is further reduced by 1 compared to the other $\chi^2$ summed over the
S/N subsamples.} 
 at $\zav = [2.3,2.6,3.0]$ 
(for the combined data using all S/N bins),
respectively implying probabilities of $P=[52\%, 59\%, 92\%]$ of getting
larger values\footnote{These $\chi^2$ values are very small for the degrees of freedom, suggesting that 
we may have overestimated the size of our systematic errors, but as we shall see 
this does not affect our ability to place constraints on $\gamma$ and merely makes
our conclusions rather conservative.}.
At the higher redshifts best-fitting mean-transmission for the $\gamma=1.6$ case is pushed to 
significantly discrepant values with respect to the fiducial
\citet{becker:2013} values (Figure~\ref{fig:chisq_vs_fmean}).  

The $\gamma = 1.3$ model also provide acceptable fits to the models,
with $\chi^2 = [71, 73, 58]$ for 70 d.o.f. ($P = [43\%, 40\%, 84\%]$) at $\zav = [2.3, 2.6, 3.0]$, 
but at the two higher redshift bins this requires $\fmeanlya$ values that are increasingly 
discrepant compared to \citet{becker:2013} ($+2.3\sigma$ and $+5\sigma$ respectively at $\zav =[2.6,3.0]$ ).  
The isothermal $\gamma =
1.0$ models are disfavored at the two lower redshift bins,
with best-fit values of $\chi^2=[98, 97]$ for 70 d.o.f. ($P = [2\%, 2\%]$)
at $\zav = [2.3, 2.6]$, whereas at $\zav = 3$, the error bars on the PDF are sufficiently large
that acceptable fits are obtainable using $\gamma = 1.0$, 
with $\chi^2 = 68$ for 70 d.o.f. ($P=54\%$). However, this 
requires a $+5\sigma$ discrepancy in $\fmeanlya$ with respect to \citet{becker:2013}.  
{In Figure~\ref{fig:pdfgamma_bestfit}, one sees that
fitting for $\fmeanlya$ allows the $\gamma=1.0$ models to be in good
agreement with the data in the $F > 0.7$ portion of the PDF, but
gives rise to discrepancies in the $0.4 \lesssim F \lesssim 0.7$ range which limits the
goodness-of-fit, and cannot easily be compensated by modifying the metals or LLS model.}

From Figure~\ref{fig:chisq_vs_fmean}, it is clear that as we move to
higher redshifts, we require increasingly higher $\fmeanlya$ relative
to the fiducial \citet{becker:2013} values in order to agree with the
data: at $\zav = 2.3$, our best-fit mean-transmission for the $\gamma=1.6$ model agrees 
with \citet{becker:2013}, but at $\zav = 3$ there is a significant
deviation of $+2\sigma$ with respect to the \citet{becker:2013}
measurement.  The same trend is true for the best-fit $\gamma = 1.3$ and
$\gamma=1.0$ models, but these require even greater discrepancies with
respect to the fiducial $\fmeanlya$.
 
\begin{figure}
\includegraphics[width=0.5\textwidth]{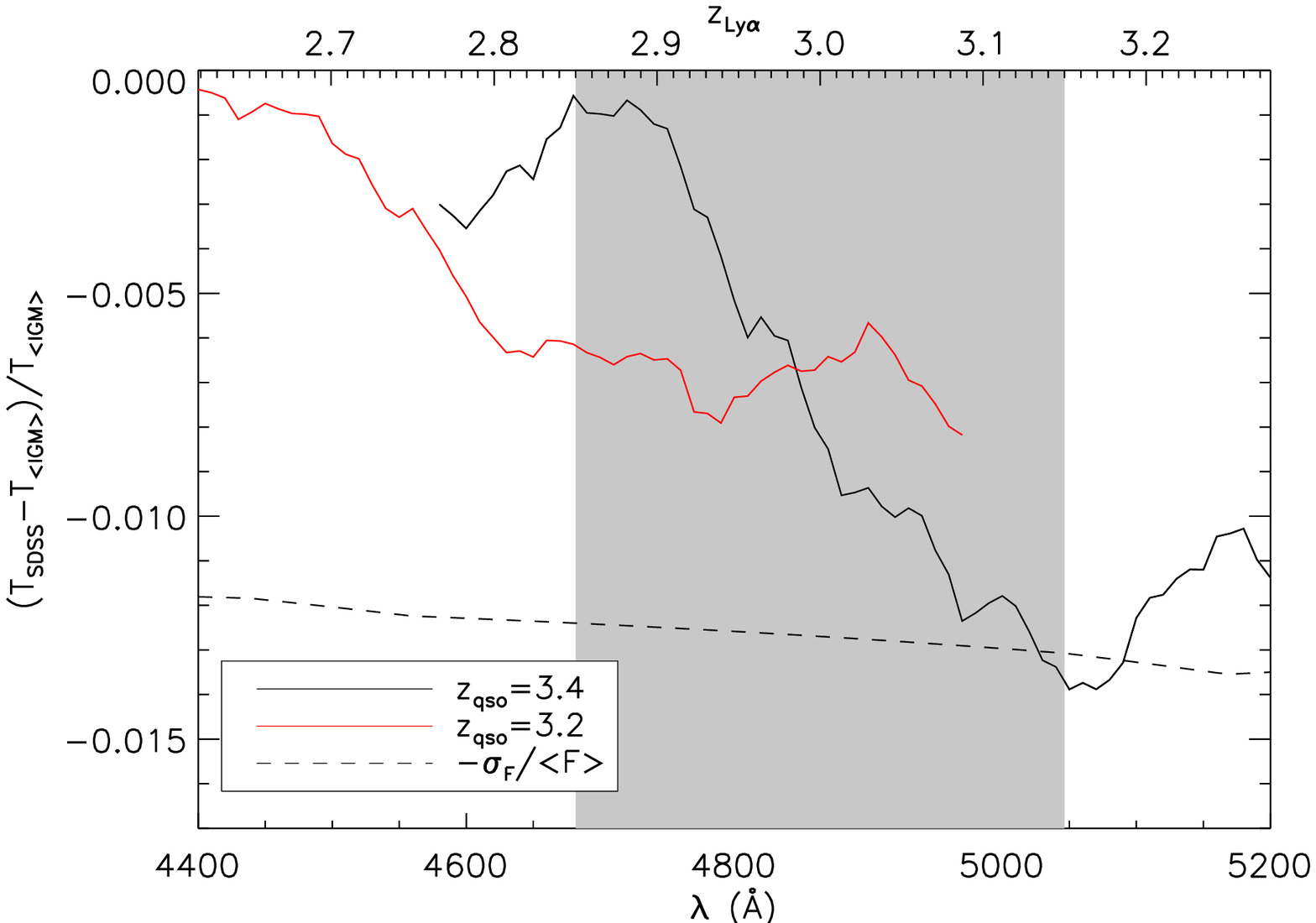}
\caption{\label{fig:igmtrans}
The red and black curves show the excess \lya\ absorption expected from sightlines of $\zq = 3.2$ and $\zq=3.4$ quasars, respectively, relative to the mean IGM transmission.  
This is caused by the SDSS selection bias described in \citet{worseck:2011a}, which yield above-average numbers of intervening LLS. 
These are derived from the same curves shown 
in Figure~17 of \citet{worseck:2011a}, but replotted as ratios smoothed by a boxcar
function over 12 pixels for clarity. The top axis labels the \lya\ absorption redshift corresponding to each wavelength,
while the shaded region indicates the wavelength range of our $\zav = 3.0$ bin.
The dashed-line shows, for comparison, the relative errors on the \lya\ forest mean transmission estimated by \citet{becker:2013}.
The discrepancy due to the SDSS bias is significant compared to the \citet{becker:2013} errors.
}
\end{figure}

One possible explanation for this discrepancy is the effect on the
\citet{becker:2013} measurement of $u$-band selection bias in the SDSS
quasars. This was first noted by \citet{worseck:2011a}, who found that
the color-color criteria used to select SDSS quasars preferentially
selected quasars, specifically in the redshift range $3 \lesssim \zq
\lesssim 3.5$, that have intervening Lyman-breaks at $\lambrest <
912\,\ang$. The $3 \lesssim \zq \lesssim 3.5$ SDSS quasars are thus
more likely to have intervening LLS in their sightlines, yielding an
additional contribution to the \lya\ absorption and hence causing
\citet{becker:2013} to possibly underestimate \fmeanlya\ when stacking the
impacted quasars.  \citet{becker:2013} mentioned this effect in their
paper but argued that it was much smaller than their estimated errors
by referencing theoretical IGM transmission curves estimated by
\citet{worseck:2011a} (Figure~17 in the latter paper).

Dr.\ G.~Worseck has kindly provided us with these transmission curves,
$T_\mathrm{IGM} (\lambda)$, which were generated for both the average
IGM absorption and that extracted from SDSS quasars affected by the
color-color selection bias.  In Figure~\ref{fig:igmtrans} we plot
the relative  difference between the biased \lya\ transmission deduced from 
$\zq = 3.2$ and $\zq = 3.4$ quasars and the true mean IGM transmission,  
using the \citet{worseck:2011a} transmission curves. 
It is clear that at \lya\ absorption redshifts of $\zabs \approx 3$,
the excess LLS picked up from such quasars contribute an additional
$\sim 1\%$ compared to the mean IGM decrement, a discrepancy that
is of the same magnitude as the error bars in the \citet{becker:2013}
measurement, indicated by the dashed line.

This could partially explain the higher \fmeanlya\ required to make our $\zav =
3$ models fit the data in Figure~\ref{fig:chisq_vs_fmean}.  Note that
we expect this UV color selection bias to be much less significant in
our BOSS data, since we have selected bright quasars in the top 5th
percentile of the S/N distribution. Given that such quasars have high
signal-to-noise ratio photometry, their colors separate much more
cleanly from stellar contaminants.  Furthermore, such bright quasars
are much more likely to have been selected with multi-wavelength data
\citep[e.g., including near-IR and radio in addition to optical
  photometry see][]{ross:2012}. For both of these reasons, we expect
our quasars to be much less susceptible to biases in color-selection
related to the presence of an LLS. A careful accounting of this bias
is beyond the scope of this paper, but from now on we will inflate by
a factor of two the corresponding errors on $\fmeanlya$ at $\zav=3$ to
account for this possible bias in the mean transmission measurements (dashed
vertical lines in bottom panel of Figure~\ref{fig:chisq_vs_fmean}).

{
Another possibility that could explain a bias in the \fmeanlya\ measured by \citet{becker:2013}
is their assumption that the metal contamination of the \lya\ forest does not evolve with redshift. 
While there are few clear constraints on the aggregate metal contamination within the forest, 
assuming that the metals actually decrease with increasing redshfit \citep[e.g., in the case of \ion{C}{4},][]{cooksey:2013}, 
then the assumption of an unevolving metal contribution calibrated at $z\approx 2.3$ would
lead to an underestimate of $\fmeanlya$ at higher redshifts, which could explain the trend we seem to be seeing.
}

It is clear from the previous discussion that there is some degeneracy between $\gamma$ and
$\fmeanlya$ in our transmission PDFs. However, we are primarily interested in $\gamma$,
while the $\fmeanlya$ has been extensively measured over the years allowing
strong priors to be placed. 
In the next section, we will therefore marginalize over $\fmeanlya$ in order to obtain 
our final results.

\section{Results} \label{sec:margmf}
Due to the uncertainties in \fmeanlya\ described in the previous sub-section, for a better comparison
between transmission PDFs, $p$, from models with different $[\gamma, T_0]$
we will marginalize the model likelihoods, $\mathcal{L} = \exp(-\chi^2/2)$,
over the \lya\ forest mean-transmission, \fmean:
\beq \label{eq:margmf}
\mathcal{L}(p\,|\gamma, T_0) = \int^{\infty}_{-\infty} \mathcal{L}(p\,|\gamma, T_0, \fmean)\; 
A(\fmean)\;\mathrm{d}F,
\eeq
where $A(\fmean)$ is the prior on $\fmean$ (for clarity in these equations, \fmean\ is used as a shorthand for
\fmeanlya). We assume a Gaussian prior:
\beq
A(\fmean) = \frac{1}{\sigma_F \sqrt(2 \pi)}\exp\left[-\frac{(\fmean - \langle F \rangle_\mathrm{B13})^2}{2 \sigma^2_F} \right],
\eeq
where $\langle F \rangle_\mathrm{B13}$ and $\sigma_F$ are the optically-thin \lya\ forest
mean-transmission and associated errors, respectively, estimated from \citet{becker:2013}. 
Note that for $\zav = 3$, we have decided to dilate the error bars by a factor of $2$ to account
for the suspected quasar selection bias discussed in the previous section.

For each model, we generate transmission PDFs with different $\fmeanlya$ (similar to Figure~\ref{fig:chisq_vs_fmean})
and evaluate the combined $\chi^2$ summed over different S/N.
We interpolate the $\chi^2$ over $\fmeanlya$ to obtain a finer grid, which then allows us to  
numerically integrate Equation~\ref{eq:margmf} using five-point Newton-Coates quadrature.

At this stage, we also analyze models with different IGM temperatures at mean-density, $T_0$. Hitherto, 
we have been working only with the central \tref{} model ($T_0(z=2.5)\sim 16000\,$K) , but we now also compare models from
the \thot{} and \tcold{} simulations, that have $T_0(z=2.5) \sim 11000\,$K and $T_0(z=2.5) \sim 21500\,$K, respectively.
Each of these temperature models also sample temperature-density relationships of $\gamma= [1.0, 1.3, 1.6]$ for a model grid 
of $3\times 3$ parameters at each redshift.

\begin{deluxetable}{lccc}
\tablecolumns{4}
\tablewidth{0.45\textwidth}
\tablecaption{Marginalized $\chi^2$ for $\nu = 71$ d.o.f.}
\startdata
\cutinhead{ $ \zav = 2.3$ }
\noalign{\vskip 0.15em} $\gamma$ & \tcold{} & \tref{} & \thot{} \\ \noalign{\vskip 0.15em} 
\noalign{\vskip 0.15em}  \hphantom{he} & ($T_0=13000$K) & ($T_0=18000$K) & ($T_0=23000$K) \\ \noalign{\vskip 0.15em} 
\tableline \noalign{\vskip 0.15em} 
1.6 &    87.7 &  72.9 &  79.5 \\
1.3 &   103.4 &  76.0 &  71.8 \\
1.0 &   174.2 & 105.5 &  88.4 \\
 \tableline \noalign{\vskip 2em}
 \cutinhead{ $ \zav = 2.6$ }
\noalign{\vskip 0.15em}  $\gamma$ & \tcold{} & \tref{} & \thot{} \\ \noalign{\vskip 0.15em} 
\noalign{\vskip 0.15em}  \hphantom{he} & ($T_0=11000$K) & ($T_0=16000$K) & ($T_0=21500$K) \\ \noalign{\vskip 0.15em} 
\tableline \noalign{\vskip 0.15em} 
1.6 &    88.8 &  72.0 &  71.4 \\
1.3 &   118.0 &  82.6 &  91.8 \\
1.0 &   203.3 & 127.3 & 111.1 \\
 \tableline \noalign{\vskip 2em}
 \cutinhead{$ \zav = 3.0 $ }
\noalign{\vskip 0.15em}  $\gamma$ & \tcold{} & \tref{} & \thot{} \\ \noalign{\vskip 0.15em} 
\noalign{\vskip 0.15em}  \hphantom{he} & ($T_0=9000$K) & ($T_0=14000$K) & ($T_0=19000$K) \\ \noalign{\vskip 0.15em} 
\tableline \noalign{\vskip 0.15em} 
1.6 &    61.7 &  65.1 &  62.5 \\
1.3 &    77.6 &  72.7 &  63.8 \\
1.0 &   119.5 &  77.7 &  85.8 
\enddata
\label{tab:marg_chisq}
\end{deluxetable}

\begin{figure} 
\centering
\includegraphics[width=0.46\textwidth]{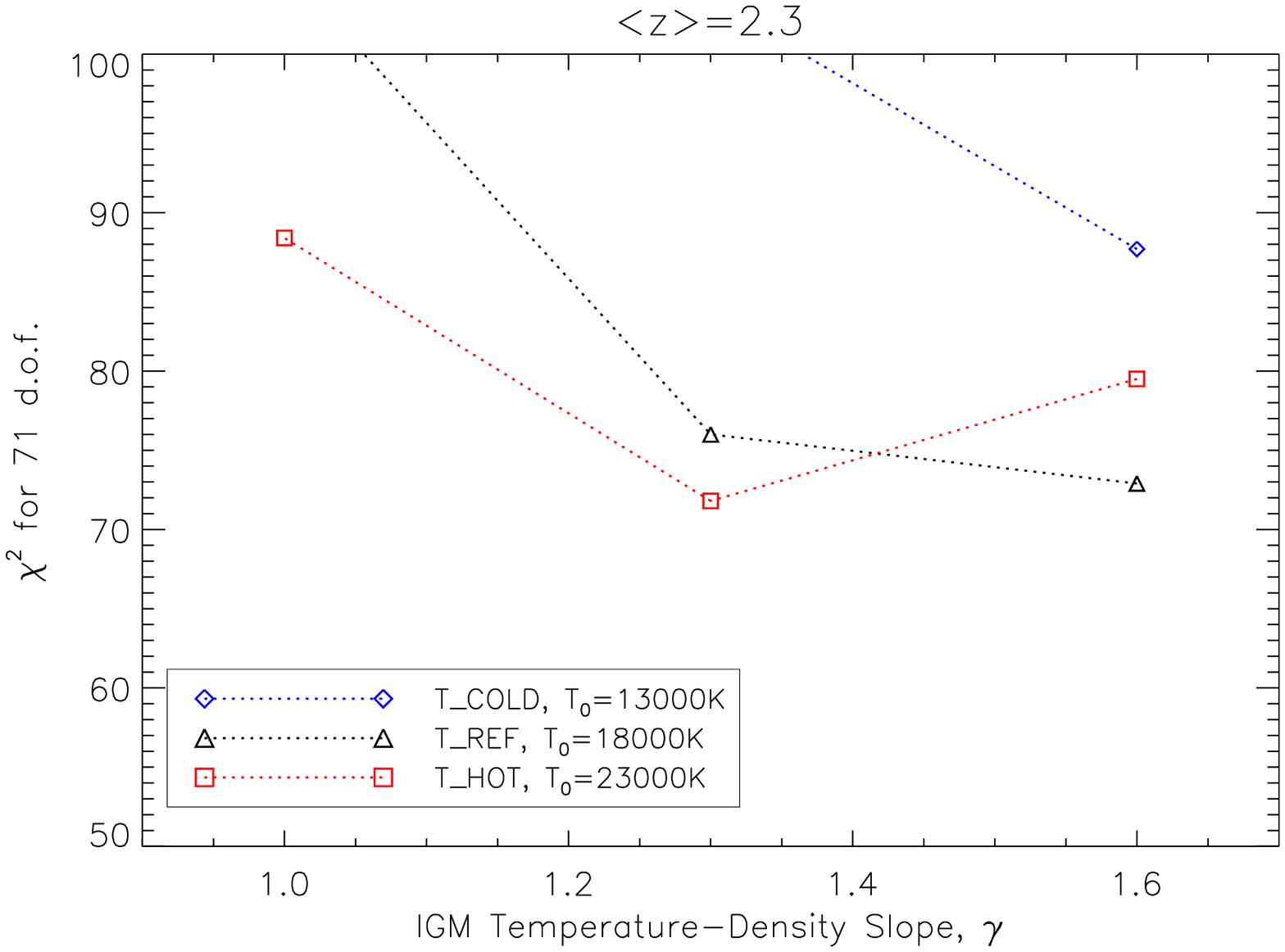} \\
\includegraphics[width=0.46\textwidth]{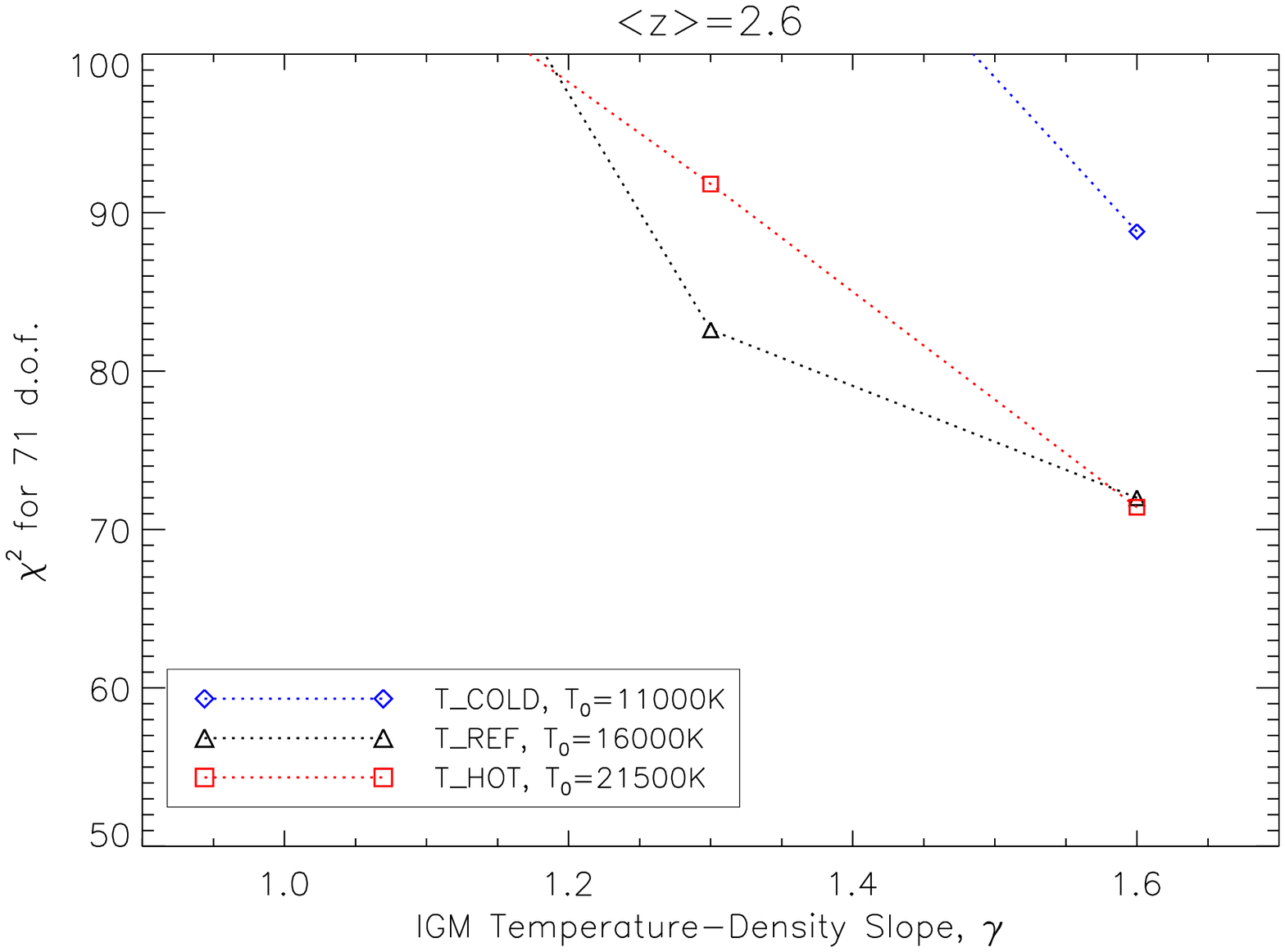} \\
\includegraphics[width=0.46\textwidth]{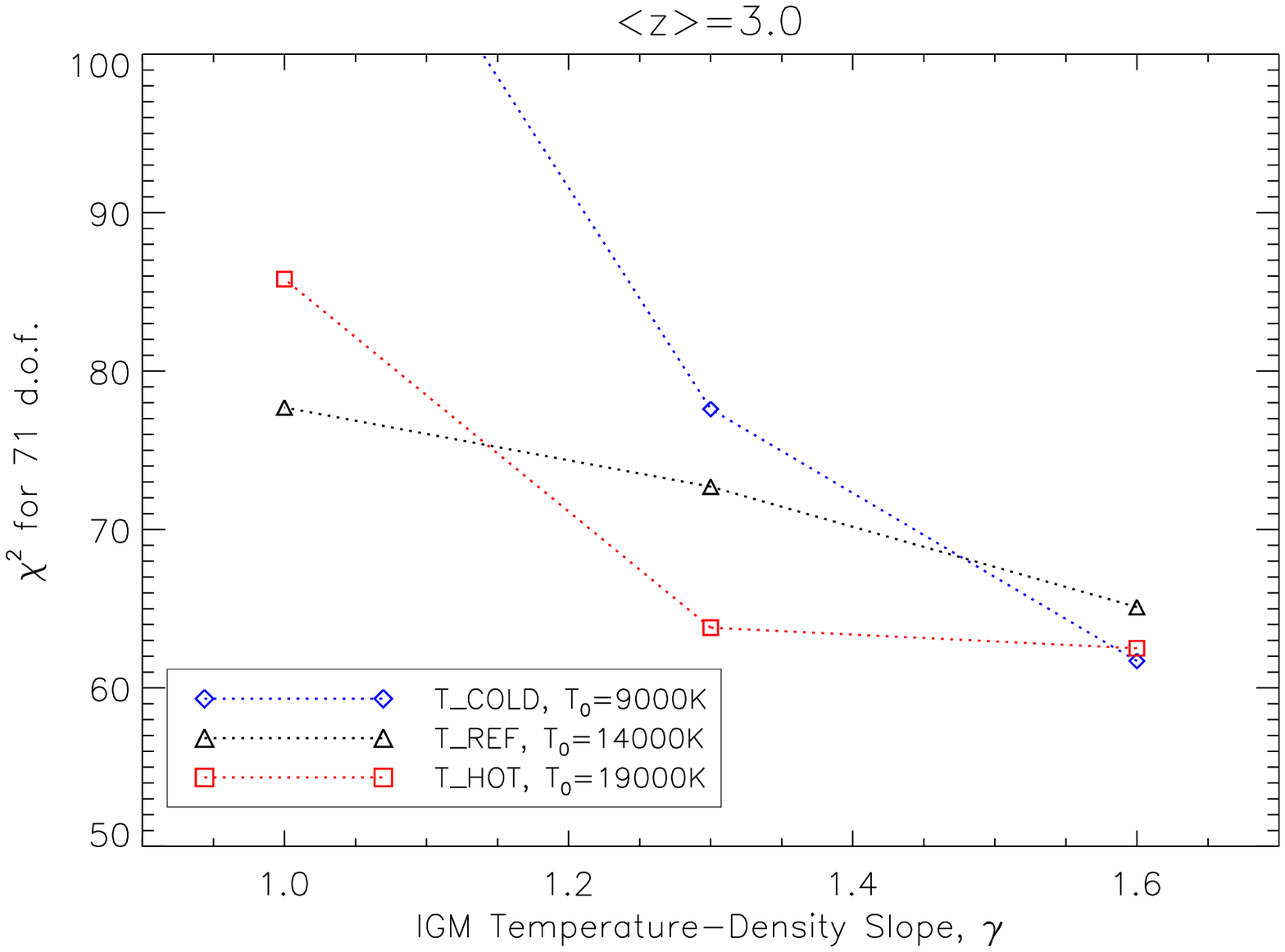}
\caption{\label{fig:marg_chisq}
$\chi^2$ values (for 71 d.o.f.) from models with different $\gamma$ and $T_0$ at different redshifts, after marginalizing over
uncertainties in the mean-transmission $\fmean$ of the \lya\ forest. Models with $\gamma=1.6$ are generally favored, 
although $\gamma=1.3$ with the \thot{} model is also acceptable at $\zav = 2.3$. The same quantities are 
also tabulated in Table~\ref{tab:marg_chisq}.
}
\end{figure}

The marginalized $\chi^2$ values for all the models are tabulated in Table~\ref{tab:marg_chisq}, and plotted as a function of 
$\gamma$ in Figure~\ref{fig:marg_chisq}. 
In general, the \tref{} models with $\gamma=1.6$ provide the best agreements with the data at all
redshifts with 
$\chi^2 \approx 60-70$ for 69 d.o.f..
The \thot{} models (with higher IGM temperatures at
mean density)
provide fits of comparable quality, and indeed at $\zav = 2.3$ the \thot{} model
with $\gamma=1.3$ gives essentially the same goodness-of-fit as the $\gamma=1.6$ 
\tref{} model.
The cooler \tcold{} models are less favored by the data, and at $\zav = 2.6$ give unreasonable
fits to the data with $\chi^2 = 89$ for 69 d.o.f. ($P = 5\%$), but at other redshifts they are 
acceptable fits to the data. In other words, the transmission PDF does not show a strong sensitivity for $T_0$, which we
shall show later is due to degeneracy with our LLS model in the low-transmission end of the transmission PDF.

The more important question to address is the possibility of
isothermal or inverted temperature-density relationships ($\gamma \leq1$) 
as suggested by some studies on the transmission PDF of high-resolution,
high-S/N echelle quasar spectra \citep[e.g.,][]{bolton:2008,
  viel:2009, calura:2012}.  It is clear from
Table~\ref{tab:marg_chisq} and Figure~\ref{fig:marg_chisq} that for all
$T_0$ models the isothermal, $\gamma=1.0$ models disagree strongly with the
BOSS data. The closest match for
an isothermal IGM is the \tref{} model at $\zav = 3.0$, which yields $\chi^2
= 78$ for 69 d.o.f., or a probability of 21\% of obtaining the
data from this model.  However, relative to the $\gamma=1.6$ model at
$\zav=3.0$ which gives the minimum $\chi^2$ at that redshift, 
we find $\Delta \chi^2 \approx 16$ for the isothermal
model, i.e. a $\sqrt{\Delta \chi^2} = 4\sigma$ discrepancy from the best-fit model.  
The isothermal model is
also strongly disfavored at the other redshifts, where we find
$\Delta \chi^2 \approx [15, 40]$ at $\zav = [2.3,2.6]$ or 
$\sqrt(\Delta \chi^2) \approx [3.9\sigma, 6.3\sigma]$.  Since the
shape of the transmission PDF varies continuously as a function of $\gamma$
\citep[see, e.g.,][]{bolton:2008, lee:2012}, these results imply that
inverted ($\gamma<1$) IGM temperature-density slopes are even more
strongly ruled out.
 
\section{Discussion} \label{sec:discussion}

In this paper, we have studied the $\zav = 2.3 - 3$ \lya\ forest transmission probability distribution function (PDF)
 from 3373 BOSS
DR9 quasar spectra. 
Although this is a relatively small subsample selected to be in the top 95th percentile in terms of S/N, they provide
 2 orders-of-magnitude 
larger \lya\ forest path length than high-resolution, high-S/N data sets previously used for this purpose, providing unprecedented
statistical power for transmission PDF analysis.

In order to ensure accurate characterization and allow subsequent modelling
of the spectral noise, we have introduced a novel, probabilistic
method of combining the multiple exposures that comprise each BOSS
observation, using the raw sky and calibration data.  This method
significantly improves the accuracy of the noise estimation, and
additionally allows us to generate mock spectra with noise properties
tailored to each individual BOSS spectrum, but self-consistently for
different \lya\ forest realizations.  We believe that our noise
modeling --- which yields noise estimates accurate to $\sim 3\%$ 
across the relevant wavelength range ---
is the most careful treatment of spectral noise in multi-object fiber
spectra to-date, and we invite readers with similarly stringent
requirements in understanding the BOSS spectral noise to contact the
authors.  In the future, the spectral extraction algorithm described
by \citet{bolton:2010} may solve some of the issues which affected
us, but this has yet to be implemented.

For the continuum estimation, we used the mean-flux
regulated/principal component analysis (MF-PCA) method introduced in
\citet{lee:2012}. This method, which reduces the uncertainty in the
continuum estimation to $\sigcont \lesssim 5\%$, fits for a continuum
such that the resulting \lya\ forest has a mean-transmission $\fmean$
matched to external constraints, for which we use the precise
measurements by \citet{becker:2013}.  While MF-PCA does require
external constraints for $\fmean$, we argue that so long as both the
real quasars and mock spectra are continuum-fitted in exactly the same
way, the shape of the transmission PDF retains independent information on the
\lya\ forest mean-transmission.
 
To compare with the data, we used the
detailed hydrodynamical simulations of \citet{viel:2013}, that explore a range of 
IGM temperature-density slopes ($\gamma \approx 1.0 - 1.6$) and temperatures at mean
density ($T_0 (z=2.5) \approx [11000, 16000, 21500]\,$K). 
We processed the simulated spectra to take account the characteristics of the 
individual BOSS spectra in our sample, such as spectral resolution, pixel noise, 
and continuum fitting errors. 
We also incorporate the effects of astrophysical `nuisance' parameters such as 
Lyman-limit systems (LLSs) and metal contamination.
The LLSs are modeled by adding $10^{16.5} \,\persqcm \lesssim \nhi \lesssim 10^{20.3}\,\persqcm$ absorbers into 
our mock spectra, based on published measurements of the observed incidence $l_\mathrm{LLS}(z)$ 
\citep{ribaudo:2011} and \ion{H}{1}
column density distribution $f(\nhi)$ \citep{prochaska:2010}.
Meanwhile, contamination from lower-redshift metals are modeled in an empirical fashion by inserting 
$\lambrest > 1216\,\ang$ absorbers observed in lower-redshift SDSS/BOSS quasars into the same observed wavelengths 
of our mock spectra.

Our initial models did not provide satisfactory agreement with the transmission PDF measured from the BOSS spectra,
with discrepancies at both the high-transmission and low-transmission bins. 
However, the differences between data and models were consistent across the different S/N subsamples,
 indicating that our noise modelling is robust.
To resolve the discrepancies at the low-transmission end of the PDF, we explored various modifications to our LLS model.
Firstly, we steepened the column-density distribution slope of partial LLS 
($16.5 < \log_{10}(\nhi) < 17.5$ systems) to $\beta_\mathrm{LLS}=-2$ a value
 suggested from the mean-free path of ionizing photons \citep{prochaska:2010}.
 This change relieved the tension between model and data in the $F \approx 0.1-0.4$ bins, but
 implies increasing the number of pLLS by nearly an order of magnitude, but this is not unreasonable given 
 the current uncertainties on this population \citep{janknecht:2006, prochaska:2010}.
 We believe that the necessity of a pLLS distribution with $\beta_\mathrm{LLS} \approx -2$ to fit the
 BOSS \lya\ transmission PDF supports the claims of \citet{prochaska:2010} regarding the column-density distribution of this population.

 However, after adding pLLSs a major discrepancy remained in the saturated $F \approx 0$ bins, 
 which we addressed by adjusting the effective $b$-parameter assumed in all the optically-thick systems in our
 model. We found that an effective value of $b = 45\,\kms$ gave the best-fit to our 
 model\footnote{Note that we have quoted an \emph{effective} $b$-parameter, which must not be confused with the 
 $b$ from individual kinematical components, which is often quoted by workers carrying out Voigt profile analysis
  of high-resolution spectra.}.

 At the high-transmission ($F \gtrsim 0.6$) end of the model transmission PDFs, we
 found that modifying the \lya\ forest mean-transmission in the
 simulations, \fmeanlya, allowed much better agreement with the BOSS
 data.  At $\zav = [2.3, 2.6]$, the $\fmeanlya$ that gave the
 best-fitting model PDFs were within $1\sigma$ of the
 \citet{becker:2013} measurements, but at $\zav = 3$ we required a
 value that was $\sim 2\sigma$ larger.  We argue that this discrepancy
 could be due to a color-color selection bias in the $3 \lesssim \zq
 \lesssim 3.5$ SDSS quasars used by \citet{becker:2013}, which
 preferentially selected sightlines with intervening LLS, giving rise
 to additional \lya\ absorption (and thus lower \fmeanlya) at a level
 comparable to the errors estimated by \citet{becker:2013}.  Our BOSS
 spectra, on the other hand, should be comparatively unaffected on
 account of being the brightest quasars in the survey, hence they they
 separate more cleanly from the stellar locus in color-space, and
 were more likely to have been selected with additional criteria
 (radio, near-IR, variability etc) beyond color-color information
 \citep{ross:2012}.
 
 To deal with these uncertainties, we decided to marginalize over the mean-transmission in our $\chi^2$ analysis.
 At $\zav = 2.3$, the preferred model is for a hot IGM with ($T_0 = 23000\,$K) along 
 with $\gamma=1.3$ ($P \approx 45\%$), 
 although the intermediate-temperature model ($T_0 = 18000\,$K) with $\gamma=1.6$ is nearly as good a
 fit with $P\approx 82\%$. 
 The preferred models at $\zav = [2.6, 3.0]$ are for $\gamma = 1.6$ at temperatures at mean-density of 
 $T_0=[21500, 9000]\,$K ($P = [46\%, 78\%]$, respectively.
We find that the isothermal ($\gamma=1$) temperature-density relationship is strongly disfavored
at all redshifts regardless of $T_0$, with discrepancies of 
 $\sqrt{\Delta \chi^2} \sim 4-6\sigma$ compared to the best-fit models.
 
 \begin{figure}
 \includegraphics[width=0.48\textwidth]{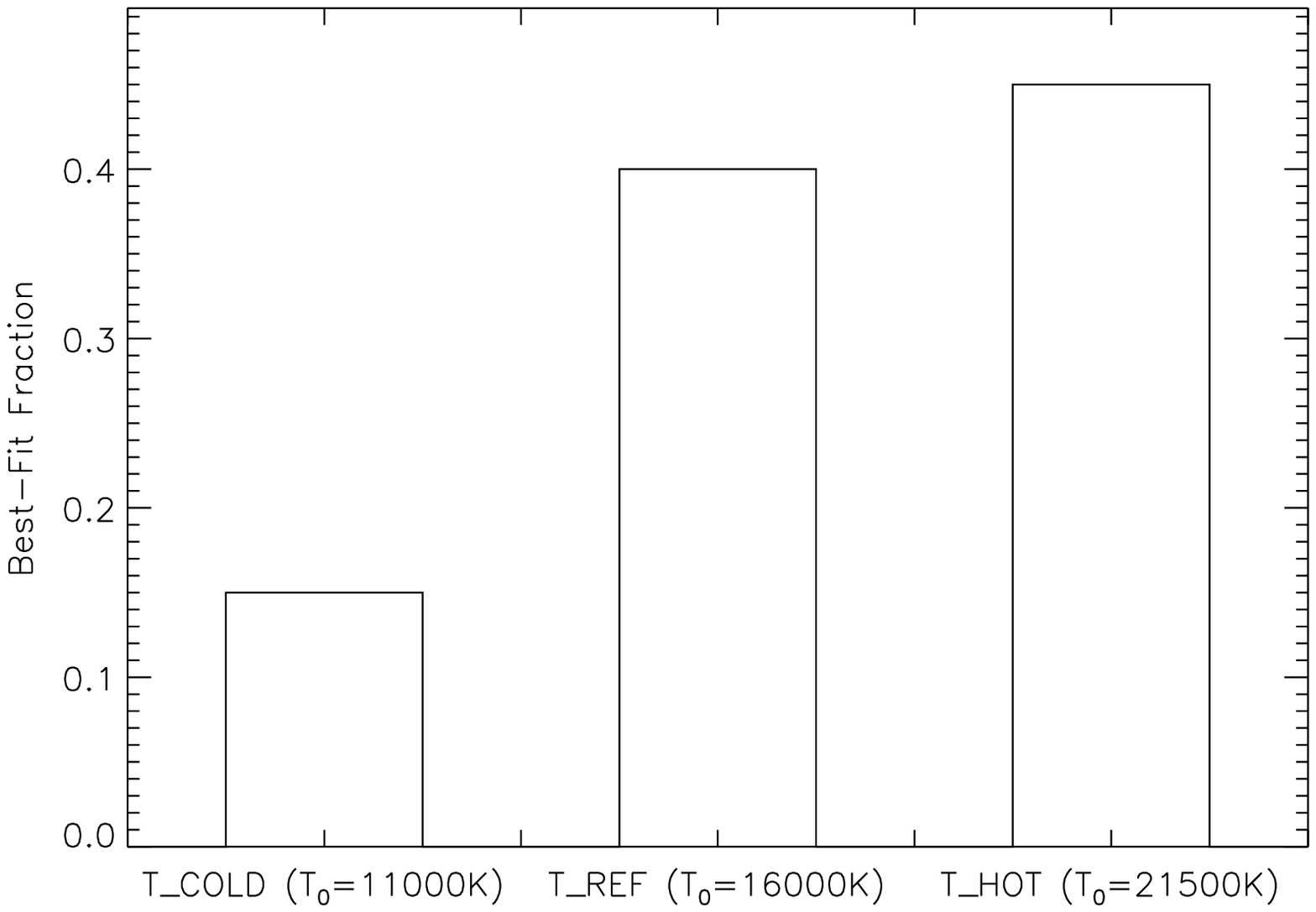} \\
 \includegraphics[width=0.48\textwidth]{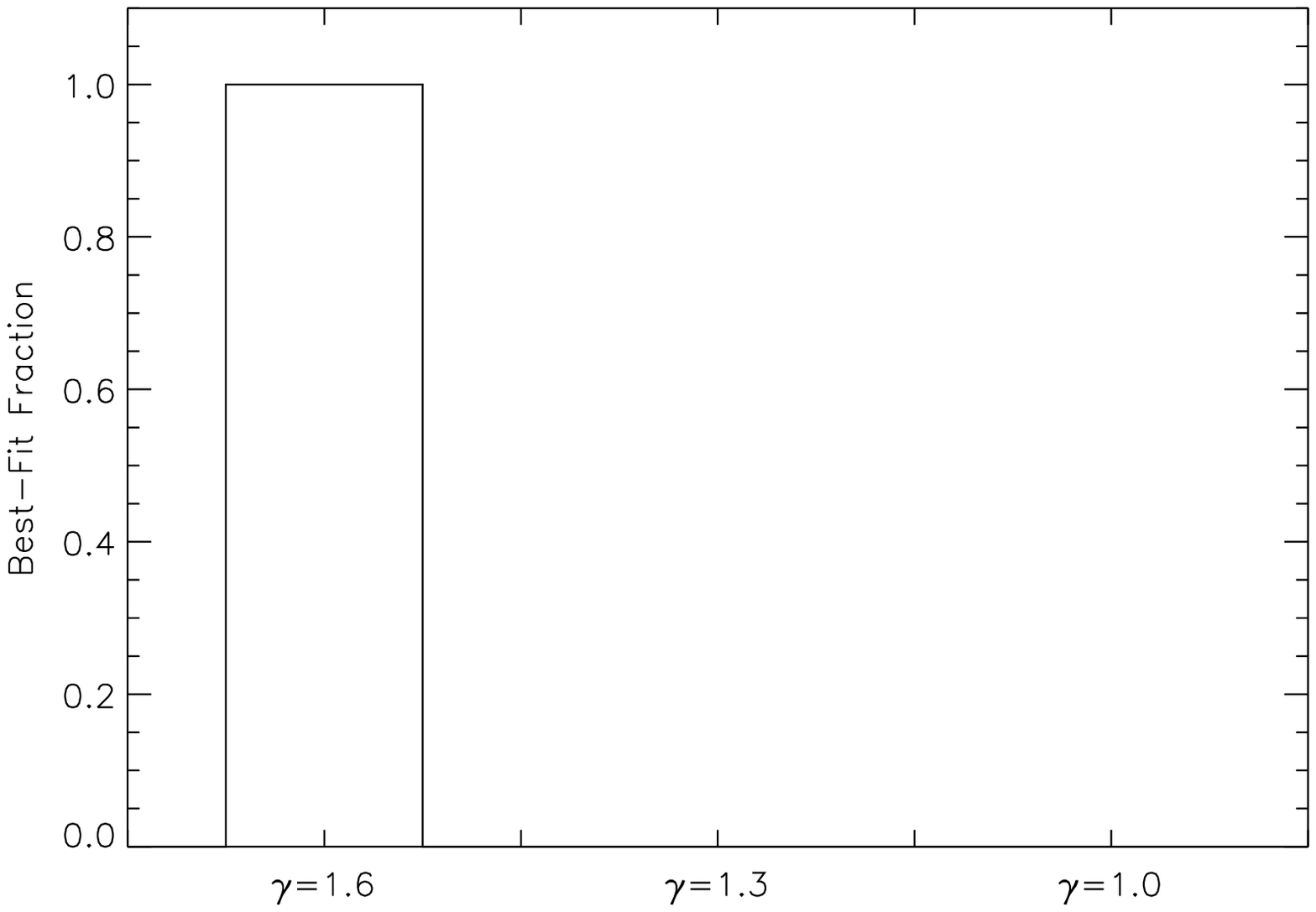} 
 \caption{\label{fig:chisq_hist}
 Histogram indicating the fraction of times a given $T_0$ (top) or $\gamma$ (bottom) model is favored for the 
 $\zav=2.3$, S/N=8-10 transmission PDF when the systematics levels in the model are randomly sampled 20 times.
 While different systematics could lead to different best-fitting models for $T_0$, the models
 with $\gamma=1.6$ are always preferred. This indicates some degeneracy in our systematics
 model with $T_0$, but our conclusions on $\gamma$ are robust. 
 }
\end{figure}
  
One might be skeptical of the results given the various assumptions we had to make in modelling
 astrophysical nuisance parameters. To test the robustness of our results to systematics, we generated 
 20 iterations of model transmission PDFs sampling all nine of our $[T_0, \gamma]$ models (i.e. 180 PDFs in total)
 in the $\zav = 2.6$, S/N=8-10 bin, 
 where each iteration has a random realization of the systematics (LLS, metals, continuum errors etc) drawn
 in the same way as our Monte-Carlo estimate of systematic uncertainty (\S\ref{sec:systematics}). 
 We then asked how many times each $T_0$ or $\gamma$ model gave the lowest $\chi^2$ when compared 
 with the data. For this test we only evaluated the $\chi^2$  at the fiducial $\fmeanlya$ without marginalization.

 The results of this test is shown in Figure~\ref{fig:chisq_hist}. 
 In the top panel, the \tref{} and \thot{} models are favored $\sim 40\%$ of the time but the \tcold{} has $\sim 15\%$ 
of being favored depending on the (random) choice of systematics. In other words, there is significant degeneracy 
between our systematics model and $T_0$.
 We suspect this is driven largely by the choice of the LLS $b$-parameter, which changes 
 the shape of the transmission PDF in a similar way to $T_0$ (compare Figure~\ref{fig:pdf_bpar}
 with Figure~\ref{fig:pdf_temperature}). In contrast, the bottom panel of Figure~\ref{fig:chisq_hist}
 shows that whatever systematics we choose, $\gamma=1.6$ is always favored indicating a
 robust constraint. 
 
 {There is however some degeneracy between $\gamma$ and the \lya\ forest mean transmission, $\fmeanlya$.
 While we marginalize over the latter quantity, the choice of prior can, in principle, affect the results.
 However, at $\zav = [2.3, 2.6]$, the chi-squared minimum of the $\gamma=1.0$ PDF model as a function of $\fmeanlya$ 
is $\chi^2 \approx 100$ for 71 d.o.f. (Figure~\ref{fig:chisq_vs_fmean}), which has a probability of $P \approx 1\%$. In other words, 
 even if we fine-tuned $\fmeanlya$ in an attempt to force the isothermal model as the best-fit model at these redshifts, 
 it would still be an unacceptable fit, and the $\gamma=1.3$ model would still be preferred over it. 
 This is less clear-cut at $\zav = 3$, where the error bars are large enough to permit a reasonable
minimum chi-squared of $\chi^2 \approx 70$ for 71 d.o.f. using the $\gamma=1$ model, but this requires a value of 
$\fmeanlya=0.71$, which is $5\sigma$ discrepant from the value reported by \citet{becker:2013}. 
While this \fmeanlya\ measurement is dependent on corrections for metals and LLS absorption (and indeed we argue that they have
neglected a subtle bias related to SDSS quasar selection), they have attempted to incorporate these uncertainties into
their errors and we have no particular reason to believe that they have underestimated this by a factor of $>5$.
A quick survey of the available measurements on the forest mean-transmission from the past decade yield 
$\fmeanlya(z=3) \approx 0.65-0.69$ \citep{kim:2007,faucher-giguere:2008a,dallaglio:2008}, albeit with larger errors.
The use of any of these measurements as priors for our analysis would therefore disfavor an IGM with $\gamma \leq 1$, 
(which requires $\fmeanlya(z=3)\geq 0.71$), 
unless all the available literature in the field have significantly underestimated the mean-transmission.
} 

{There are several cosmological and astrophysical effects that we did not model, that could in principle 
affect our conclusions on $\gamma$. Since the \lya\ forest transmission PDF essentially measures the contrast between
high-absorption and low-absorption regions of the IGM, this can be degenerate with the underlying amplitude of matter
fluctuations which is specified by a combination of $\sigma_8$ and $n_s$, 
the matter fluctuation variance on $8\,\mpc$ scales and the slope of the amplitude power spectrum, respectively.
While these parameters are increasingly well-constrained \citep[e.g.,][]{planck-collaboration:2013}, there is still some uncertainty
regarding the level of the fluctuations on the sub-Mpc scales relevant to the \lya\ forest which could be degenerate with our 
$\gamma$ measurement. \citet{bolton:2008} explored this degeneracy between $\sigma_8$ and $\gamma$
in the context of transmission PDF measurements from high-resolution spectra, and found that the PDF is less sensitive to 
plausible changes in $\sigma_8$ compared to $\gamma$, e.g. modifying $\sigma_8$ by $\Delta \sigma_8 \pm 0.1$, 
affected the shape of the PDF less than a modification of $\Delta \gamma \pm 0.25$ (Figure~2 in their paper). 
This degeneracy is in fact further weakened when an MCMC analysis of the full parameter space is considered, as shown
by the likelihood contours in \citet{viel:2009}. 
}

{The astrophysical effects that could be degenerate with $\gamma$ include galactic winds and inhomogeneities in the
background UV ionizing field. The injection of gas into the IGM by strong galaxy outflows could in principle modify \lya\ forest
statistics at fixed $\gamma$; this was studied using 
hydrodynamical simulations by \citet{viel:2013b}, who concluded that the effect on the PDF is small compared to the uncertainties
in high-resolution PDF measurements. Our BOSS measurement has roughly the same errors as those from 
high-resolution spectra once systematic uncertainties are taken into account, therefore it seems unlikely that 
galactic winds could significantly bias our conclusions on $\gamma$.
Meanwhile, fluctuations in the UV ionizing background, $\Gamma$, that are correlated with the overall density field could also 
be degenerate with the temperature-density relationship (c.f.\ Equation~\ref{eq:fgpa}). 
This effect was studied by \citet{mcdonald:2005} in simulations using an extreme model that considered only UV background contributions from highly-biased AGN, which maximizes the inhomogeneities. 
They concluded that while these UV fluctuations affected forest transmission statistics 
at $z \sim 4$, the effect was small at $z \lesssim 3$, the redshift range of our measurements.
}

{Various observational and systematic effects could also, in principle, affect our constraints
on $\gamma$. For example, our modeling of the BOSS spectral resolution assumes a Gaussian smoothing kernel
which might affect our constraints if this were untrue. However, in their analysis of the 1D forest transmission power spectrum, 
\citet{palanque-delabrouille:2013a} examined the BOSS smoothing kernel and did not find significant deviations from Gaussianity.
There are also possible systematics caused by our simplified modeling of LLS and metal contamination in the data, 
for example in our assumption of
a single $b$-parameter for all LLSs and our neglect of very weak metal absorbers. However, we believe that the
test performed in Figure~\ref{fig:chisq_hist} samples larger differences in the transmission PDF than those 
caused by our model simplifications, e.g. it seems unlikely that going from a single LLS $b$-parameter
 to a finite $b$-distribution could 
cause to greater differences in the flux PDF than varying the single $b$-parameter by $\pm 50\%$ as was done in 
Figure~\ref{fig:chisq_hist}. 
As for continuum-estimation, we carry out the exact same continuum-fitting procedure
on the mock spectra as on the real quasar spectra, which leads no overall bias since in both cases the resulting forest
transmission field is forced to have the same overall transmission, $\fcont$. 
The only uncertainty then relates to the distribution of $c'/c-1$, i.e. the per-pixel error of the estimated continuum, 
$c'$, relative to the true continuum, $c$. In reality the shape of this distribution could be different between the data and mocks, whereas 
within our mocks framework we could only explore overall rescalings of the distribution width. 
Again, we find it unlikely that differences in the transmission PDF caused by the true shape of the $c'/c-1$ distribution 
could be so large as to be comparable to the effect caused by varying the width of the continuum error distribution, that we have examined.
}

{
While we do not think that the effects described in the previous few paragraphs qualitatively
affect our conclusion that the BOSS data is inconsistent with isothermal or inverted IGM temperature-density relationships 
($\gamma\leq 1$),
when taken in aggregate these systematic uncertainties do weaken our
formal $4-6\,\sigma$ limits against $\gamma \leq 1$ and need to explicitly considered in future analyses.
}
 
 \subsection{Astrophysical Implications}
  
How does this compare with other results on the thermal state of the IGM? 
\citet{mcdonald:2001} analyzed the transmission PDF from 8 high-resolution, 
high S/N spectra and compared with now-obsolete hydrodynamical simulations. 
They found the data to be consistent with a temperature-density
relationship (TDR) with the expected values of $\gamma \approx 1.5$ \citep{hui:1997a}.
More recently, \citet{bolton:2008} and \citet{viel:2009} carried out analyses of the transmission PDF measured 
from a larger sample (18 spectra) of \lya\ forest sightlines measured by \citet{kim:2007}
and found evidence for an inverted TDR ($\gamma < 1$).
\citet{viel:2009} found that at $z \approx 3.0$, the temperature-density relation was highly
inverted ($\gamma \approx 0.5$), and remained so as low as $z \approx 2.0$ although 
at the lower redshifts the data was marginally consistent with an isothermal 
IGM. They suggested the difference between their results and those
 of \citet{mcdonald:2001} was due to the now-obsolescent cosmological parameters
 and less-detailed treatment of intervening metals in the earlier study. 
However, \citet{lee:2012} then pointed out that there is a sensitivity of the measured values of $\gamma$
from the transmission PDF on continuum-fitting. 
Since continuum-fitting of high-resolution data generally involves manually placing the continuum at \lya\ forest transmission peaks 
which do not necessarily reach the true continuum, it is conceivable that continuum biases combined with
underestimated jacknife errors bars \citep[e.g.,][]{rollinde:2013} could have
led \citet{bolton:2008} and \citet{viel:2009}
to erroneously deduce an inverted temperature-density relation (see \citealt{bolton:2014} for a detailed discussion on
this point).
In our analysis we have fitted our continua using an automated process that is free from the same continuum-fitting
bias, although it does require an assumption on the underlying \lya\ forest transmission which we have marginalized 
over in our analysis.

Most recent measurements of the transmission PDF from high-resolution data have continued to favor
an isothermal or inverted $\gamma$ ---
 \citet{calura:2012} analyzed the transmission PDF from a
sample of $z \approx 3.3-3.8$ quasars and also found an isothermal TDR at $z =
3$, although combining with the \citet{kim:2007} data drove the
estimated $\gamma$ to inverted values at $z < 3$. 
However, \citet{rollinde:2013} carried out a re-analysis of the transmission PDF
from various high-resolution echelle data sets, which included significant overlap
with the \citet{kim:2007} data.  
They argue that previous analyses have underestimated the error on the transmission PDF, and 
found the observed transmission PDF to be
consistent with simulations that have $\gamma \approx 1.4$ over $2 < z < 3$ --- this discrepancy is probably
also driven by a different continuum-estimation from the \citet{kim:2007} measurement.

The use of other statistics on high-resolution spectra have however tended to disfavor an isothermal or
inverted TDR. \citet{rudie:2012}
analyzed the lower-end of the $b$-$\nhi$ cutoff from individual
\lya\ forest absorbers measured in a set of 15 very high-S/N quasar
echelle spectra, and estimated $\gamma \approx 1.5$ at $z = 2.4$.
\citet{bolton:2014} compared the \citet{rudie:2012} measurements to hydrodynamical 
simulations and corroborated their determination of the TDR slope. 

\citet{garzilli:2012} analyzed the \citet{kim:2007} sample and found that while the transmission PDF supports an isothermal
or inverted TDR,
a wavelet analysis favors $\gamma > 1$. 
Note, however, that the $b$-$\nhi$ cutoff and the transmission PDF are sensitive to different
density ranges, with the PDF probing gas densities predominantly below the mean \citep[e.g.,][]{bolton:2014}.

Our result of $\gamma \approx 1.6$ at $\zav = [2.3, 2.6, 3.0]$ are thus in rough agreement with 
measurements that do \emph{not} involve the transmission PDF from high-resolution \lya\ forest spectra 
\citep[with the exception of][]{rollinde:2013}. 
Our value of $\gamma$ at $\zav = 3$ is somewhat unexpected because one expects a flattening of the TDR 
close to the \heii\ reionization epoch at $z \sim 3$ (\citealt{furlanetto:2008}; \citealt{mcquinn:2009};
but see \citealt{gleser:2005, meiksin:2012}), but $\gamma=1.3$ is not strongly disfavored ($\sqrt{(\Delta \chi^2)} \sim 2.6$)

Taken at face value, the TDR during \heii\ reionization can be
made steeper by a density-independent reionization and/or a lower heating rate in the IGM \citep{furlanetto:2008},
which could be reconciled with an extended \heii\ event \citep{shull:2010, worseck:2011}.

Our constraints on $\gamma$ 
appear to be in conflict with the prediction of the theories of
\citet{broderick:2012} and \citet{chang:2012}, who elucidated a relativistic pair-beam channel for plasma-instability heating
of the IGM from TeV gamma-rays produced by a population of luminous blazars.
This mechanism provides a uniform volumetric heating rate,
which would cause an inverted 
 TDR in the IGM \citep{puchwein:2012} since voids would experience
a higher specific heating rate compared with heating by \heii\ reionization alone.
This picture has been challenged by the recent study of
\citet{sironi:2014}, who dispute the amount of heating this mechanism could provide, 
since they found that the momentum dispersion of such relativistic pair beams  
allows $\ll 10\%$ of the beam energies to be deposited into the IGM.

However, in this paper we have assumed relatively simple TDRs in which the
bulk of the IGM in the density range $0.1 \lesssim \Delta \lesssim 5$ follows a relatively tight power-law. 
We have therefore not studied more complicated $T-\Delta$ relationships, e.g., with a spread of temperatures at fixed
density \citep[e.g.,][]{meiksin:2012, compostella:2013} that might be caused by \ion{He}{2} reionization or other phenomena. 
It is therefore possible that such complicated TDRs could result in \lya\ forest transmission PDFs that mimic
the $\gamma \approx1.6$ power-law; this is something that needs to be examined in more detail in future work.

\subsection{Future Prospects}
Looking forward, the subsequent BOSS data releases will significantly
enlarge our sample size, e.g., DR10 \citep{ahn:2014} is nearly double
the size of the DR9 sample used in this paper, while the final BOSS
sample (DR12) should be three times as large as DR9.  In particular,
the newer data sets should be sufficiently large for us to analyze the
transmission PDF and constrain $\gamma$ during the epoch of
\heii\ reionization at $z > 3$. This would be a valuable measurement,
since high-resolution spectra are particularly affected by
continuum-fitting biases at these redshifts \citep{faucher-giguere:2008a, lee:2012}.

The analysis of the optically-thin \lya\ forest transmission PDF from these
expanded data sets will have vanishingly small sample errors, and the
errors will be dominated by systematic and astrophysical
uncertainties. At the high-transmission end, our uncertainties are dominated
by the scatter of the continuum-fitting, which is dominated by the
question of whether our quasar PCA templates, derived from
low-luminosity low-redshift quasars \citep{suzuki:2005}, or
high-luminosity SDSS quasars \citep{paris:2011}, respectively, are an
accurate representation of the BOSS quasars. This uncertainty should
be eliminated in the near-future by PCA templates derived
self-consistently from the BOSS data (Nao Suzuki et al. 2014, in
prep). The modelling of metal contamination could also be improved in
the near future by advances in our understanding of how metals are
distributed in the IGM \citep[e.g.,][]{zhu:2014}, although metals
are a comparatively minor contribution to the uncertainty in our transmission
PDF.

We also aim to improve on the rather
\emph{ad hoc} data analysis in this paper, in which we accounted
for some uncertainties in our modelling by incorporating them into our
error covariances (e.g., LLSs, metals, continuum errors), 
while $\fmeanlya$ was marginalized over a fixed grid. In future analyses, 
it would make sense to carry out a full Markov Chain Monte Carlo treatment
of all these parameters which would rigorously account for all the uncertainties 
and allow straightforward marginalization over nuisance parameters.

Since this paper was initially focused on modelling the BOSS
spectra, for the model comparison we used only simulations sampling a
very coarse $3 \times 3$ grid in $T_0$ and $\gamma$ parameter space,
and were unable to take account for uncertainties in other
cosmological ($\sigma_8$, $n_s$ etc) and astrophysical (e.g.,
  Jeans' scale, \citealt{rorai:2013}; or galactic winds, \citealt{viel:2013b}) parameters in our analysis.  However,
methods already exist to interpolate \lya\ forest statistics from
hydrodynamical simulations given a set of IGM and cosmological
parameters \citep[e.g.,][]{viel:2006,borde:2014,rorai:2013}. In the
near future we expect to do joint analyses using other \lya\ forest
statistics in conjunction with the transmission PDF, such as new measurements
of the small-scale ($k \gtrsim 0.2\, \mathrm{s\;km^{-1}}$) 1D transmission
power spectrum (Walther et al. 2014, in prep.), moderate-scale ($
0.002\, \mathrm{s\;km^{-1}} \lesssim k \lesssim 0.2\,
\mathrm{s\;km^{-1}}$) transmission power spectrum in both 1D
\citep[e.g.,][]{palanque-delabrouille:2013a} and 3D \citep[from
  ultra-dense \lya\ forest surveys using high-redshift star-forming galaxies,][]{lee:2014,lee:2014a}, the
phase angle probability distribution function determined from close
quasar pair sightlines \citep{rorai:2013}, and others.  Such efforts
would require a fine grid sampling the full set of cosmological and
IGM thermal parameters in order to ensure that the interpolation
errors are small compared to the uncertainties in the data \citep[see
  e.g.,][]{rorai:2013}.  Efforts are underway to utilize
massively-parallel adaptive-mesh refinement codes \citep{almgren:2013}
to generate such parameter grids to study the IGM \citep{lukic:2014}
However, one of the findings of this paper is the importance of
correct modelling of LLS, in particular partial LLS ($10^{16.5}\,\persqcm \lesssim \nhi \lesssim 10^{17.5}\,\persqcm$), in accounting for
the shape of the observed \lya\ transmission PDF.  Since our hydrodynamical
simulations did not include radiative transfer and cannot accurately
capture optically thick systems, we had to add these in an \emph{ad
  hoc} manner based on observational constraints which are currently
rather imprecise. In the near future, we would want to use
hydrodynamical simulations with radiative transfer \citep[even if only
  in post-processing, e.g.,][]{altay:2011,mcquinn:2011c, altay:2013, rahmati:2013} 
to self-consistently model
the optically-thick absorbers in
the IGM.  With the unprecedented statistical power of the full BOSS
\lya\ forest sample, this could provide the opportunity to place
unique constraints on the column-density distribution function of partial LLS.

\section{Summary/Conclusions}
In this paper, we analyzed the probability distribution function (PDF) of the \lya\
forest transmitted flux using 3393 BOSS quasar spectra (with $\langle \mathrm{S/N} \rangle \geq 6$) from Data Release 9 of the 
SDSS-III survey. 

To rectify the inaccurate noise estimates in the standard pipeline, we
 first carried a custom co-addition of the individual exposures of each spectrum, 
 using a probabilistic procedure that also separates out the signal and CCD contributions, 
 allowing us to later create mock spectra with realistic noise properties. 
 We then estimated the intrinsic quasar continuum using a mean-flux regulated 
 technique that reduces the scatter in the estimated continua by forcing the resultant 
 \lya\ forest mean transmission to match the precise estimates of \citet{becker:2013}, 
 although we had to make minor corrections on the latter to account for our different assumptions 
 on optically-thick systems in the data. 
 This now allows us to measure the transmission PDF in the data, which we do so at
 $\zav = [2.3, 2.6, 3.0]$ (with bin widths of $\Delta z = 0.3$), and split into S/N subsamples of 
 S/N = [6-8, 8-10, 10-25] at each redshift bin.
 
 The second part of the paper describe finding a transmission PDF model which describes the data, 
 based on detailed hydrodynamical simulations of the optically-thin \lya\ forest
 that sample different IGM temperature-density
 relationship slopes, $\gamma$, and temperatures at mean-density, $T_0$ (where
 $T(\Delta) = T_0 \Delta^{\gamma-1}$). Using these simulations we generate mock spectra
 based on the real spectra. These take into account the following
 instrumental and astrophysical effects:
 \begin{description}
 \item[Lyman-Limit Systems] These are randomly added into our mock spectra
 based on published incidence rates \citep{ribaudo:2011} and column-density distributions
 \citep{prochaska:2010}, including a large population of partial LLS 
 ($10^{16.5} \,\persqcm \leq \nhi \leq 10^{17.5}\,\persqcm$) with a power-law distribution of
 roughly $f(\nhi) \propto \nhi^{-2}$. 
 We assumed an effective $b=45\,\kms$ for the velocity width of these absorbers.
\item[Metal Contamination] We measure metal absorption rom the $1260\,\ang \lesssim \lambda 
 \lesssim 1390\,\ang$ restframe region of lower-redshift quasars at the same observed wavelength,
 then add these directly into our mock spectra.
 \item[Spectral Resolution and Noise] Each mock spectrum is smoothed by the dispersion vector
 of the corresponding real spectrum (determined by the BOSS pipeline), and we apply corrections 
 which bring the spectral resolution modeling to within $\sim 1\%$ accuracy. We then introduce pixel noise based on the
 noise parameters estimated by our probabilistic co-addition procedure on the real data, which also achieves
percent level accuracy on modeling the noise. 
 \item[Continuum Errors] Since we generate a full mock \lya\ forest spectrum including the simulated quasar continuum
 (based on the continua fitted to the actual data), we can apply our continuum-estimation procedure on each mock
 to fit a new continuum. The difference between the new continuum and the underlying simulated
 quasar continuum yields an estimate of the continuum error.
 \end{description}
 
 We then compare the model transmission PDFs with the data, using an error covariance that includes
 both bootstrap errors and systematic uncertainties in the model components described above. 
 At $\zav = 3.0$ we find a discrepancy in the assumed \lya\ forest mean-transmission, $\fmeanlya$, between our
 data and that derived from \citet{becker:2013}, which we argue is likely caused by a selection bias 
 in the SDSS quasars used by the latter. 
 We therefore marginalize out these uncertainties in $\fmeanlya$ to obtain our final results.
 
The models with an IGM temperature-density slope of $\gamma=1.6$ give the best-fit 
 to the data at all our redshift bins ($\zav = [2.3, 2.6, 3.0]$). Models with an isothermal or inverted
 temperature-density relationship ($\gamma \leq 1$) are disfavored at the 
 $\sqrt{(\Delta \chi^2)} = [3.9, 6.3, 4.0]\sigma$ at $\zav = [2.3, 2.6, 3.0]$, respectively.
 Due to a degeneracy with our LLS model, we are unable to put robust
 constraints on $T_0$ but we have checked that our conclusions on $\gamma$ are robust
 to such systematics {as can be considered within our model framework. 
 There are other possible systematics we did not consider that could in principle affect our measurement,
  such as cosmological parameters ($\sigma_8$, $n_s$) and astrophysical effects (galactic winds, 
  inhomogeneous UV ionizing background), but we argue that these are unlikely to qualitatively
  affect our conclusions.
 }

\acknowledgements{
We thank Michael Strauss, J.\ Xavier Prochaska, Gabor Worseck, and Joop Schaye for
useful comments and discussion.  We also thank the members of the
ENIGMA group (\url{http://www.mpia-hd.mpg.de/ENIGMA/}) at the Max Planck
Institute for Astronomy (MPIA) for helpful discussions.
JFH. acknowledges generous support from the Alexander von Humboldt
foundation in the context of the Sofja Kovalevskaja Award. The
Humboldt foundation is funded by the German Federal Ministry for
Education and Research. The hydrodynamic simulations in this work were
performed using the COSMOS Supercomputer in Cambridge (UK), which is
sponsored by SGI, Intel, HEFCE and the Darwin Supercomputer of the
University of Cambridge High Performance Computing Service
(\url{http://www.hpc.cam.ac.uk/}), provided by Dell Inc.\ using
Strategic Research Infrastructure Funding from the Higher Education
Funding Council for England. COSMOS and DARWIN are part of the DIRAC
high performance computing facility funded by STFC.  MV is supported
by the FP7 ERC grant ``cosmoIGM'' GA-257670, PRIN-MIUR and INFN/PD51
grants.  JSB acknowledges the support of a Royal Society University
Research Fellowship. BL acknowledges support from the NSF Astronomy and Astrophsics Fellowship grant AST-1202963.

Funding for SDSS-III has been provided by the Alfred P. Sloan Foundation, the Participating Institutions, the National Science Foundation, and the U.S. Department of Energy Office of Science. The SDSS-III web site is \url{http://www.sdss3.org/}.

SDSS-III is managed by the Astrophysical Research Consortium for the Participating Institutions of the SDSS-III Collaboration including the University of Arizona, the Brazilian Participation Group, Brookhaven National Laboratory, University of Cambridge, Carnegie Mellon University, University of Florida, the French Participation Group, the German Participation Group, Harvard University, the Instituto de Astrofisica de Canarias, the Michigan State/Notre Dame/JINA Participation Group, Johns Hopkins University, Lawrence Berkeley National Laboratory, Max Planck Institute for Astrophysics, Max Planck Institute for Extraterrestrial Physics, New Mexico State University, New York University, Ohio State University, Pennsylvania State University, University of Portsmouth, Princeton University, the Spanish Participation Group, University of Tokyo, University of Utah, Vanderbilt University, University~of~Virginia, University~of~Washington, and Yale~University.   
}
\\
\bibliographystyle{apj}

\bibliographystyle{apj}
\bibliography{lyaf_kg,lss_galaxies,my_papers,apj-jour}

\appendix

In this Appendix, we describe our probabilistic procedure for combining
the multiple BOSS exposures of each spectrum\footnote{Defined as unique combinations of plate number, fiber number and MJD of observation.} 
while simultaneously estimating the noise variance in terms of a parametrized model. 
We assume the noise in each pixel can be described by
\beq 
\sigma_{\lambda i}^2 = A_1 \hat{S}_{\lambda i}\left(\mathcal{F}_{\lambda}+ s_{\lambda_i}\right) + 
 A_2 \hat{S}_{\lambda i}^2\sigma^2_{\rm RN, eff} \sigma_\mathrm{disp}(\lambda) \label{eqn:mcmc_noise2}
\eeq
where 
\beq 
\hat{S}_{\lambda i}=S_{\lambda i} \left(1 - \exp(-A_3 \lambda + A_4) \right) \label{eqn:calib_wave}.
\eeq
The true object flux $\mathcal{F}_\lambda$ and $A_{j=1-4}$ are noise parameters which 
we will determine given the individual exposure spectra $f_{\lambda,i}$, sky flux estimates $s_{\lambda,i}$, 
and calibrations vectors $S_{\lambda,i}$ (which convert between detector counts and photons).
$\sigma_{\rm RN,eff}$ is the effective read noise which we fixed to $\sigma_{\rm RN,eff}=12$; this can be
thought of as an effective number of pixels times the true read
noise of the CCD squared, which we multiplied by the spectrograph dispersion $\sigma_\mathrm{disp}(\lambda)$
to approximately account for the change in spot-size as a function of wavelength.
Equation~\ref{eqn:calib_wave} parametrizes wavelength-dependent biases in the calibration vector.

We search for the model that best describes the multiple exposure
spectra $f_{\rm \lambda i}$, where our model parameters are $A_j$
from Eq.~(\ref{eqn:mcmc_noise2}) and $\mathcal{F}_{\lambda}$ is the true flux of
the object. In what follows, we will outline a method for determining
the posterior distribution $P(A_j, \mathcal{F}_{\lambda} |f_{\lambda i})$
using a Markov Chain Monte Carlo (MCMC) method. From this 
distribution, we can obtain both an accurate model for the noise via
Eq.~(\ref{eqn:mcmc_noise2}), and our final combined spectrum. The
estimates for $A_j$ can also be used to self-consistently
generate pixel noise in mock \lya\ forest spectra.

The probability of the data given the model, or the likelihood, can be written
\begin{eqnarray}
L(A,\mathcal{F}_{\lambda}) &=& P\left(f_{\lambda i} \vert A_j, \mathcal{F}_{\lambda}\right) \nonumber \\
&=& \prod_{\lambda i}\frac{1}{\sqrt{2\pi}\sigma_{\lambda i}} 
\exp{\left(\frac{\left(f_{\lambda i}-\mathcal{F}_{\lambda}\right)^2}{2\sigma_{\lambda i}^2}\right)}.\label{eqn:lhood}
\end{eqnarray}

Note that individual exposure data $f_{\lambda i}$ are on the native
wavelength grid of each CCD exposure, whereas the BOSS pipeline interpolates and
then combines these individual spectra into a final co-added spectrum,
defined on a wavelength grid with uniform spacing.  Furthermore,
flexure and other variations in the spectrograph wavelength solution
will result in small (typically sub-pixel) shifts between the
individual exposure wavelength grids. In Eq.~(\ref{eqn:lhood}) our
model $\mathcal{F}_{\lambda}$ must be computable at every wavelength $f_{\rm
  \lambda i}$ of the individual exposures. We are free to choose the
wavelengths at which $\mathcal{F}_{\lambda}$ is represented, but this choice is
a subtle issue for several reasons. First, note that we want to avoid
interpolating the data, $f_{\lambda i}$, onto the model wavelength grid, as this would
correlate the data pixels, and require that we track covariances in
the likelihood in Eq.~(\ref{eqn:lhood}), making it significantly more
complicated and challenging to evaluate. Similarly, it is undesirable
to interpolate our model $\mathcal{F}_{\lambda}$, as this would introduce
correlations in the model parameters, making it much more difficult to
sample them with our MCMC.  Finally, note that $\mathcal{F}_{\lambda}$ also
represents our final co-added spectrum, so we might consider opting for
a a uniform wavelength grid, similar to what is done by the BOSS
pipeline. Our approach is to simply determine the model flux
$\mathcal{F}_{\lambda}$ at each wavelength of the individual exposures
$f_{\lambda i}$. Shifts among the individual exposure wavelength grids
result in a more finely sampled model grid.  For the reasons explained
above, we use nearest grid point (NGP) interpolation, so that the
$f_{\lambda i}$ are evaluated on the $\mathcal{F}_{\lambda}$ grid (and vice
versa) by assigning the value from the single nearest pixel.

In our MCMC iterations, we use the standard Metropolis-Hastings criterion to
sample the parameters $A_j$, with trials drawn from a uniform prior. For
the $\mathcal{F}_{\lambda}$, we exploit an analogy with Gibbs
sampling, which dramatically simplifies MCMC for likelihood functions
with a multivariate Gaussian form. Gibbs sampling exploits the fact
that given a multivariate distribution, it is much simpler to sample from
conditional distributions than to integrate over a joint
distribution. To be more specific, the likelihood in Eq.~(\ref{eqn:lhood}) 
is proportional to the joint probability distribution of the noise parameters $A_j$ and 
$\mathcal{F}_{\lambda}$, but it is also proportional to the conditional probability 
distribution of the $\mathcal{F}_{\lambda}$ at fixed $A_j$. With $A_j$ fixed the
probability of $\mathcal{F}_{\lambda}$ is then
\be
P(\mathcal{F}_{\lambda}| A, f_{\lambda i}) \propto   
\prod_{\lambda i}\frac{1}{\sqrt{2\pi}\sigma_{\lambda i}} 
\exp{\left(\frac{\left(f_{\lambda i}-\mathcal{F}_{\lambda}\right)^2}{2\sigma_{\lambda i}^2}\right)} \label{eqn:gauss}, 
\ee
which is very nearly a multivariate Gaussian distributions for $\mathcal{F}_{\lambda}$ 
with a diagonal covariance matrix. The equation above slightly deviates from a 
Gaussian because the $\sigma_{\lambda i}$ depend on $\mathcal{F}_{\lambda}$ via 
Eq.~(\ref{eqn:mcmc_noise2}). In what follows, we ignore this small deviation, 
and assume that the conditional PDF of the $\mathcal{F}_{\lambda}$ (at fixed $A_j$) is Gaussian. 

Given that Eq.~\ref{eqn:gauss} is a multivariate Gaussian with
diagonal covariance, the Gibbs sampling of the $\mathcal{F}_{\lambda}$ becomes
trivial.  Since, Eq.~(\ref{eqn:gauss}) can be factored into a product
of individual Gaussians, we need not follow the standard Gibbs sampling
algorithm, whereby each parameter is updated sequentially holding the others
fixed.  Instead we need only hold $A_j$ fixed (since the likelihood is
not Gaussian in these parameters), \emph{and we can sample all of the $\mathcal{F}_{\lambda}$
simultaneously}. This simplification, which dramatically speeds up the
algorithm, is possible because the conditional distribution
for $\mathcal{F}_{\lambda}$ can be factored into a product of Gaussians
for each pixel $\mathcal{F}_{\lambda}$, thus the conditional distribution
at any wavelength is completely independent of all the others. 

Completing the square in Eq.~(\ref{eqn:gauss}) we can then write
\be 
P(\mathcal{F}_{\lambda}| A_j, f_{\lambda i}) \propto \prod_{\lambda} 
\exp{\left(\frac{\left(f_{{\rm opt},\lambda}-\mathcal{F}_{\lambda}\right)^2}{2\sigma_{{\rm opt},\lambda}^2}\right)} 
\ee 
where
\be
f_{{\rm opt},\lambda} \equiv 
\frac{1}{\sigma^2_{{\rm opt},\lambda}} \sum_{i} \frac{f_{\lambda i}}{\sigma_{\lambda i}^2}
\,\,\,\,\,{\rm and}\,\,\,\,\,\,
\frac{1}{\sigma_{{\rm opt},\lambda}^2} \equiv \sum_i 
\frac{1}{\sigma_{\lambda i}^2}\label{eqn:opt}.
\ee
The expressions above for $f_{\rm opt,\lambda}$ and $\sigma^2_{\rm opt,\lambda}$ 
simply represent the \emph{optimally} combined flux estimator and the 
resulting variance. Thus one can think of our MCMC algorithm as performing
an optimal combination of the individual exposure spectra $f_{\lambda i}$, whereby the noise is simultaneously 
determined via an iterative procedure. 

\smallskip
Thus the basic steps of our algorithm can be summarized as follows:

\begin{itemize}
\item Initialize, by creating a model $\lambda$ grid from all unique 
  wavelengths in the individual exposures, and use NGP interpolation to assign 
  a $f_{\rm \lambda i}$ to this grid for each exposures. 

\item Choose a starting guess for noise parameters $A_j$. For the starting $\mathcal{F}_{\lambda}$
  use $\mathcal{F}_{\lambda} = f_{\rm opt,\lambda}$ from Eq.~(\ref{eqn:opt}),
  but with the model $\sigma_{\lambda i}$ replaced by the noise
  delivered by the pipeline
    
\item Begin the MCMC loop:

\begin{enumerate} 

\item Use the current values of $A_j$ and $\mathcal{F}_{\lambda}$ to compute the
  variance $\sigma^2_{\lambda i}$ for each exposure via Eq.~(\ref{eqn:mcmc_noise2}). 

\item Compute $f_{\rm opt,\lambda}$ and $\sigma^2_{\rm opt,\lambda}$ from Eq.~\ref{eqn:opt}. 

\item Take a Gibbs step for each wavelength of $\mathcal{F}_{\lambda} = f_{\rm opt,\lambda} + g_{\lambda}\sigma_{\rm opt,\lambda}$ simultaneously, where $g_{\lambda}$ is a 
vector of unit variance Gaussian deviates. 

\item Use NGP to interpolate the model $\mathcal{F}_{\lambda}$ onto each individual
  exposure $f_{\lambda i}$ wavelength grid. 

\item Compute the likelihood $L(A_j, \mathcal{F}_{\lambda})$ according to
  Eq.~(\ref{eqn:lhood})

\item Take trial steps in the $A_j$ according to $A_{j,\rm try} =  A_j + g_j dA_j$, 
where $dA_j$ is a stepsize and $g_j$ is a Gaussian deviate between zero and one, 
drawn for each individual noise parameter $A_j$.

\item Compute the likelihood at $L(A_{j,\rm try}, \mathcal{F}_{\lambda})$

\item Apply the Metropolis-Hastings criteria to the likelihood difference. 
  If it is satisfied then accept the values of $A_j$ as part of the Markov 
  chain. If not, then use the previous values. Note that the $\mathcal{F}_{\lambda}$
  are always accepted, because they are Gibbs sampled. 
\end{enumerate}

\item Use only the second half of the chain for the posterior 
  distributions, as the first half is the burn in phase. 

\end{itemize}

Our MCMC algorithm directly determines the posterior 
distribution $P(A_j, \mathcal{F}_{\lambda}| f_{\lambda i})$, 
which provides all the information we need to construct mock spectra using
Eq.~(\ref{eqn:mcmc_noise2}) as described in \S\ref{sec:noisemodel}. 

The distribution of $P(\mathcal{F}_{\lambda}|f_{\lambda i})$, on the other hand, contains 
everything we need to know about the combined spectrum. Namely, we can define 
\be
{\bar{\mathcal{F}}_{\lambda}} \equiv \int P(\mathcal{F}_{\lambda}|f_{\lambda i})
\mathcal{F}_{\lambda} d\mathcal{F}_{\lambda} 
\ee 
as the combined spectrum, and 
\be
\sigma^2_{\lambda} \equiv \int P(\mathcal{F}_{\lambda}|f_{\lambda
  i})(\mathcal{F}_{\lambda} - {\bar{\mathcal{F}}_{\lambda}})^2 d\mathcal{F}_{\lambda}  \label{eqn:extract}
\ee 
as its variance. If the formal noise returned by BOSS pipeline were
actually the true noise in the data, then our ${\bar{\mathcal{F}}_{\lambda}}$ in
Eq.~(\ref{eqn:extract}) would be equivalent to the optimally combined
noise and our variance the optimal variance, i.e.  according to
Eq.~(\ref{eqn:opt}). In practice, the BOSS pipeline does not return
the true noise and so our ${\bar{\mathcal{F}}_{\lambda}}$ is optimal whereas the
pipeline flux is sub-optimal, and our $\sigma^2_{\lambda}$ is an empirical
estimate of the actual noise in the data. 

\end{document}